\documentclass{article}
\usepackage{float}
\usepackage{amssymb}
\usepackage[utf8]{inputenc}
\usepackage{blindtext}
\usepackage{amsmath}
\usepackage{verbatim}
\usepackage{graphicx}
\usepackage{caption}
\usepackage{subcaption}
\usepackage{xcolor}

\providecommand{\p}{\partial}

\providecommand{\si}{\sigma_i}
\providecommand{\g}{g^{1/4}}

\providecommand{\braket}[1]{\langle #1\rangle}
\providecommand{\com}[1]{\left[#1\right]}

\newcommand{\address}{\\ \textit{Instituto de Ciencias Nucleares}\\  \textit{Universidad Nacional Aut\'onoma de M\'exico,  CDMX 04510, M\'exico}
\\ $^*$ \textit{sergio1@ciencias.unam.mx} \\
$^{**}$ \textit{javier.cano@correo.nucleares.unam.mx}\\ 
$^{***}$ \textit{joshua.davy@correo.nucleares.unam.mx}\\
$^\dag$ \textit{jascao90@gmail.com}\\
$^\ddag$\textit{vergara@nucleares.unam.mx}}
\newcommand{\hstry}{\\ Received (Day Month Year)\\
Revised (Day Month Year)}
\usepackage[left=2.5cm,right=2.5cm,top=2cm,bottom=2.5cm]{geometry}
\usepackage{titling}
\begin{document}

\markboth{Davy-Castillo, Cano-Arango, Ju\'arez, Austrich-Olivares, Vergara}
{Quantum Geometry of the parameter space: A proposal for curved materials }



\title{\textbf{Quantum geometry of the parameter space: a proposal for curved systems}}

\author{JOSHUA DAVY-CASTILLO$^{***}$, JAVIER A. CANO-ARANGO$^{**}$,
\\ SERGIO B. JU\'AREZ$^*$,  JOAN A. AUSTRICH-OLIVARES$^\dag$, and  J. DAVID VERGARA$^\ddag$\\
\address\\
\hstry}
\date{}

\maketitle
\vspace{-1cm}
\begin{abstract}
In this paper, we extend the quantum geometric tensor for parameter-dependent curved spaces to higher dimensions, and introduce an equivalent definition that generalizes the Zanardi, et al, formulation of the tensor. The parameter-dependent metric modifies the behavior of both the quantum metric tensor and Berry curvature in a purely geometric way. Our focus is on understanding the distinctions in higher dimensions that emerge when using the generalized tensor compared to the conventional one. Through a comparative analysis, illustrated with examples in two dimensions, we highlight unique quantum geometric properties for both the quantum metric tensor and the Berry curvature. Additionally, we explore differences between analytical and perturbative approaches in solving the problems. 
\end{abstract}

\textbf{Keywords}: Quantum Geometric Tensor; Parameter dependent curvature; Berry Curvature; Quantum Information Geometry

\section{Introduction}

In recent years, the geometric properties of quantum information have gained relevance as a tool to analyse certain quantum phenomena, such as quantum phase transitions \cite{Carollo2020,Sachdev}, entanglement \cite{Marmo1,Somma}, separability  of quantum systems \cite{Simon2000,Peres1996,HORODECKI1997} and quantum metrology \cite{Toth,Liu2020}. One of the core concepts used to explore all these properties is the quantum geometric tensor (QGT), which contains all the information related to the parameter space geometry of a physical system. It is composed of two parts: the real part is the quantum metric tensor (QMT) \cite{Provost,Zanardi2007Information,Sarkar2012}, and the imaginary one is proportional to the Berry curvature, which is related to quantum interference \cite{Berry1985,chruscinski2012geometric,GuReview}.  

Particularly for quantum phase transitions, which occur at absolute zero temperature by varying a physical parameter (e.g. an external magnetic field), the QGT has become an exceptional tool in their detection, since it exhibits divergent behavior at points of the parameter space where energy-level crossings occur \cite{Zanardi2007Information,Niu2017,Sondhi}.

 Experimentally, the QGT has been completely determined \cite{Tan2019,Yu2020}, using it to characterize the geometry and topology of tunable superconducting circuits \cite{Tan2019} and has enabled an evaluation of the quantum Cramér-Rao bound \cite{Cappellaro2022}.

On the other hand, there are systems where the physical geometry depends on certain parameters, such that distinct configurations change the properties of these systems,  particularly those corresponding to the electronic structure. One example is the magic-angle twisted bilayer graphene (MATBG), which consists of two sheets of graphene stacked on top of each other with a slight misalignment of approximately $1.1^\circ$. In this setup, MATBG exhibits superconducting properties \cite{Oh2021,Jaoui2022,Tarnopolsky2019}. The bilayer graphene undergoes a quantum phase transition in configurations close to this magic angle as it transforms from a weakly correlated Fermi liquid to a strongly correlated two-dimensional electron system, making it highly sensitive to its parameters \cite{Andrei2020}. As a result, MATBG is an ideal material for investigating strongly correlated phenomena \cite{Cao2018}. Curvature-induced quantum phase transitions can also occur in electron-hole systems. Introducing periodic curvature to these structures makes it possible to transition from a topologically trivial state to a non-trivial state \cite{Siu2018}.

In a recent publication \cite{Joan2022}, the authors construct a generalization of the QGT to a parameter-dependent curved space by considering a metric that depends on the system's parameters. Then, the QGT acquires additional terms due to modifying the inner product. These terms not only change the real part of the quantum geometric tensor but also include a modification of Berry's curvature. This article introduces additional examples of calculating the quantum metric tensor and the Berry curvature in systems with parameter-dependent metrics in higher dimensions. In addition, we show how it is possible to obtain the equivalent of a perturbative formula for the QGT \cite{Zanardi2007Information}, in this context.

This paper is divided into five sections. Section 2 describes the formulation of the QGT in a parameter-dependent curved space, and we obtain the novel perturbative form for the QGT in this setting. Section 3 introduces the general groundwork for quantization in a curved space. Section 4 considers several examples, including (i) the ground state of symmetrically coupled Toda oscillators, studied both in an analytical and perturbative way; (ii) an anharmonic oscillator coupled with a Toda oscillator; and (iii) a system with non-vanishing Berry curvature, corresponding to a Toda oscillator coupled to a gauge potential. We conclude in Section 5.

\section{The quantum geometric tensor in parameter-dependent curved space}\label{pera}

The quantum geometric tensor in a curved space background is a complex tensor exceptionally well-suited for describing systems with parameter-dependent geometries. In the general case, we consider an $N$-dimensional configuration space $(x^\mu), \ \mu=1,\dots, N$ with an $m$-dimensional parameter space $(\lambda^i), \ i=1,\dots, m$. Then, in this space, the  inner product of quantum mechanics is given by
\begin{equation}\label{curvedinnerprod}
	\braket{\g\phi|\g\psi} \equiv \int_{Vol}d^N x \left(\g(x,\lambda)\phi(x,\lambda)\right)^* \left(\g(x,\lambda)\psi(x,\lambda)\right), 
\end{equation}
where $g(x,\lambda)=\det[g_{\mu\nu}(x,\lambda)]$ is the determinant of the $N$-dimensional configuration space metric.\\
As mentioned in \cite{Joan2022},  when considering only pure states, the components of the quantum geometric tensor are 
\begin{equation}\label{QGT}
	\begin{split}
		\mathcal{G}_{i j} = & \braket{\partial_i  \psi|\partial_j \psi}-\braket{\partial_i  \psi|\psi}\braket{\psi|\partial_j \psi}-\frac{1}{4}\braket{\psi|\sigma_i |\partial_j \psi}\\
		& -\frac{1}{4}\braket{\partial_i  \psi|\sigma_j |\psi}+\frac{1}{4}\braket{\sigma_i }\braket{\psi|\partial_j \psi}+\frac{1}{4}\braket{\sigma_j }\braket{\partial_i  \psi|\psi} \\
		& +\frac{1}{16}\braket{\sigma_i \sigma_j } -\frac{1}{16}\braket{\sigma_i }\braket{\sigma_j }.
	\end{split}
\end{equation}
where the factors of $g^{1/4}$ have been omitted and $\si$ is defined as\footnote{The expectation value must be read as: $\braket{\si}=\braket{\g\psi|\si|\g\psi}$.}
\begin{equation} \label{eq:sigma}
    \si = g_{\mu\nu}\p_i g^{\mu\nu}.
\end{equation}
This quantity, which we shall call the \textit{deformation vector}, arises solely due to the curvature of the spatial metric and is responsible for the extra terms that appear in the generalization of the QGT.\\

It is essential to note that the derivative with respect to the parameters of the normalization condition, $\braket{\g\psi|\g\psi} = 1$, is modified in the following manner:
\begin{equation}\label{eq:norm_cond_mod}
	\p_i\braket{\g\psi|\g\psi}  = \braket{\g\p_i\psi|\g\psi} + \braket{\g\psi|\g\p_i\psi} - \frac{1}{2}\braket{\si} = 0,
\end{equation}
giving rise to the modified Berry connection, whose components are
\begin{equation}\label{eq:modBerryconn}
    \beta_i = -i \braket{\g\psi| \g\partial_i\psi} + \frac{i}{4}\braket{\si}.
\end{equation}
From \eqref{eq:norm_cond_mod} one can verify that $\si$ and the Berry connection, $\beta_i$, are real. First, for $\si$:
\begin{equation}
\begin{split}
   \frac{1}{2}\braket{\si} & = \braket{g^{1/4}\psi | g^{1/4}\p_i \psi} + \braket{g^{1/4}\p_i\psi | g^{1/4}\psi},        \\
        & =\braket{g^{1/4}\psi | g^{1/4}\p_i \psi} + \braket{g^{1/4}\psi | g^{1/4}\p_i \psi}^*,  \\
        & = 2\operatorname{Re}\left[\braket{g^{1/4}\psi | g^{1/4}\p_i \psi}\right].
\end{split}
\end{equation}
Second, for the Berry connection, we take the difference with respect to its complex conjugate:
\begin{equation}
\begin{split}
    \beta_i^{*} - \beta_i & = i\braket{\psi | \partial_i \psi} +  i\braket{\partial_i\psi|\psi} - \frac{i}{2}\braket{\sigma_i} , \\
        & = 0.
\end{split}
\end{equation}
Moreover, we emphasize that the Berry connection is a 1-form in parameter space due to the transformation rule of its components
\begin{equation}
	\beta'_i = \frac{\p \lambda^j}{\partial \lambda'_i}\beta_j.
\end{equation}
This connection defines the Berry curvature by taking its exterior derivative in parameter space
\begin{equation}
	\mathcal{F} = \mathrm{d}\beta,
\end{equation}
whose components are $\mathcal{F}_{i j} = \partial_i  \beta_j  - \partial_j \beta_i $, we must notice that in this case, the Berry curvature corresponds to an Abelian group.

Another definition of the quantum geometric tensor was proposed as an expansion by Zanardi et al. \cite{Zanardi2007Information}; this definition is quite convenient to compute the QGT on a perturbative basis. Then, it is interesting to generalize this definition to the case in question, where the space metric depends explicitly on the system's parameters. To do so, we need to note that after the modification of the inner product \eqref{curvedinnerprod}, and  Schrödinger equation
\begin{equation}\label{eq:curvedevalue}
\sqrt{g}\hat{H}|\psi_m\rangle  = E_m\sqrt{g}|\psi_m\rangle,
\end{equation}
and the identity operator takes the following form
\begin{equation}\label{eq:curvedidentity}
    \mathbb{I} = \sum_{m}\left| g^{1/4}\psi_m\right\rangle\left\langle g^{1/4} \psi_m\right|.
\end{equation}

Now, due to the dependence of the metric on the variables of the configuration space, the time evolution of the metric determinant may not be constant, i.e. $\com{\sqrt{g},\hat{H}}\neq 0$. The same applies to the deformation vector $\si$. Therefore, taking into account \eqref{eq:curvedevalue} and \eqref{eq:curvedidentity}, the Zanardi, et al expansion for the QGT becomes:
\begin{equation}
\begin{aligned}
\mathcal{G}^{(n)}_{i j}  =  \sum_{\substack{m,\\ m\neq n}}&\frac{1}{(E_m-E_n)^2}\bigg[\left(\langle\psi_n|\sqrt{g}(\partial_i\hat{H})|\psi_m\rangle - \langle \partial_i\psi_n|[\sqrt{g},\hat{H}]|\psi_m\rangle\right)\\
&\times \left(\langle\psi_m|\sqrt{g}(\partial_j\hat{H})|\psi_n\rangle + \langle\psi_m|[\sqrt{g},\hat{H}]|\partial_j\psi_n\rangle\right)\\
&  +\frac{1}{4}\left(\langle \psi_n|[\sqrt{g}\sigma_i,\hat{H}]|\psi_m\rangle\right)\left(\langle\psi_m|\sqrt{g}(\partial_j\hat{H})|\psi_n\rangle + \langle\psi_m|[\sqrt{g},\hat{H}]|\partial_j\psi_n\rangle\right) \\
& -\frac{1}{4}\left(\langle \psi_m|[\sqrt{g}\sigma_j,\hat{H}]|\psi_n\rangle\right) \left.\left(\langle\psi_n|\sqrt{g}(\partial_i\hat{H})|\psi_m\rangle - \langle\partial_i\psi_n|[\sqrt{g},\hat{H}]|\psi_m\rangle\right)\right.\\
&  -\frac{1}{16}\langle \psi_n|[\sqrt{g}\sigma_i,\hat{H}]|\psi_m\rangle\langle \psi_m|[\sqrt{g}\sigma_j,\hat{H}]|\psi_n\rangle\bigg].
\end{aligned}\label{zanardi_joshua}
\end{equation}
It is essential to remark that, in this case, the derivatives with respect to the parameters of the eigenfunctions are present, which were not included in the original formulation. We must note that the expression \eqref{zanardi_joshua} reduces to the formulation of Zanardi et al. \cite{Zanardi2007Information}, in the case that $[\sqrt{g}\sigma_i, H]=0$, and $[\sqrt{g}, H]=0$, i. e. when the metric is independent of the coordinates and parameters as it should.  The expression \eqref{zanardi_joshua} retains the core property of Zanardi's expansion, meaning that it can be evaluated perturbatively by choosing a suitable basis to rewrite the eigenfunctions $|\psi_m\rangle$ making it extremely useful for numerical computations.

\section{General setting to quantum mechanics in a curved space}

From \eqref{QGT} we can see that the construction of the QGT is strongly dependent on the knowledge of the wave function of our system. In general, we begin with a Lagrangian of the form
\begin{equation}
	\mathcal{L}= \frac{1}{2}g_{\mu\nu}(x,\lambda)\dot{x}^\mu \dot{x}^\nu - V(x,\dot{x},\lambda),
\end{equation}
which describes the motion of a particle in curved space. We can divide the potential as
\begin{equation}\label{potve}
    V(x,\dot{x},\lambda) = F_\mu(x,\lambda) \dot{x}^\mu + V_1(x,\lambda). 
\end{equation}
This could allow us to incorporate an electromagnetic interaction naturally.
Then, the momenta are
\begin{equation}
    p_\mu = g_{\mu\nu}\dot{x}^\nu - F_\mu(x,\lambda),
\end{equation}
and the Hamiltonian takes the form 
\begin{equation}
	\mathcal{H}= \frac{1}{2}g^{\mu\nu}(p_{\mu}+F_\mu)(p_{\nu}+F_\nu) + V_1(x,\lambda) .
\end{equation}
In the coordinate representation, the quadratic part in the momenta of the Hamiltonian is given directly in terms of the Laplace-Beltrami operator \cite{jost2017},
\begin{equation}\label{Laplace_Beltrami}
	g^{\mu\nu}p_{\mu}p_{\nu}  \to  \nabla^2 \psi = \frac{1}{\sqrt{g}}\frac{\partial}{\partial x^\nu}\left(\sqrt{g}g^{\mu\nu}\frac{\partial \psi}{\partial x^\mu} \right). 
\end{equation}
On the other hand, for the linear part, we need to fix an ordering prescription to obtain a complete and normalizable solution for the system \cite{Margalli2015,Margalli2020}. We will explicitly see how to do so on Sec. \ref{sec:generalizado_exponencial}, but first, we will illustrate our work with two other examples that do not require this condition, aiming to be able to construct a complete Toda chain \cite{Toda1975}.   

\section{Examples}

\subsection{Symmetrically coupled Toda oscillators}\label{sec:osc_acop_expo}

For our first example, we will study a system that is a first approach to a Toda model \cite{Poland2016,2020conformal} in a two-dimensional configuration space $\vec{x}=(x, y)$ and a four-dimensional parameter space $\vec{\lambda}=(k,\kappa,\lambda,\beta)$. This model is described by the Lagrangian
\begin{equation}\label{eq:Sym_Toda_L}
    L(\vec{x},\vec{\lambda})=\frac{1}{2}[g_{\mu\nu}\dot{x}^\mu \dot{x}^\nu-k(e^{-2\lambda x} + e^{-2\beta y})-\kappa (e^{-\lambda x} - e^{-\beta y})^2],
\end{equation}
and the Hamiltoninan
\begin{equation}\label{eq:Sym_Toda_H}
H(\vec{x},\vec{\lambda})=\frac{1}{2}[g^{\mu\nu}p_\mu p_\nu +k(e^{-2\lambda x} + e^{-2\beta y})+\kappa(e^{-\lambda x} - e^{-\beta y})^2],
\end{equation}
whose metric is
\begin{equation}\label{eq:Sym_Toda_metric}
g_{\mu \nu}=\left(\begin{array}{cc}
\lambda^2 e^{-2 \lambda x} & 0\\
0 & \beta^2 e^{-2 \beta y} 
\end{array}\right).
\end{equation}

Given that the metric \eqref{eq:Sym_Toda_metric} explicitly depends on the parameters $\lambda$ and $\beta$, we are able to obtain its deformation vector \eqref{eq:sigma}, 

\begin{equation}\left(\begin{array}{c}
\sigma_k \\
\sigma_\kappa \\
\sigma_\lambda\\
\sigma_\beta
\end{array}\right)=\left(\begin{array}{c}
0 \\
0 \\
-2\lambda^{-1}+2x \\
-2\beta^{-1}+2y
\end{array}\right).
\end{equation}

To explicitly see the quantum evolution of this system, we obtain the time-independent Schrödinger equation from the Laplace-Beltrami operator \eqref{Laplace_Beltrami}:
\begin{align} \label{eq:Schrodinger_sym_exp_osc}
     &E_n \psi_{n} (\vec{x}) =\\ \nonumber
     &\left[-\frac{1}{2\sqrt{g}}\frac{\partial}{\partial x^\nu}\left(\sqrt{g}g^{\mu\nu}\frac{\partial}{\partial x^\mu} \right) +\frac{k}{2}(e^{-2\lambda x} + e^{-2\beta y})+\frac{\kappa}{2}(e^{-\lambda x} - e^{-\beta y})^2\right]\psi_{n} (\vec{x}). 
\end{align}

To solve \eqref{eq:Schrodinger_sym_exp_osc}, we introduce a change of variables:
\begin{equation}\label{eq:canonical_u1u2}
    U_1 = e^{-\lambda x}, \qquad U_2 = e^{-\beta y}, \qquad P_1 = -\frac{e^{\lambda x}}{\lambda} p_1, \qquad P_2 = -\frac{e^{\beta y}}{\beta}p_2.
\end{equation}
It can be easily verified that they correspond to a canonical transformation. In addition, we introduce the normal coordinates given by
\begin{equation}\label{eq:can_trans_upum}
    U_{\pm} = \frac{1}{\sqrt{2}}\left(U_1 \pm U_2  \right), \qquad  \qquad P_{\pm}=\frac{1}{\sqrt{2}}\left(P_1 \pm P_2 \right),
\end{equation}
yielding the Hamiltonian of two uncoupled harmonic oscillators 
\begin{equation}
    H=1/2\left[P_+^{2} + P_-^2 + \omega_+^2 U_+^2 + \omega_-^2 U_-^2\right],
\end{equation}
with frequencies
\begin{equation}
	\omega_{+}^2=k,
\end{equation}
\begin{equation}
	\omega_{-}^2=k+ 2\kappa.
\end{equation}
Thus, the solution for each uncoupled system is given by
\begin{equation}
	\psi_{n_\pm}(U_\pm) = \mathcal{N}_{n_\pm} \exp\left(-\frac{\omega_\pm  U_\pm^2}{2\hbar}\right)H_n\left(\sqrt{\frac{\omega_\pm}{\hbar}}U_\pm\right),
\end{equation}
where $H_n(\xi)=(-1)^n e^{\xi^2} \frac{d^n}{d \xi^n} e^{-\xi^2}$ are the Hermite polynomials, and $\mathcal{N}_n$ is a normalization constant which will be derived later for the ground state. Moreover, each of the energy eigenvalues is the same as those for the two-dimensional harmonic oscillator:
\begin{equation}\label{eq:En-osc}
	\mathrm{E}_{n_+,n_-} = \hbar\omega_+\left(n_+ +\frac{1}{2}\right) + \hbar\omega_-\left(n_- +\frac{1}{2}\right).
\end{equation}
Therefore, the solution for time-independent Schrödinger equation \eqref{eq:Schrodinger_sym_exp_osc} is
\begin{equation}\label{eq:eigenfunc-osc}
    \psi_{n_+,n_-}(U_+,U_-)=\psi_{n_+}(U_+)\psi_{n_-}(U_-),
\end{equation}
where $n_\pm = 0,1,2,...$ are the quantum numbers for each uncoupled oscillator. For this case, the Berry curvature is automatically zero since there is no imaginary term in the quantum states $\psi_n(\vec{x})$. Thus, the QGT becomes identical to the QMT.

This system's ground state wave function is\footnote{For more details on how to get this equation, see the Appendix A.} 
\begin{equation}
    \psi_0(x,y) = \mathcal{N}_0\exp\left(-\frac{\omega_1 e^{2\lambda x} + \omega_2 e^{2\beta y}}{2} - \gamma e^{\lambda x+\beta y}\right),
\end{equation}
with the normalization constant given by
\begin{equation}\label{eq:OAESA_psi0}
    \mathcal{N}_0=   \left[\frac{[k(k+2\kappa)]^{1/4}}{\arctan\left([\frac{2\kappa}{k} + 1]^{1/4}   \right)}  \right]^{1/2}
\end{equation}
and $\gamma =\left( \sqrt{k}-\sqrt{k+2\kappa}\right)/2$, $\omega_1 = \omega_2= \left(\sqrt{k} + \sqrt{k+2 \kappa}\right)/2=\left(\omega_+ + \omega_-\right)/2$.
It is essential to highlight that while the system's energies are equivalent to those of the system composed of two coupled harmonic oscillators, its parameter dependence is widely different. This distinction translates into notable differences in the behavior of the quantum geometric tensor. Additionally, Appendix A shows that the integration region and the normalization constant are also modified.

We will begin by presenting the complete analytical solution for the QGT, followed by the perturbative approximation solution up to second order. This sequence vividly illustrates the distinctions between the two methods and unveils the system's non-perturbative characteristics.

\subsubsection{Analytical solution}

From the generalization of the QGT given by \eqref{QGT}, we can see that in addition to knowing the expectation values of the deformation vector \eqref{eq:sigma}, it is required to calculate several integrals with respect to the parameters that involve the derivatives of \eqref{eq:OAESA_psi0}.  Most of the integrals used to compute the QGT of this system are of the form
\begin{equation}\label{eq:integralTodaModel}
    \int\limits_0^\infty\int\limits_0^\infty dxdy \ e^{-ay+bxy-cy^2}f(x,y)\log[x]\log[y]
\end{equation}
with $f(x,y)$ a polynomial function of $x$ and $y$. To obtain a solution it is necessary to use the identity
$ \log[z]=\lim_{\epsilon\to\infty}\epsilon(z^{1/\epsilon}-1)=\lim_{\varepsilon\to 0}(z^\varepsilon-1)/\varepsilon $
which transforms them into
\begin{equation}
    \int\limits_0^\infty\int\limits_0^\infty \ dx dy \lim_{\epsilon\to 0}\lim_{\varepsilon\to 0}\frac{e^{-ay+bxy-cy^2}F(x,y)}{\epsilon\varepsilon}.
\end{equation}
\begin{figure}[!h]
\begin{subfigure}{.22\textwidth}
  \centering
  \includegraphics[width=.8\linewidth]{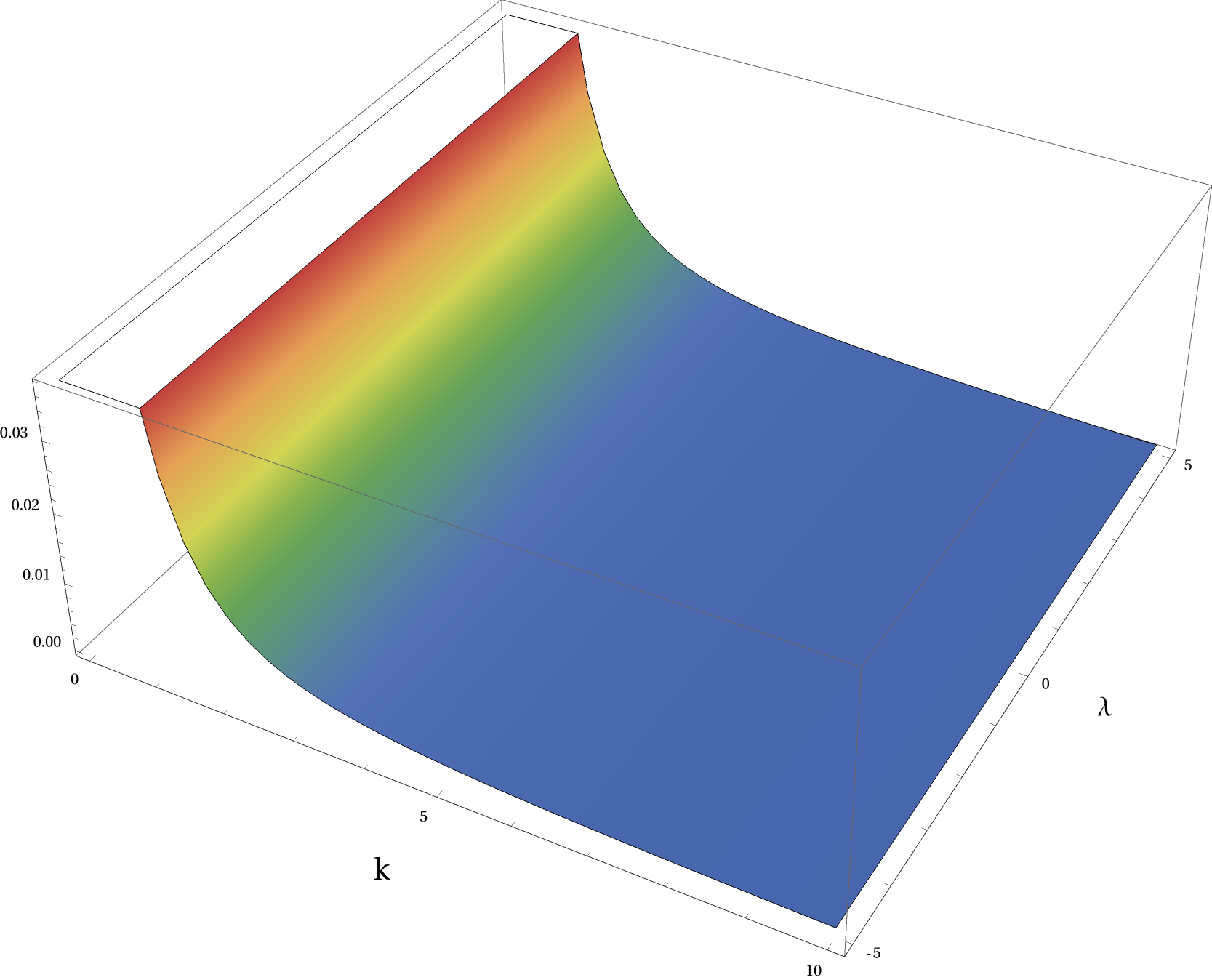}
  \caption{$G_{kk}$}
  \label{fig:acople1}
\end{subfigure}
\hfill
\begin{subfigure}{.22\textwidth}
  \centering
  \includegraphics[width=.8\linewidth]{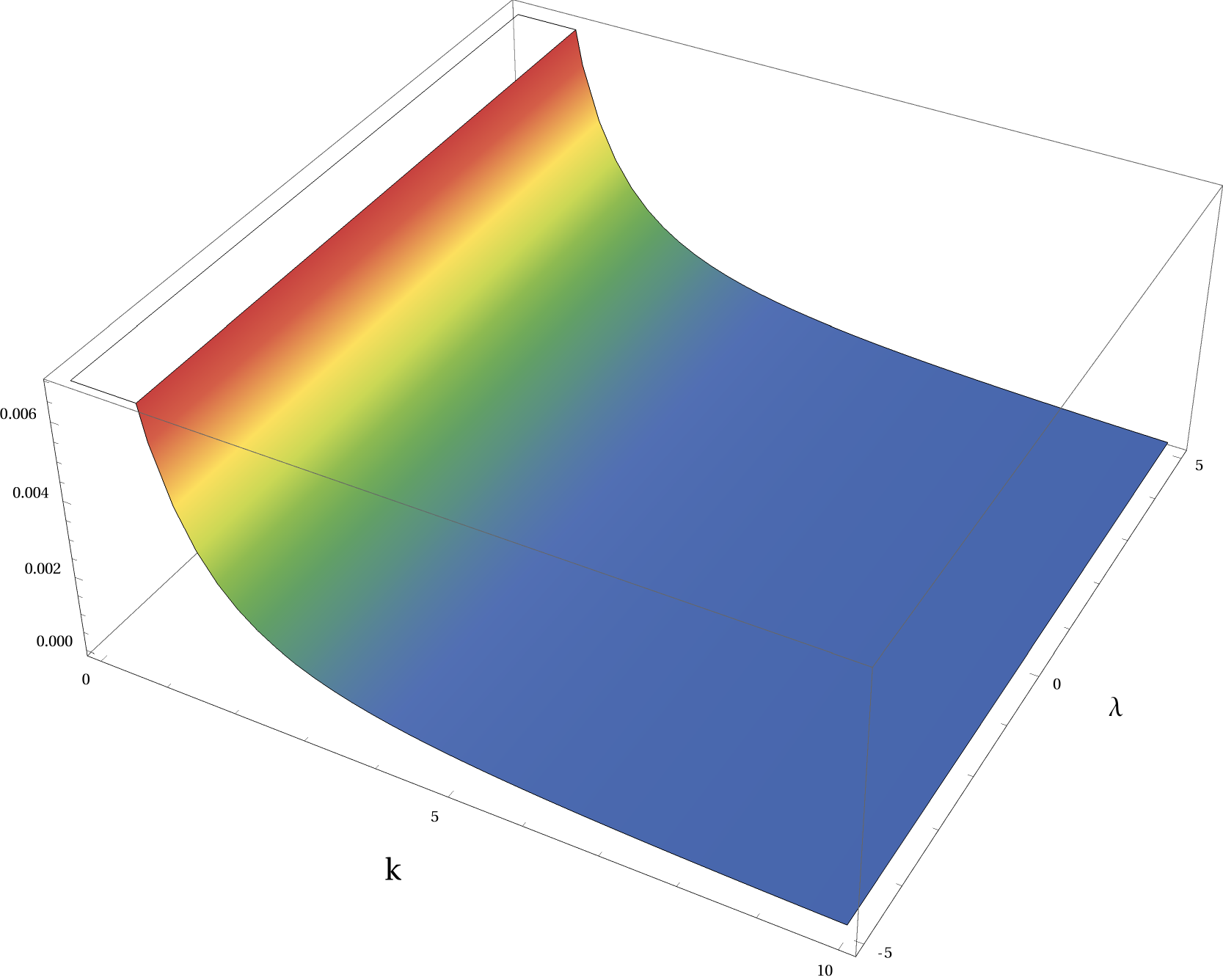}
  \caption{$G_{k\kappa}$}
  \label{fig:acople2}
\end{subfigure}
\hfill
\begin{subfigure}{.22\textwidth}
  \centering
  \includegraphics[width=.8\linewidth]{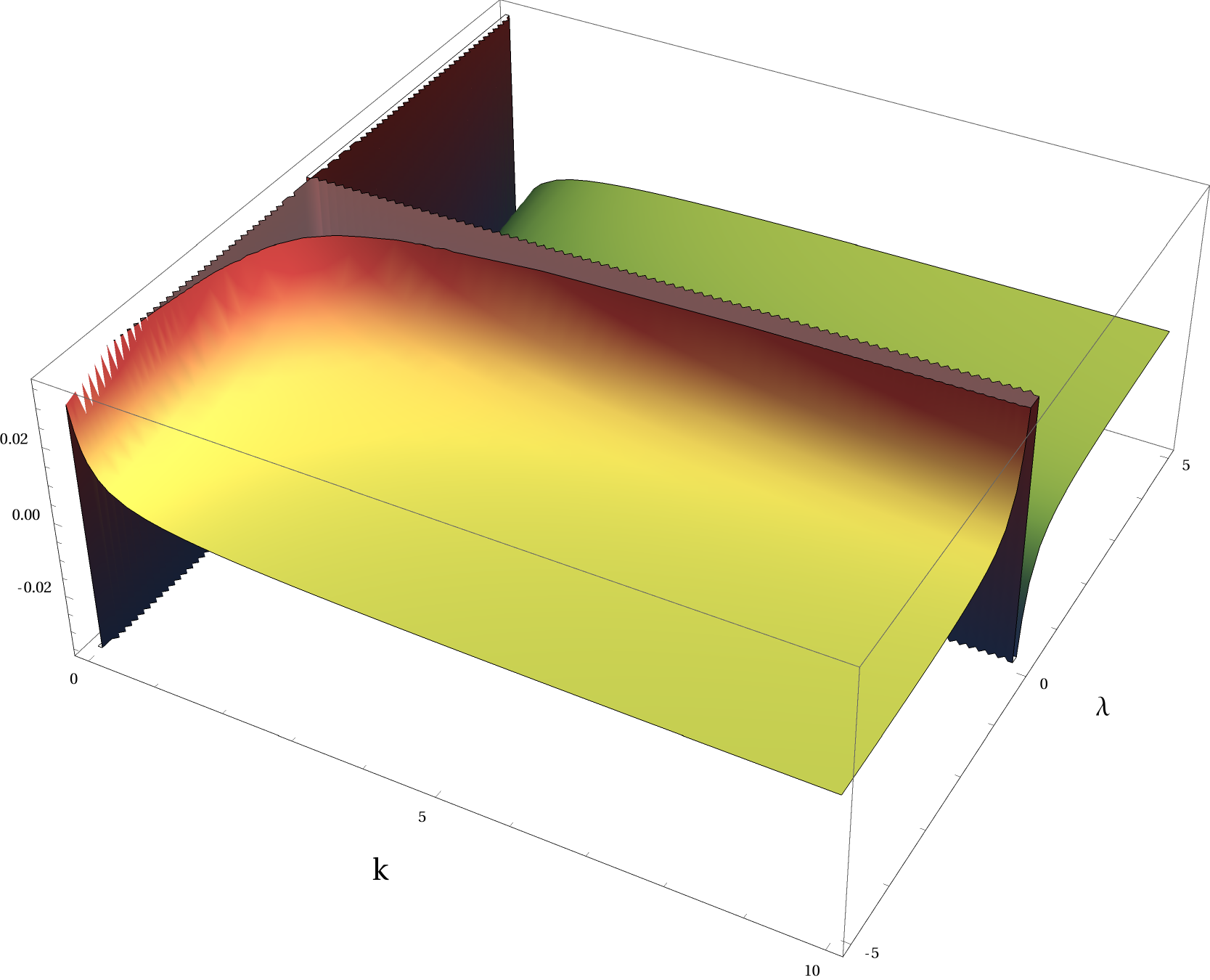}
  \caption{$G_{k\lambda}$}
  \label{fig:acople3}
\end{subfigure}
\\
\begin{subfigure}{.22\textwidth}
  \centering
  \includegraphics[width=.8\linewidth]{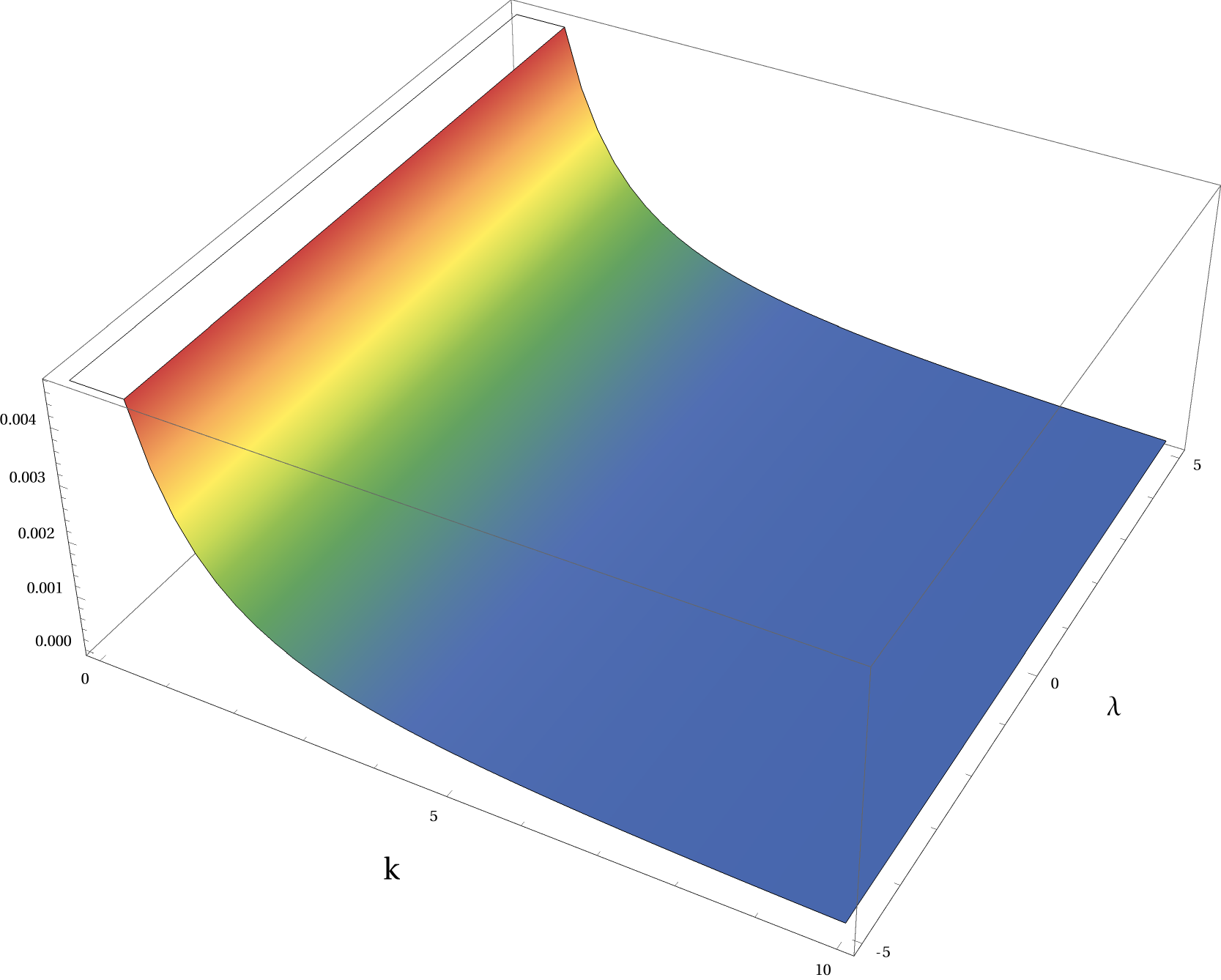}
  \caption{$G_{\kappa\kappa}$}
  \label{fig:acople4}
\end{subfigure}
\hfill
\begin{subfigure}{.22\textwidth}
  \centering
  \includegraphics[width=.8\linewidth]{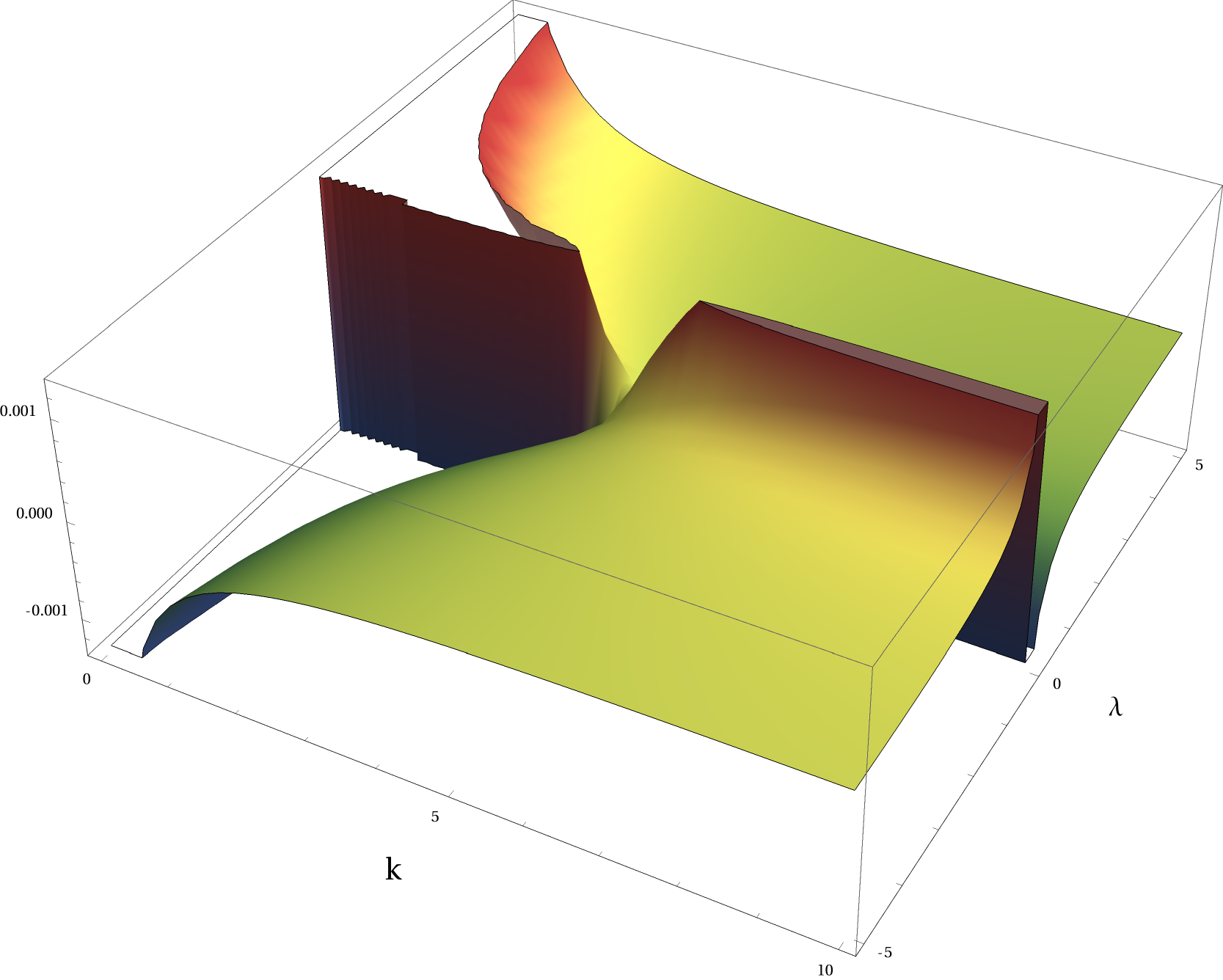}
  \caption{$G_{\kappa\lambda}$}
  \label{fig:acople5}
\end{subfigure}
\hfill
\begin{subfigure}{.22\textwidth}
  \centering
  \includegraphics[width=.8\linewidth]{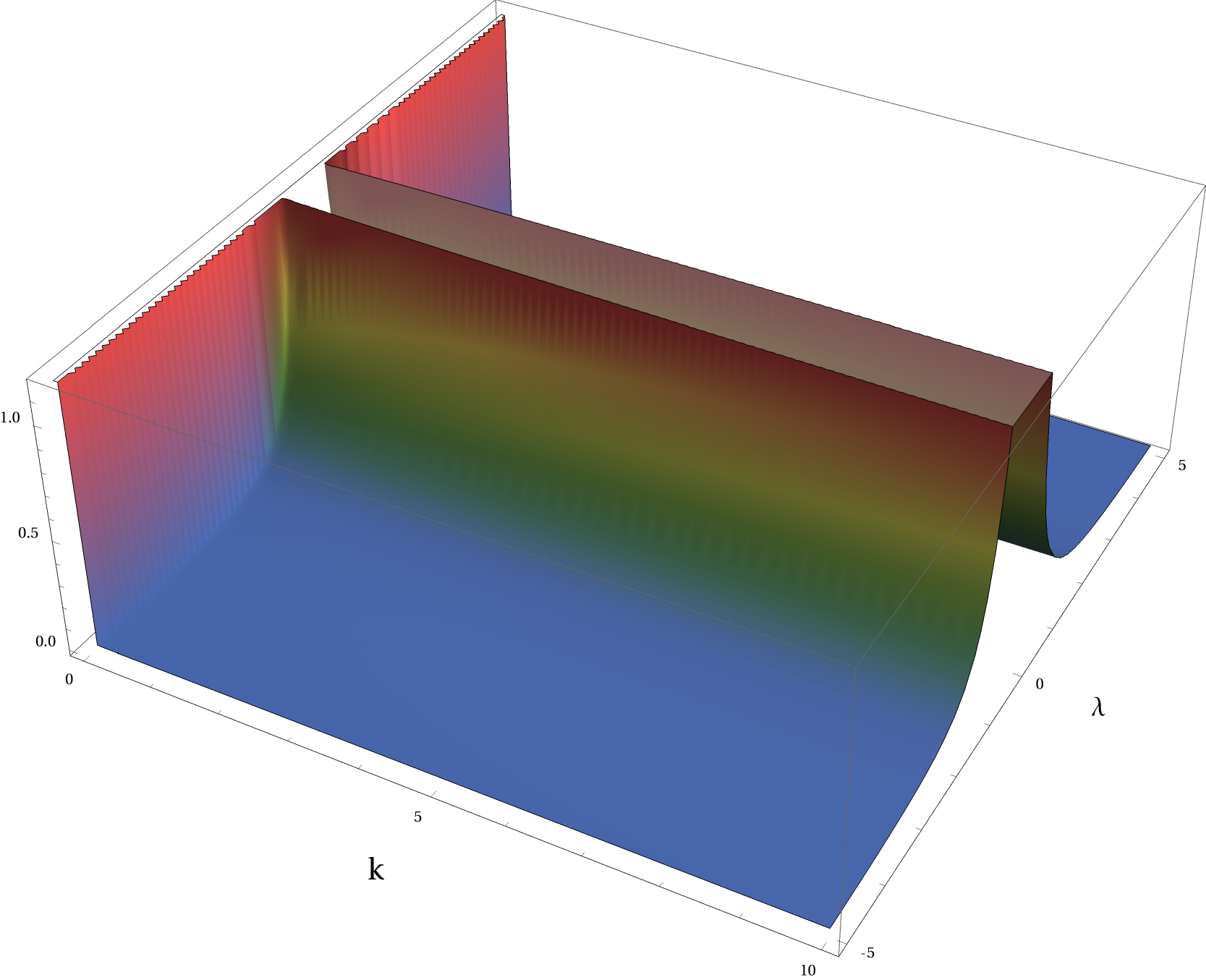}
  \caption{$G_{\lambda\lambda}$}
  \label{fig:acople6}
\end{subfigure}
\\
\begin{subfigure}{.22\textwidth}
  \centering
  \includegraphics[width=.8\linewidth]{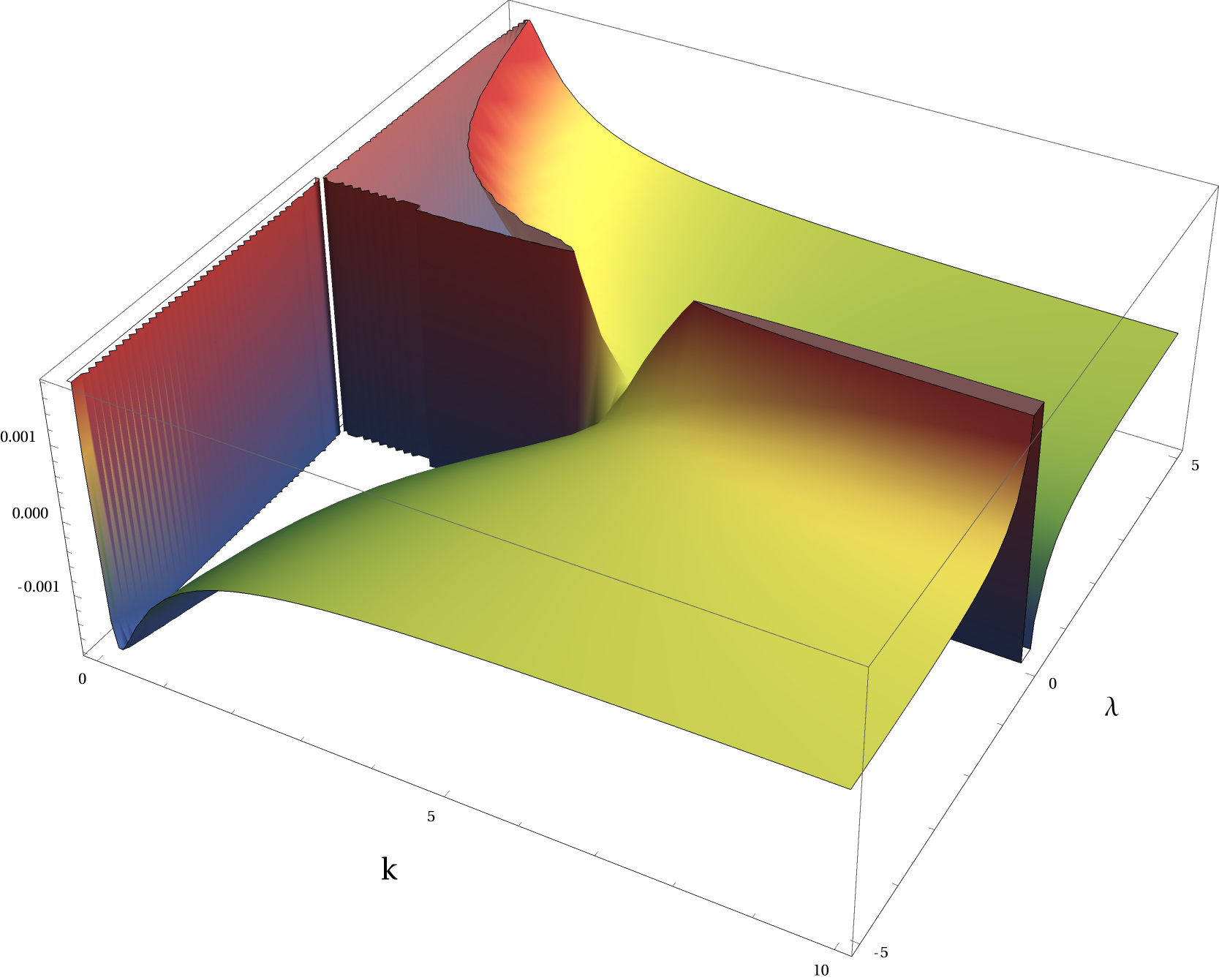}
  \caption{$G_{\lambda\beta}$}
  \label{fig:acople7}
\end{subfigure}
\hfill
\begin{subfigure}{.22\textwidth}
  \centering
  \includegraphics[width=.8\linewidth]{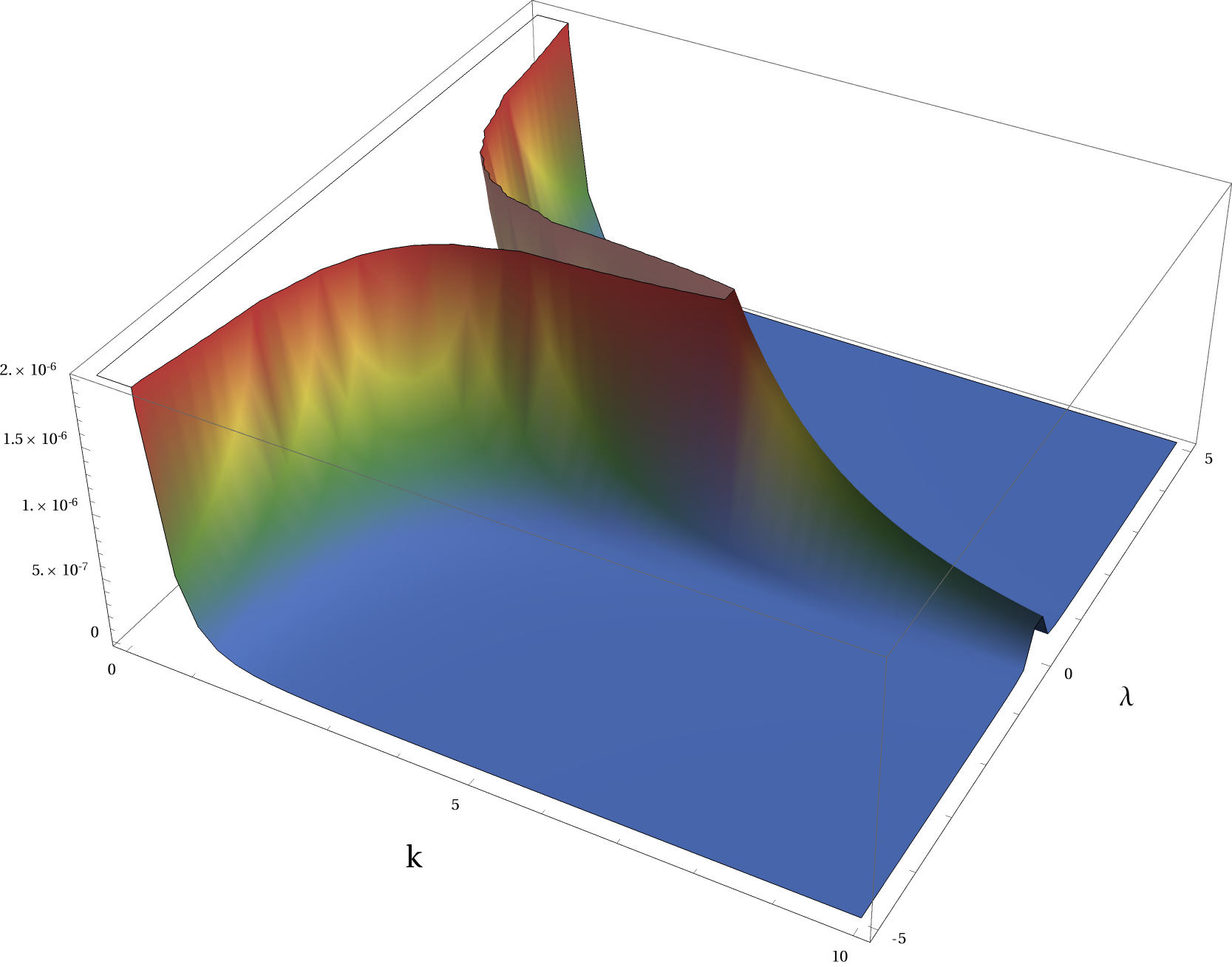}
  \caption{$\det[G]$}
  \label{fig:acople8}
\end{subfigure}
\hfill
\begin{subfigure}{.22\textwidth}
  \centering
  \includegraphics[width=.8\linewidth]{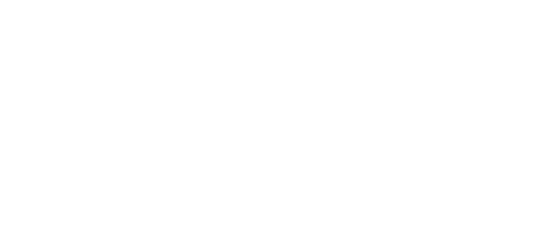}
\end{subfigure}
\caption{QGT components of the symmetrically coupled Toda oscillators in terms of $(k,\,\lambda)$, with $\kappa=\beta=1$.}
\label{fig:acopleexp}
\end{figure}
\newline
We used $\log[x]=\lim_{\epsilon\to 0}(x^\epsilon-1)/\epsilon$ and $\log[y]=\lim_{\varepsilon\to 0}(y^\varepsilon-1)/\varepsilon$.
Now $F(x,y)$ is a polynomial function allowing us to utilize the Lebesgue's dominated convergence theorem \cite{measure} to obtain:
\begin{equation}
    \lim_{\epsilon\to 0}\lim_{\varepsilon\to 0}\int\limits_0^\infty\int\limits_0^\infty dxdy \ \frac{e^{-ay+bxy-cy^2}F(x,y)}{\epsilon\varepsilon}.
\end{equation}
It should also be noted that for this particular system, it is easier to work in terms of the canonical transformation \eqref{eq:canonical_u1u2} in order to avoid dealing with the extra terms corresponding to the determinant of the metric. For more details on this, refer to Appendix A.

Although it is possible to attain the analytical solution of the QGT for every entry, the expressions are enormous and there is no point in stating the equations explicitly. Therefore, in Fig. \ref{fig:acopleexp}, we present the graphs of the projections of the entries of the QGT in terms of $(k,\lambda)$ while fixing $(\kappa=\beta=1)$. These figures visually represent the essential characteristics of the tensor and demonstrates its dependence on various parameters, which allow us to qualitatively analyze the structure of parameter space. In Appendix B we show other projections, specifically the one corresponding to $(\kappa,\lambda)$ dependence, useful for studying  entanglement in the system.

\begin{figure}[!h]
  \centering
  \includegraphics[width=0.6\linewidth]{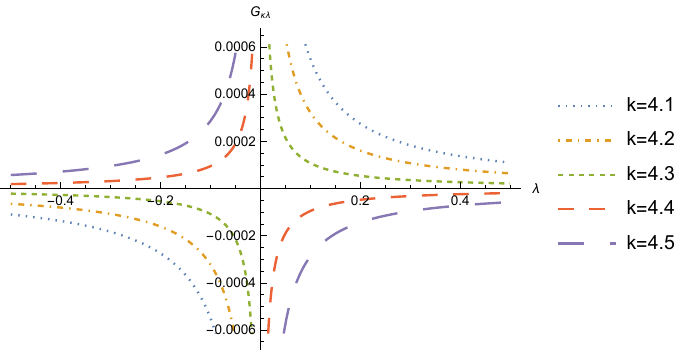}
  \caption{Projection of the entry $G_{\kappa\lambda}$ (Fig. \ref{fig:acople5}) taking $\kappa, \beta \to 1$, varying  $\lambda$ while setting $k$ to values close to the inversion of the divergence switch}
  \label{fig:2DGkplamvarlam}
\end{figure}

Let us now turn our attention to the entries $G_{\kappa\lambda}$ and $G_{\lambda\beta}$, which behave similarly. The behavior of $G_{\kappa\lambda}$ setting $\kappa,\beta\to 1$ is illustrated in Fig. \ref{fig:acople5}. We observe that although the entries diverge as $\lambda$ approaches zero, there is a reversal in the direction of the divergence. This change becomes more evident when we examine the two-dimensional projections shown in Fig. \ref{fig:2DGkplamvarlam}, where we vary $\lambda$ while keeping $k$ close to the switching value.

\begin{figure}[!h]
  \centering
    \includegraphics[width=0.6\linewidth]{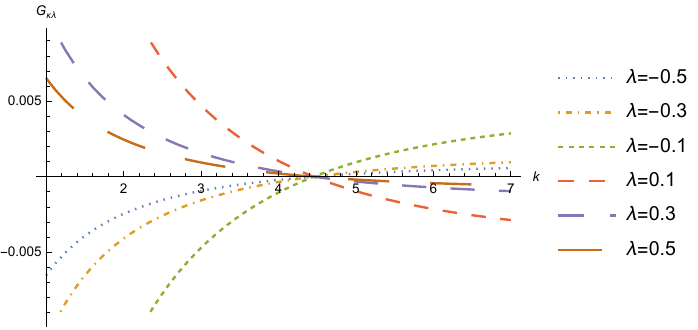}
    \caption{Projection of the entry $G_{\kappa\lambda}$ (Fig. \ref{fig:acople5}) taking $\kappa, \beta \to 1$, varying $k$ while setting $\lambda$ to values close to the inversion of the divergence.}
  \label{fig:2DGkplamvark}
\end{figure}

Conversely, when we vary $k$ and set $\lambda$ to values close to zero, as shown in Fig. \ref{fig:2DGkplamvark}, we can observe that the value $k=4.452452\dots$ becomes the focal point of interest.

It should be noted that when $\lambda\to 0$, the system changes drastically in resemblance to a quantum phase transition. In this limit, the Lagrangian given by \eqref{eq:Sym_Toda_L} then transforms into
\begin{equation}\label{eq:lambda_0_L_SymToda}
    L(y,\vec{\lambda})=\frac{1}{2}[\beta^2 e^{-2\beta y}\dot{y}^2 -k(1 + e^{-2\beta y})-\kappa (1 - e^{-\beta y})^2],
\end{equation}
and after the canonical transformation \eqref{eq:canonical_u1u2} the Hamiltonian is
\begin{equation}\label{eq:lambda 0 H SymToda}
    H=\frac{1}{2}\left[(\kappa + k) U_2^2 + P_2^{2} -2 \kappa U_2  + (\kappa + k)\right],
\end{equation}
which resembles a harmonic oscillator with a linear term 
\begin{equation}
    H= 1/2 [Xq^2 + Z p^2] +Wq,
\end{equation}
where $X=(\kappa + k)$, $Z=1$ and $W=-\kappa$. The solution, taking into account our parameters, is now: 
\begin{equation}
\psi_n(U_2)=\left(\kappa + k\right)^{1 / 8} \chi_n\left[\left(U_2-\frac{\kappa }{\kappa + k}\right) \left(\kappa + k\right)^{1 / 4}\right],
\end{equation}
with $\chi_n(\xi)=\left(2^n n ! \sqrt{\pi}\right)^{-1 / 2} e^{-\xi^2 / 2} H_n(\xi)$ the Hermite functions, and $H_n(\xi)$ once again the  Hermite polynomials.

\begin{figure}[!h]
\begin{subfigure}{.3\textwidth}
  \centering
  \includegraphics[width=\linewidth]{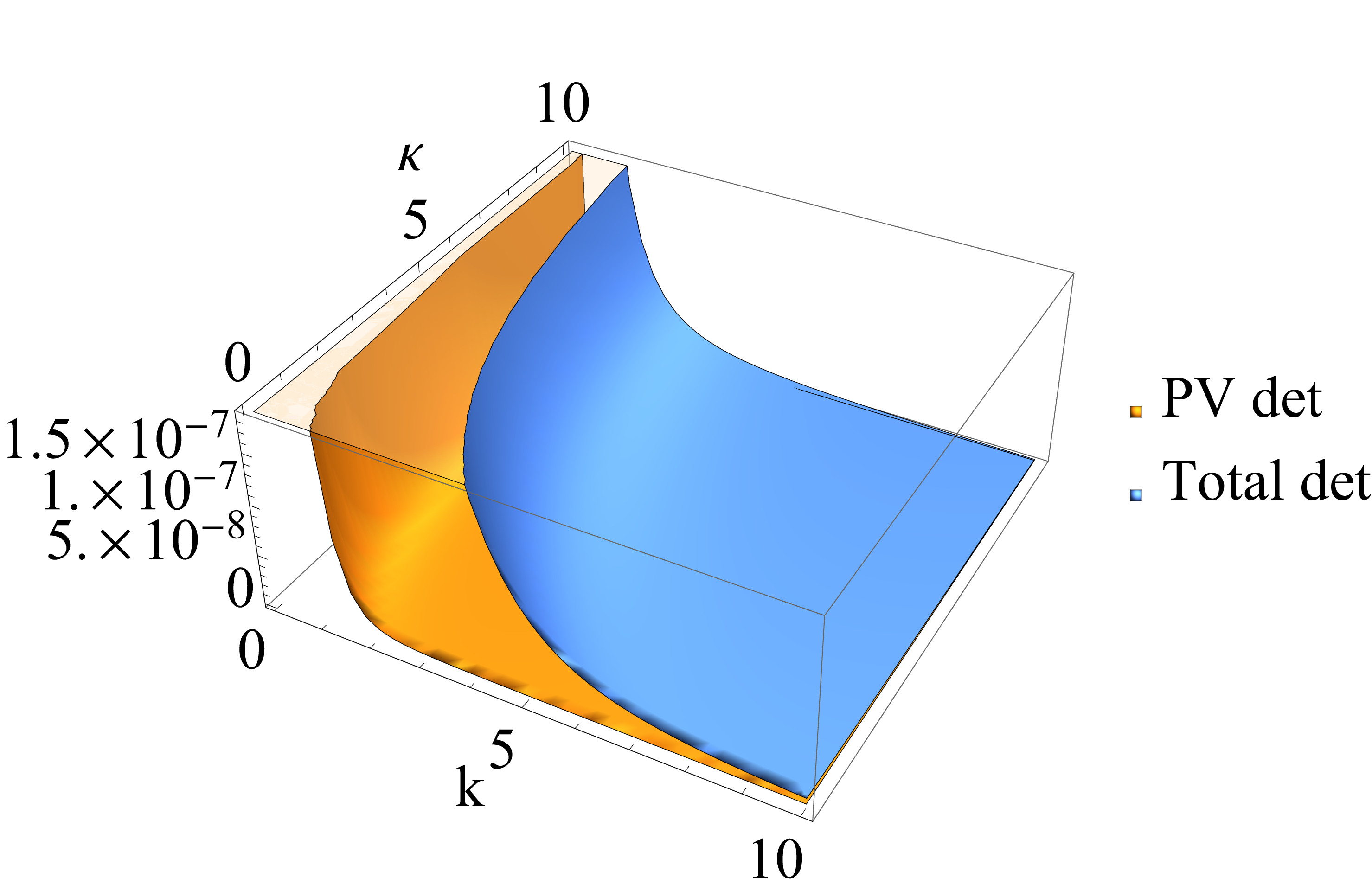}
  \caption{$\lambda, \beta \to 1$}
\end{subfigure}
\hfill
\begin{subfigure}{.3\textwidth}
  \centering
  \includegraphics[width=\linewidth]{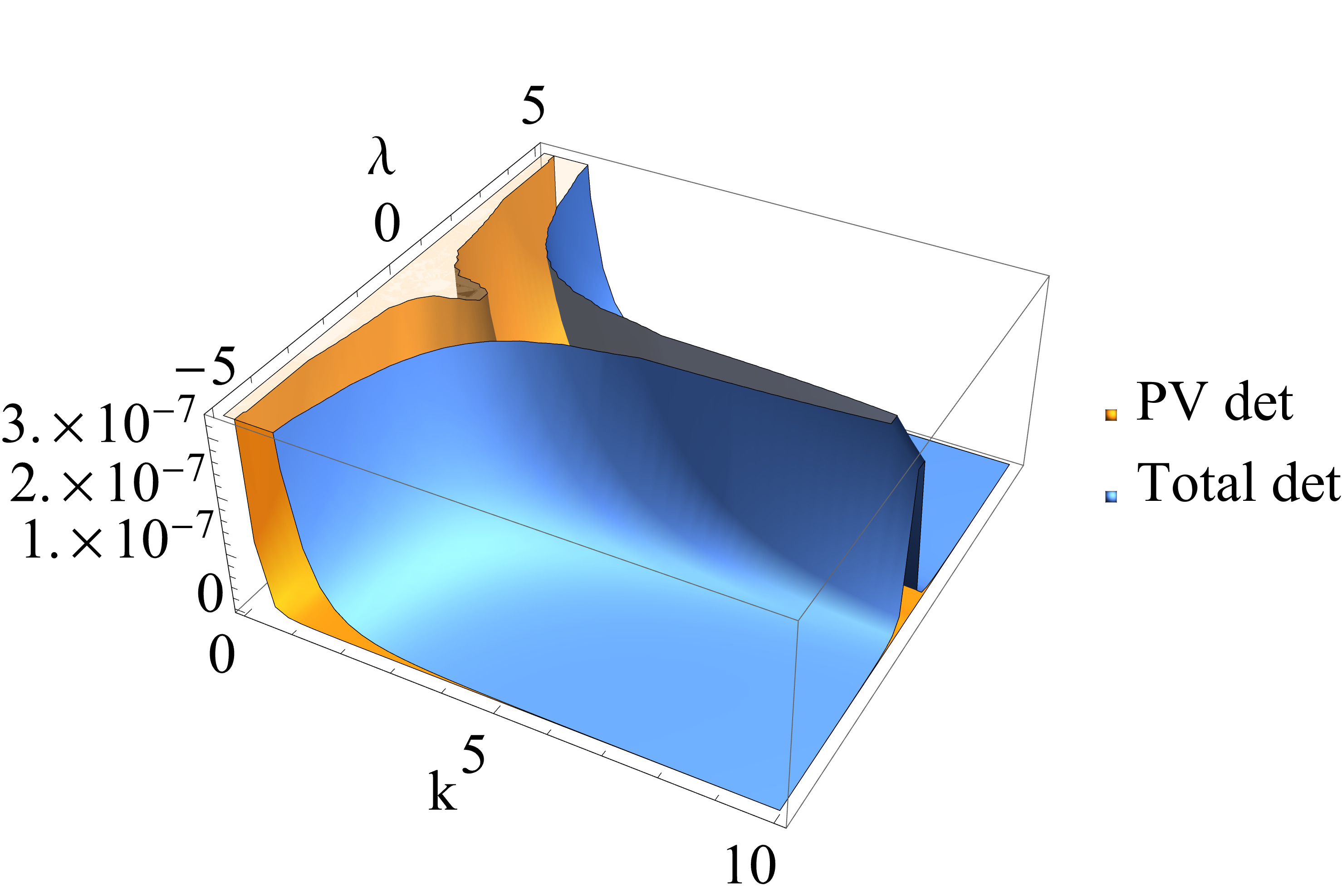}
  \caption{$\kappa, \beta \to 1$}
\end{subfigure}
\hfill
\begin{subfigure}{.3\textwidth}
  \centering
  \includegraphics[width=\linewidth]{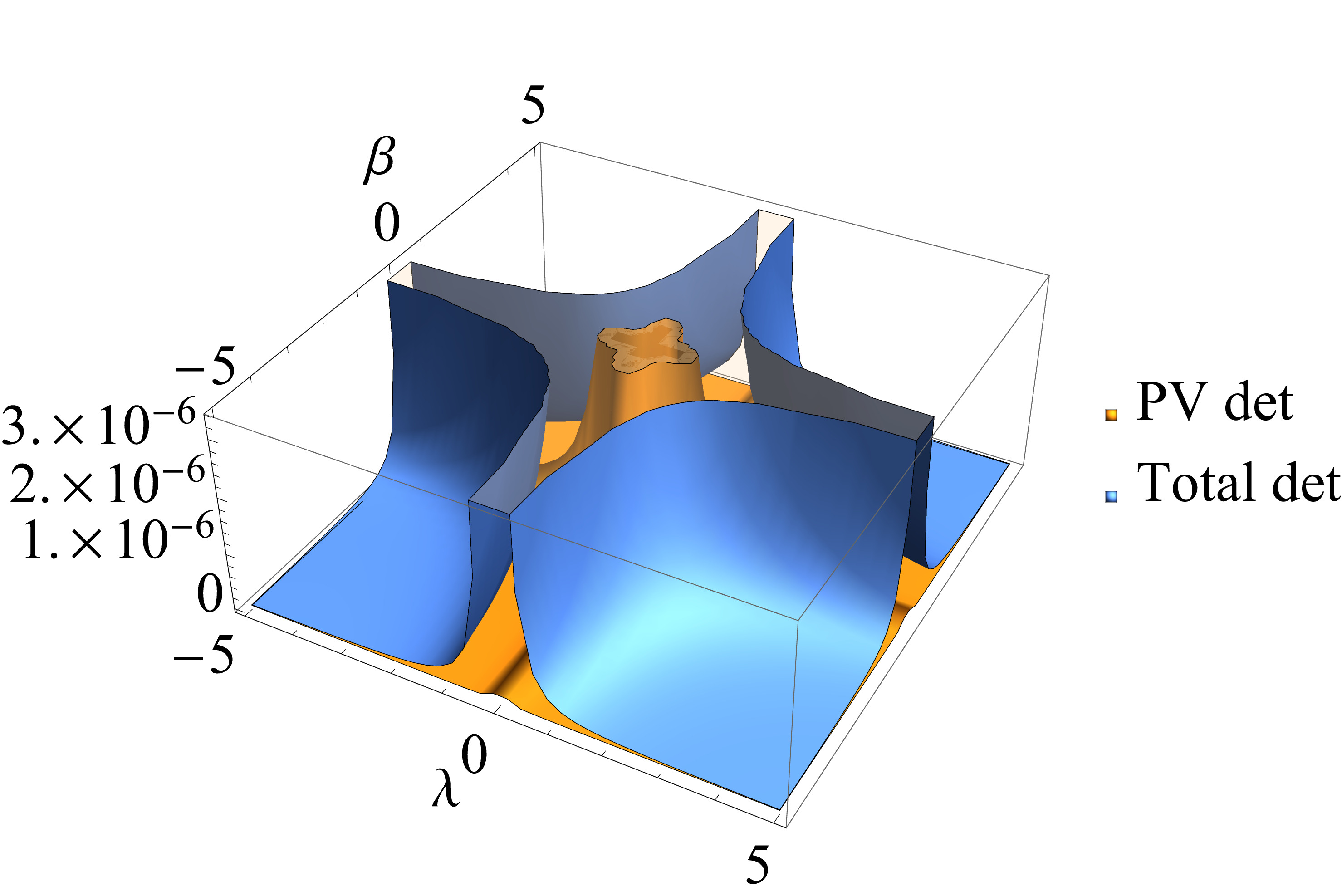}
  \caption{$k, \kappa \to 1$}
\end{subfigure}
\caption{Provost-Vallee QGT determinant vs the total QGT including the metric dependent terms of the symmetricaly coupled Toda oscillators}
\label{fig:detTodaSym} 
\end{figure}

Finally, in Fig. \ref{fig:detTodaSym}, we compare several projections of the determinants of both the usual Provost-Vallee QGT and the curved QGT, which includes the extra metric dependent terms of \eqref{QGT}. Here, we can explicitly see how the extra terms modify the tensor, emphasizing the quantum phase transitions of the system. 

\subsubsection{Analytical vs perturbative.}

Given the complexity of this system, one could be tempted to solve it perturbatively. However, as we shall see, several entries up to the second order do not behave closely to the analytical solution. 

We make a series expansion for the coupling term in the wave function \eqref{eq:OAESA_psi0}, and obtain the QGT for each perturbation order. In this way, we compare them with the analytical solution and to recognize the non-perturbative properties of the system up to the second order. 
For the first order our system becomes:
\begin{equation}
    \psi^{(1)}_0=\mathcal{N}^{(1)}_0\exp\Bigl(-\frac{\omega_1 U_1^2}{2}-\frac{\omega_2 U_2^2}{2}\Bigr)\Bigl[1-\gamma U_1U_2\Bigr]
\end{equation}
where the new normalization constant is\footnote{The upper index indicates the perturbative order.}
\begin{equation}
 \mathcal{N}^{(1)}_0=4 \sqrt{\frac{(\omega_1\omega_2)^{3/2}}{\pi  \gamma^2-8 \gamma \sqrt{\omega_1\omega_2}+4 \pi  \omega_1\omega_2}}.
\end{equation}
Analogously, for the second order, we get
\begin{equation}
    \psi^{(2)}_0=\mathcal{N}^{(2)}_0\exp\Bigl(-\frac{\omega_1 U_1^2}{2}-\frac{\omega_2 U_2^2}{2}\Bigr)\Bigl[1-\gamma U_1U_2+\frac{1}{2}\gamma^2U_1^2U_2^2\Bigr]
\end{equation}
where the normalization constant for this case is
\begin{equation}
 \mathcal{N}^{(2)}_0=16 \sqrt{\frac{(\omega_1\omega_2)^{5/2}}{9 \pi  \gamma^4-64 \gamma^3 \sqrt{\omega_1\omega_2 }+32 \pi  \gamma^2\omega_1\omega_2-128 \gamma (\omega_1\omega_2)^{3/2}+64 \pi{(\omega_1\omega_2)}^2}}
\end{equation}

\begin{figure}[!h]
\begin{subfigure}{.27\textwidth}
  \centering
  \includegraphics[width=.8\linewidth]{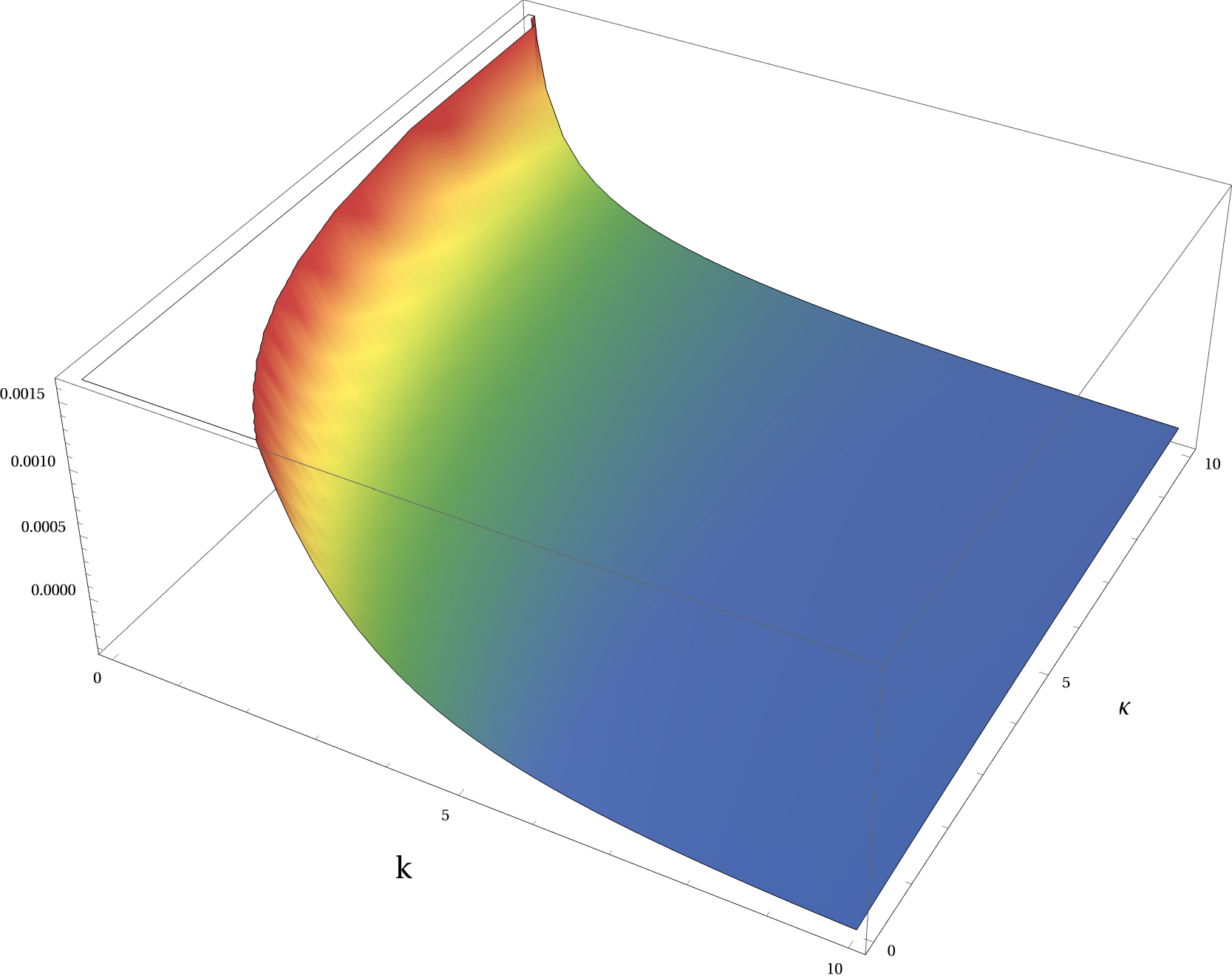}
  \label{fig:sfig1}
\end{subfigure}
\hfill
\begin{subfigure}{.27\textwidth}
  \centering
  \includegraphics[width=.8\linewidth]{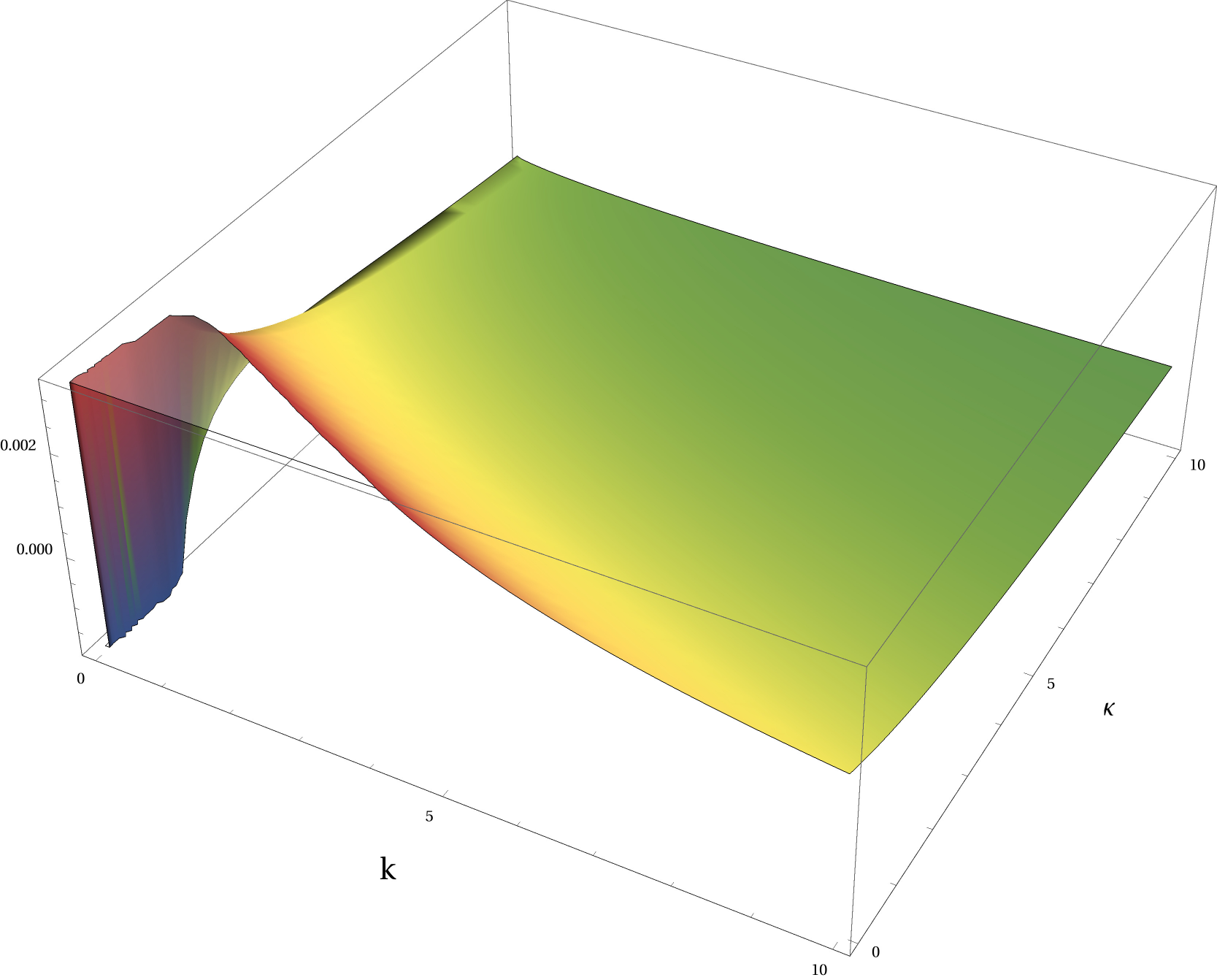}
  \caption*{$G_{\kappa\lambda}(k,\kappa)$}
  \label{fig:sfig2}
\end{subfigure}
\hfill
\begin{subfigure}{.27\textwidth}
  \centering
  \includegraphics[width=.8\linewidth]{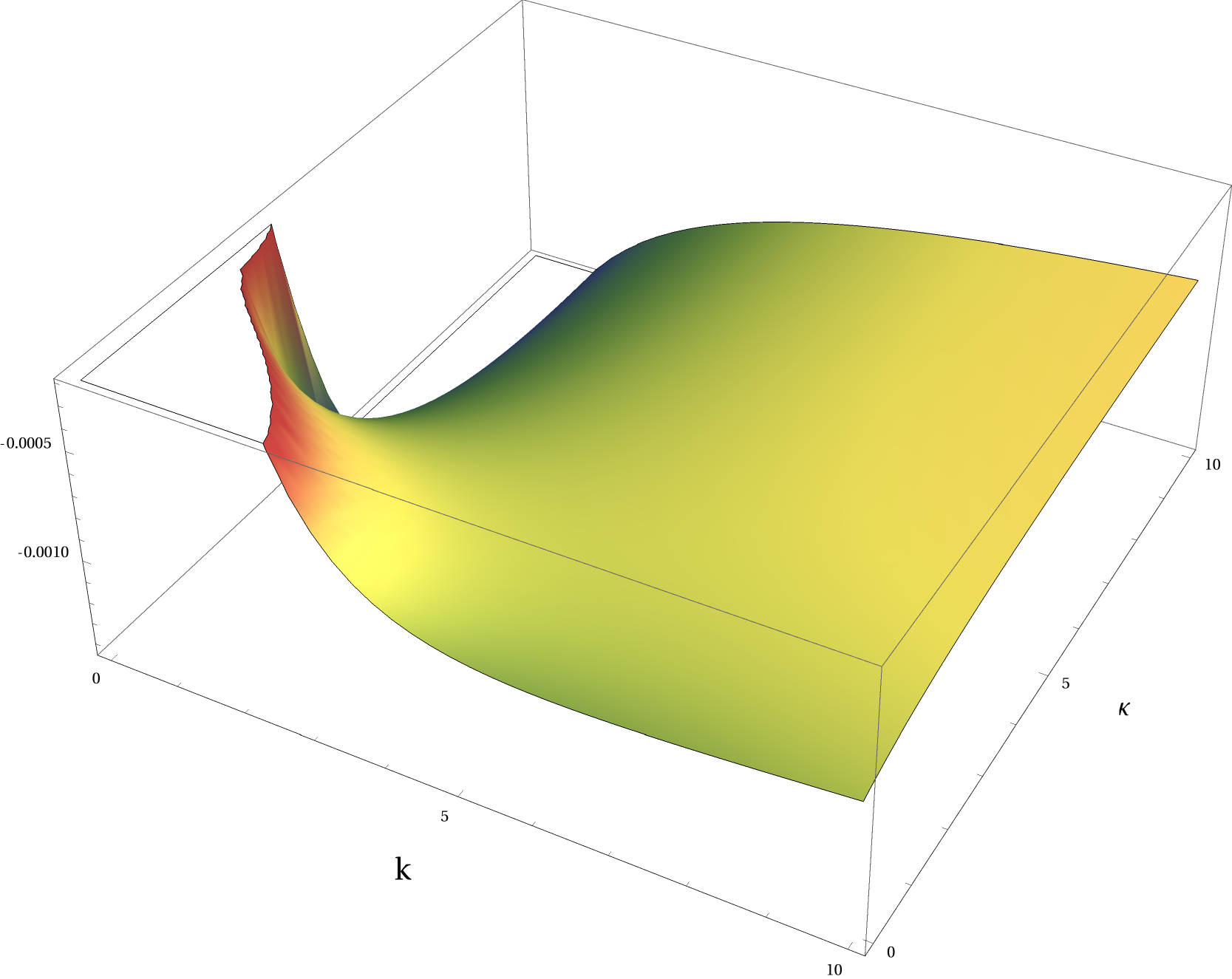}
  \label{fig:sfig3}
\end{subfigure}
\\
\begin{subfigure}{.27\textwidth}
  \centering
  \includegraphics[width=.8\linewidth]{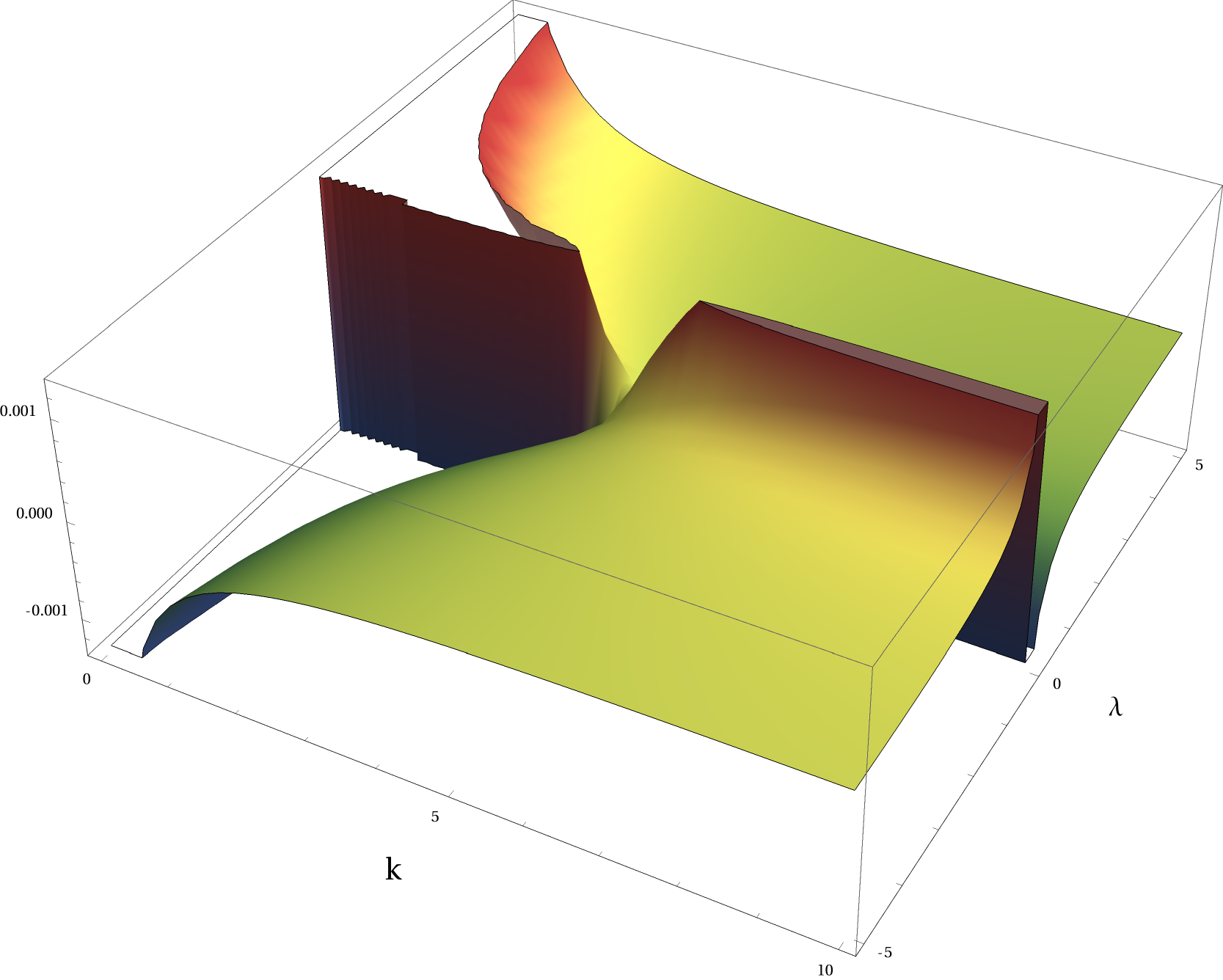}
  \label{fig:sfig4}
\end{subfigure}
\hfill
\begin{subfigure}{.27\textwidth}
  \centering
  \includegraphics[width=.8\linewidth]{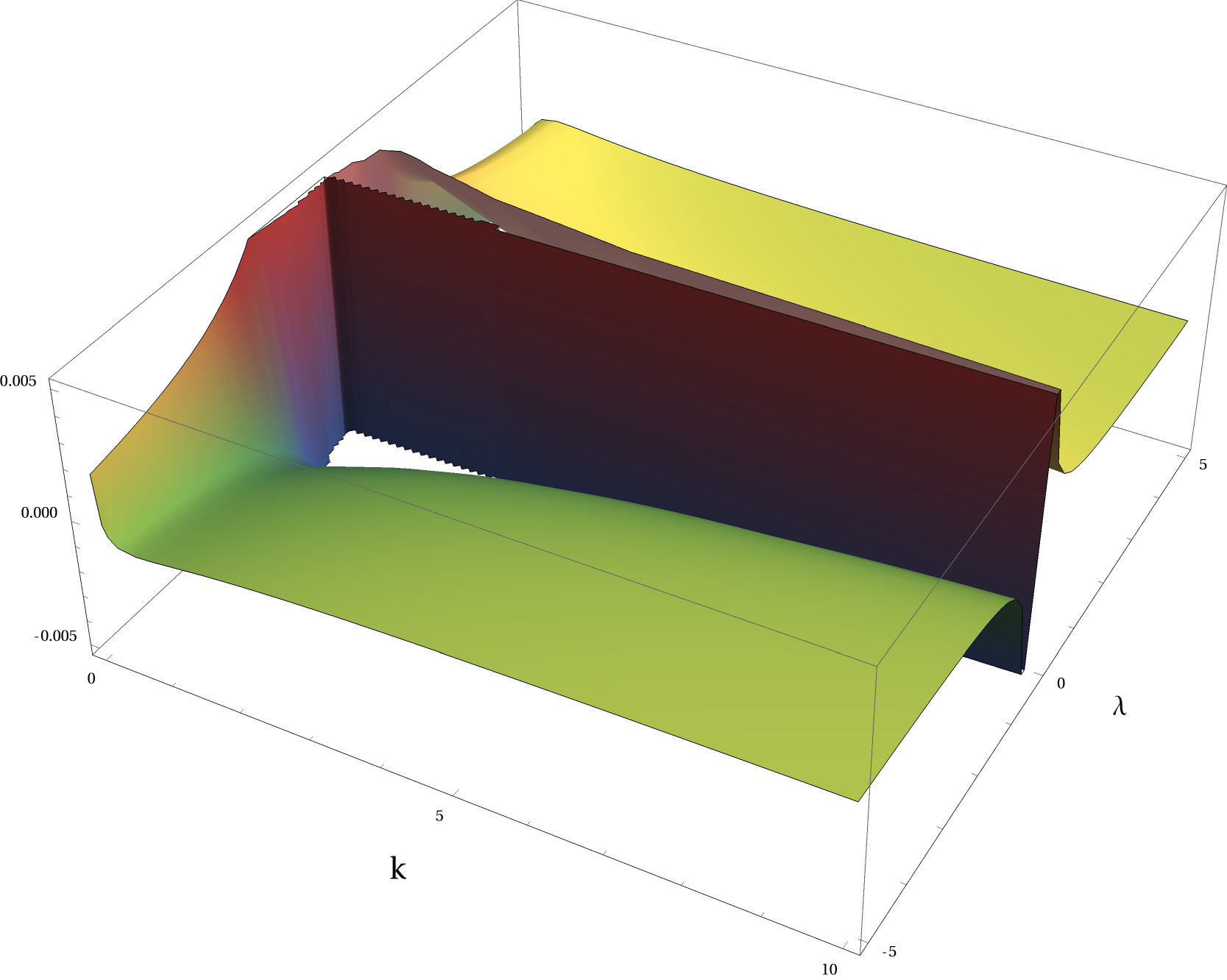}
  \caption*{$G_{\kappa\lambda}(k,\lambda)$}
  \label{fig:sfig5}
\end{subfigure}
\hfill
\begin{subfigure}{.27\textwidth}
  \centering
  \includegraphics[width=.8\linewidth]{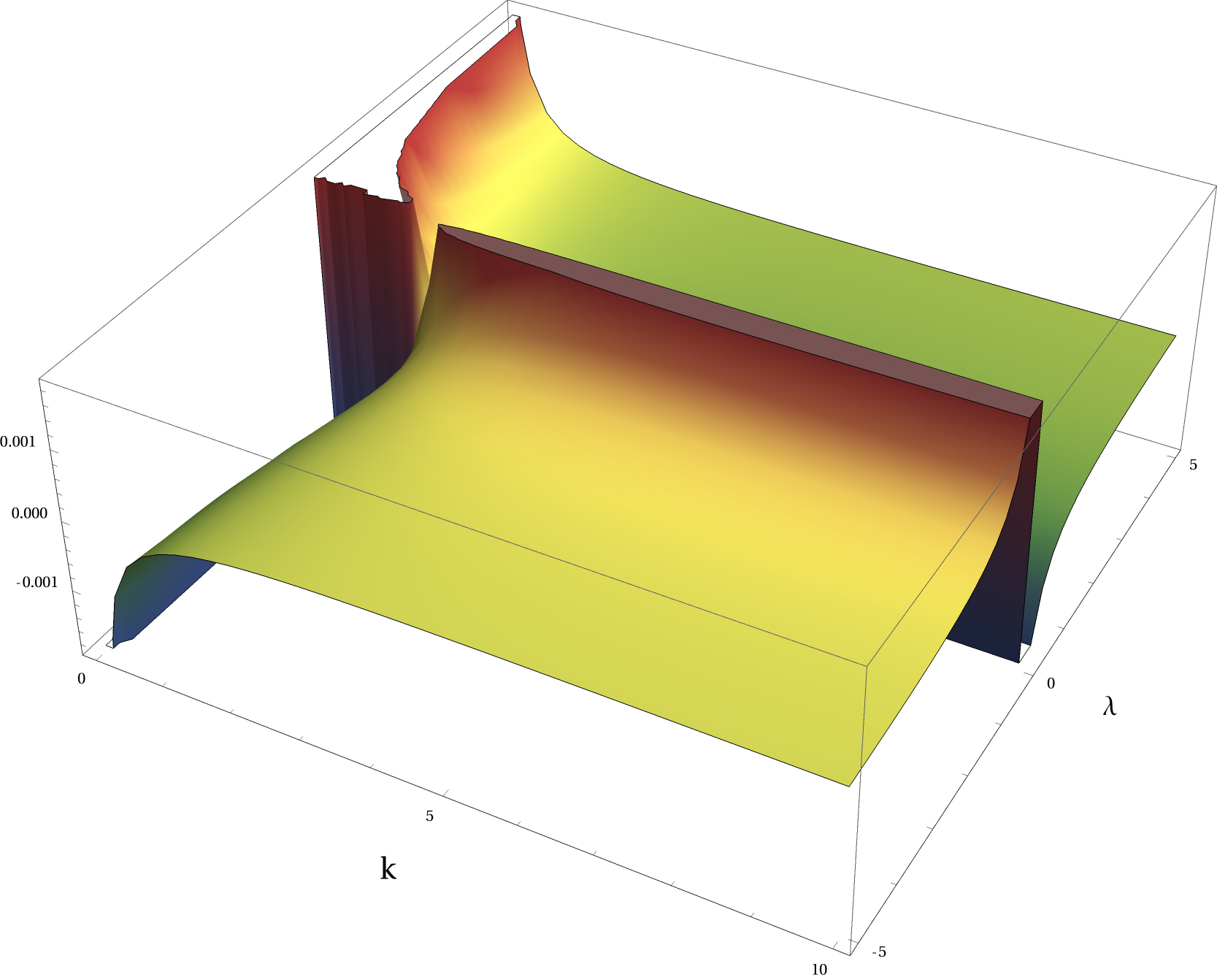}
  \label{fig:sfig6}
\end{subfigure}
\\
\begin{subfigure}{.27\textwidth}
  \centering
  \includegraphics[width=.8\linewidth]{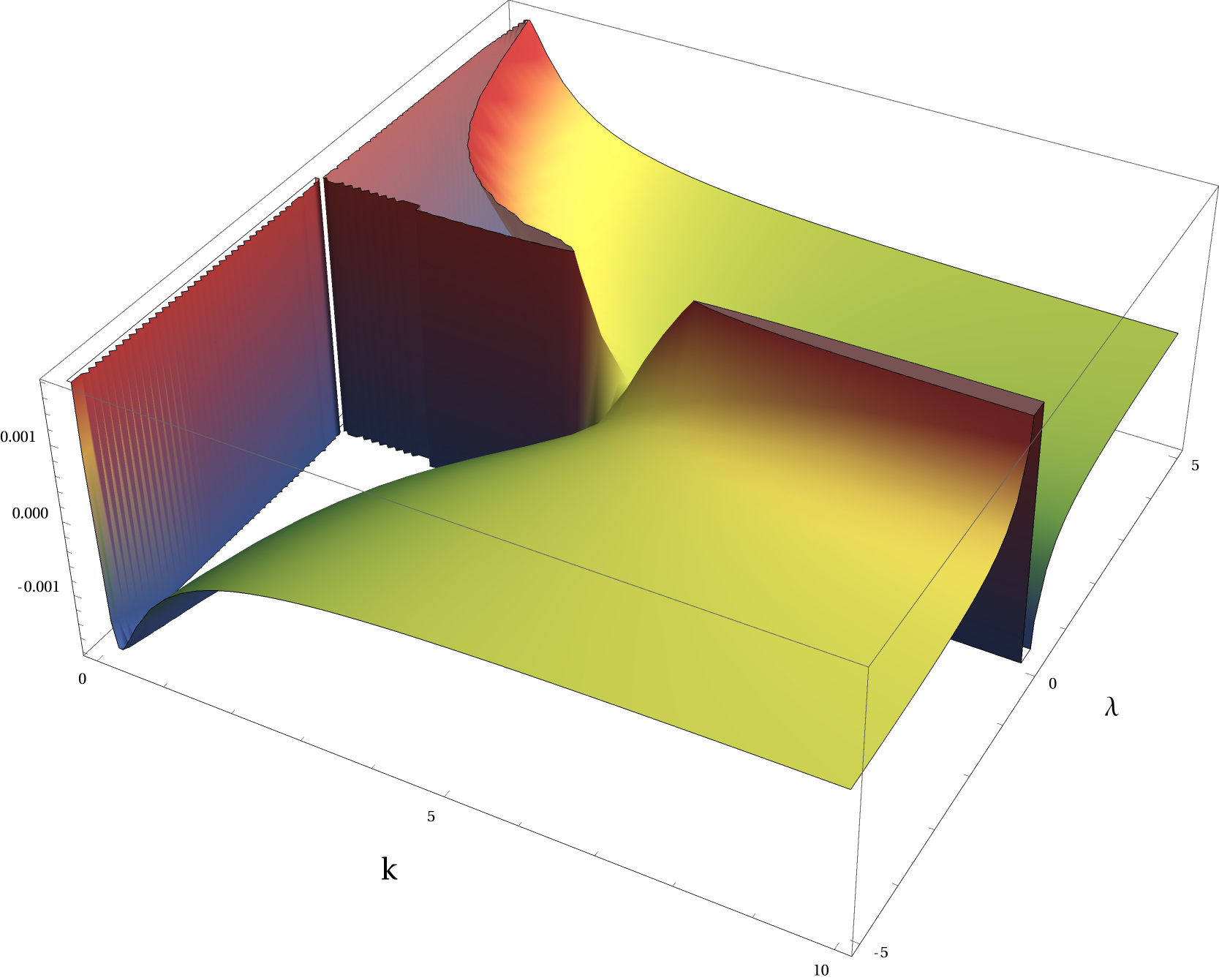}
  \label{fig:sfig7}
\end{subfigure}
\hfill
\begin{subfigure}{.27\textwidth}
  \centering
  \includegraphics[width=.8\linewidth]{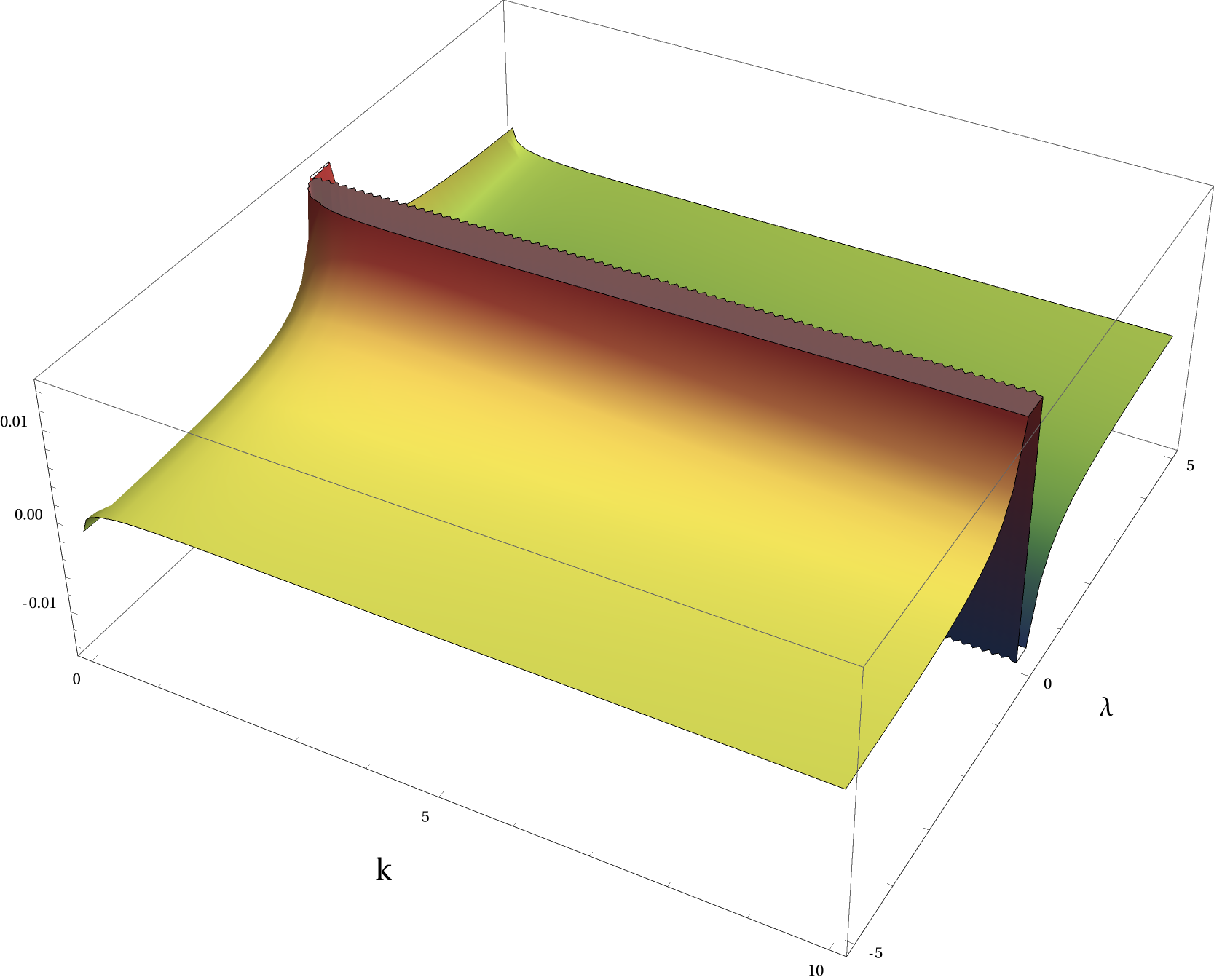}
  \caption*{$G_{\lambda\beta}(k,\lambda)$}
  \label{fig:sfig8}
\end{subfigure}
\hfill
\begin{subfigure}{.27\textwidth}
  \centering
  \includegraphics[width=.8\linewidth]{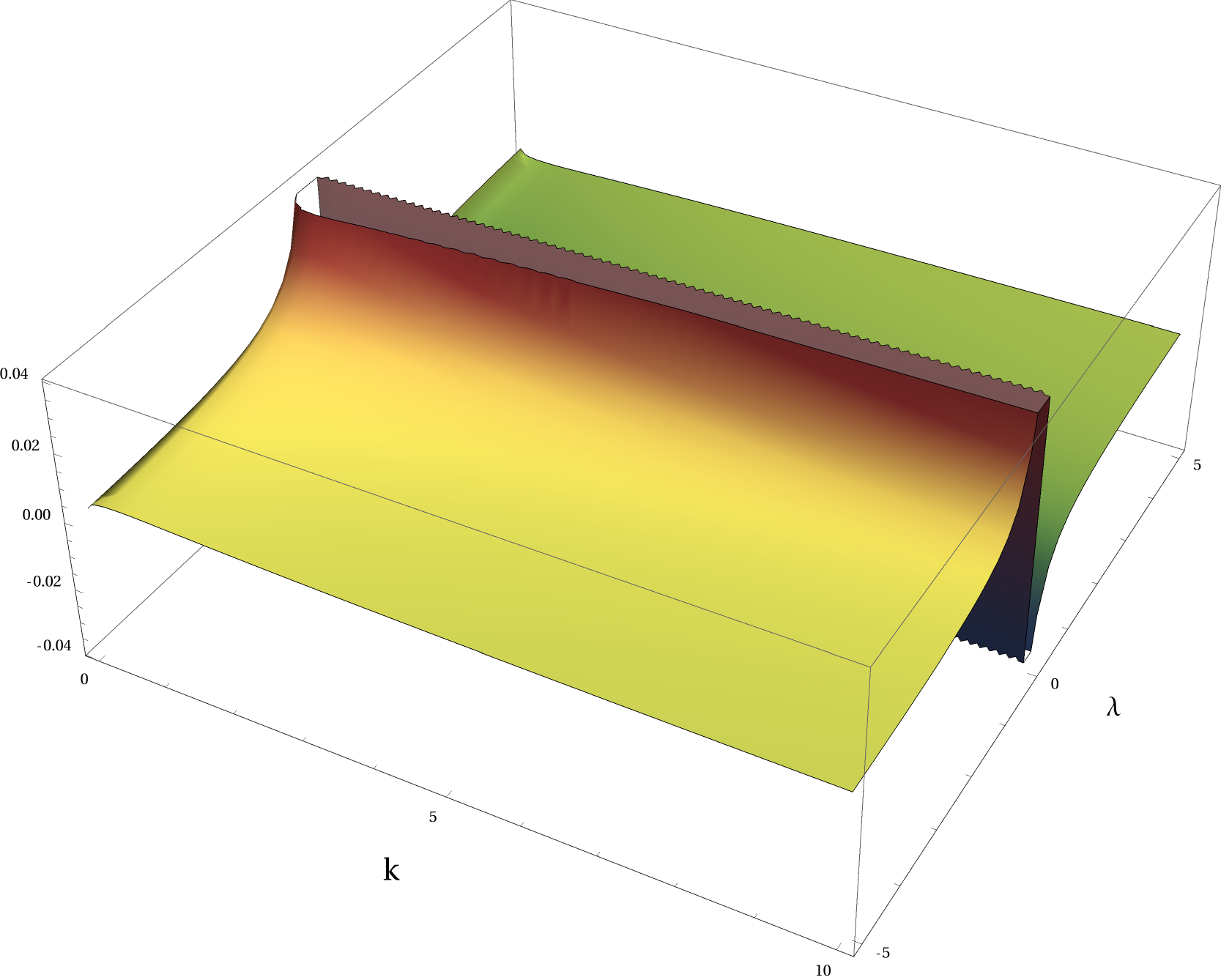}
  \label{fig:sfig9}
\end{subfigure}
\\
\begin{subfigure}{.27\textwidth}
  \centering
  \includegraphics[width=.8\linewidth]{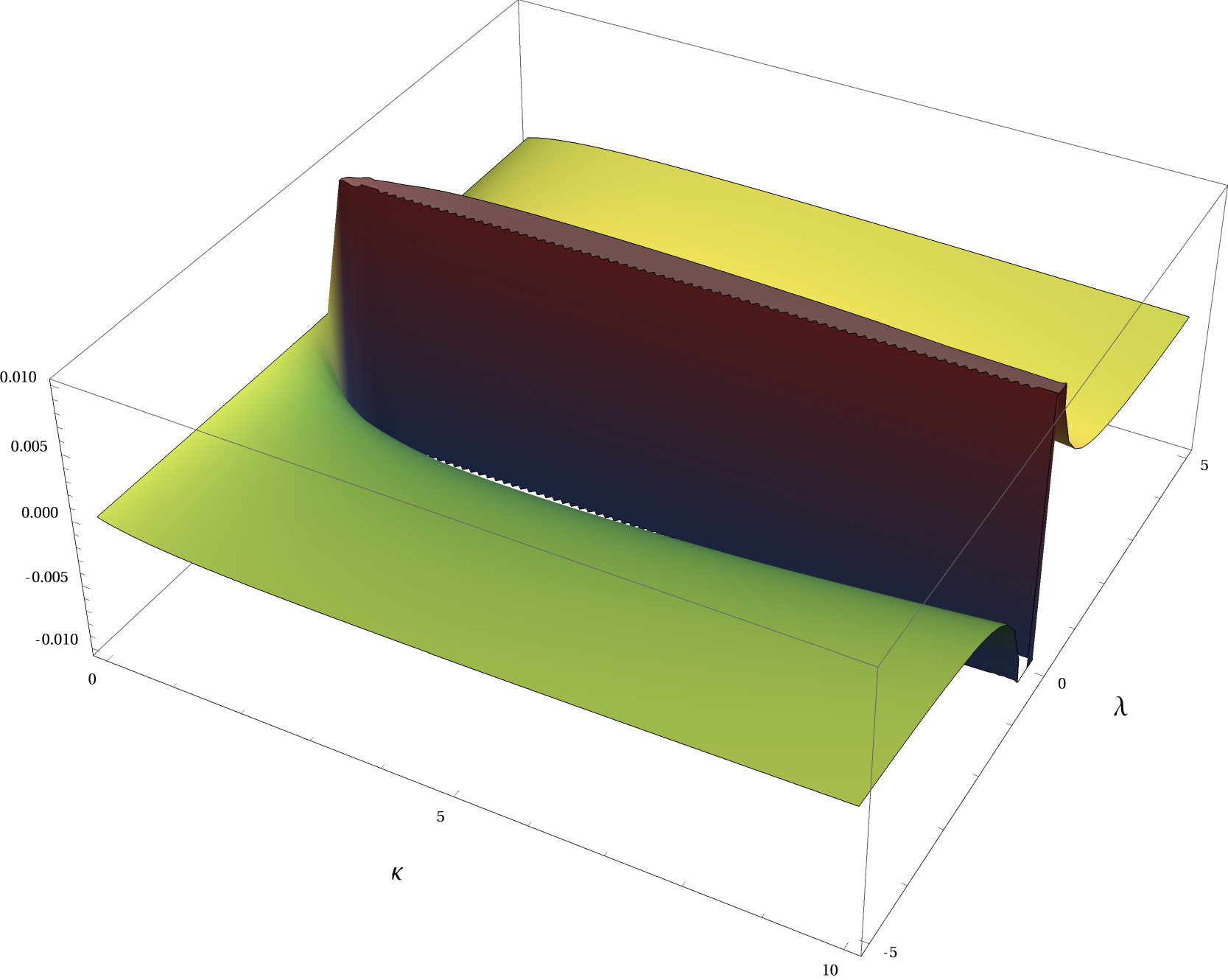}
  \label{fig:sfig10}
\end{subfigure}
\hfill
\begin{subfigure}{.27\textwidth}
  \centering
  \includegraphics[width=.8\linewidth]{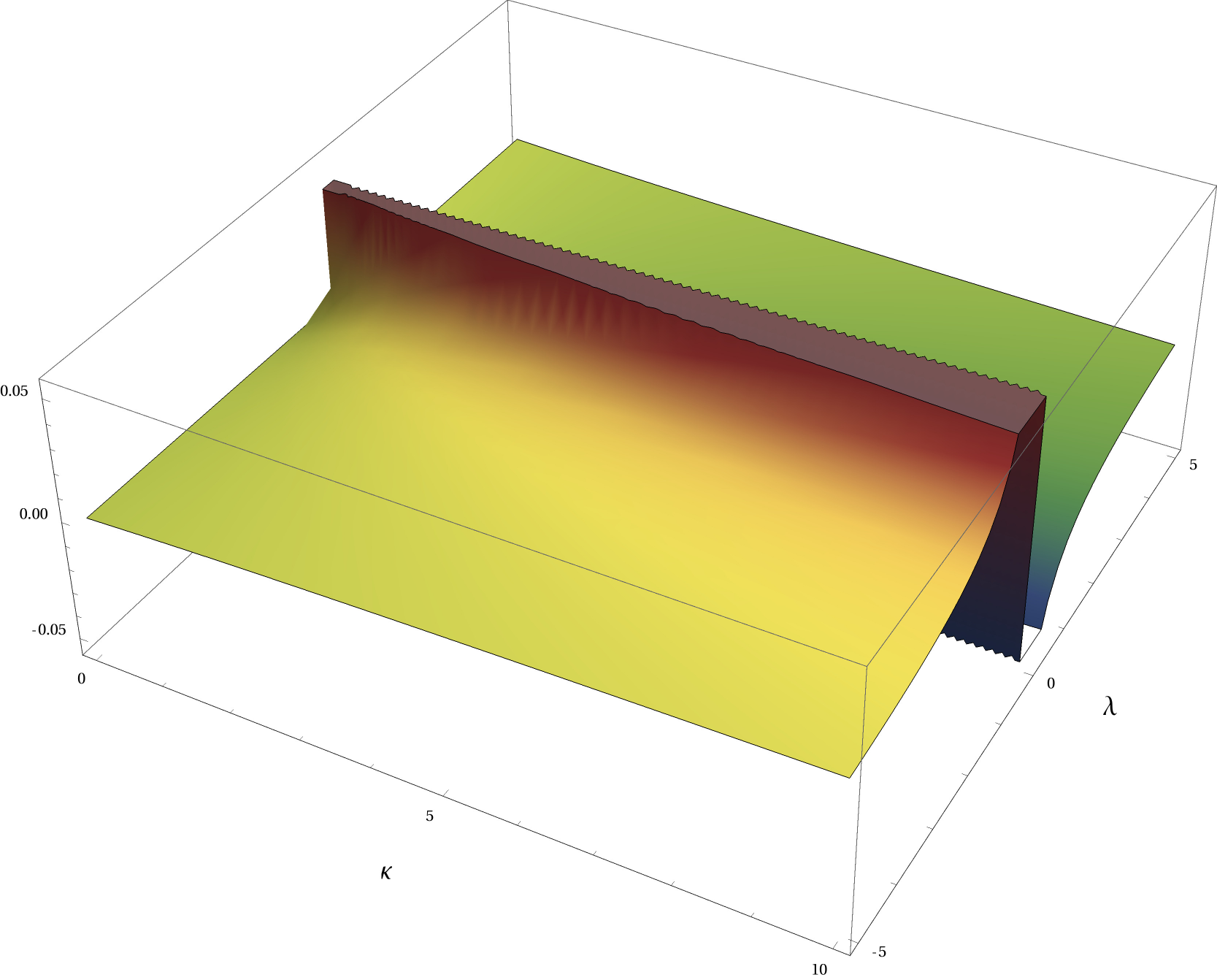}
  \caption*{$G_{\lambda\beta}(\kappa,\lambda)$}
  \label{fig:sfig11}
\end{subfigure}
\hfill
\begin{subfigure}{.27\textwidth}
  \centering
  \includegraphics[width=.8\linewidth]{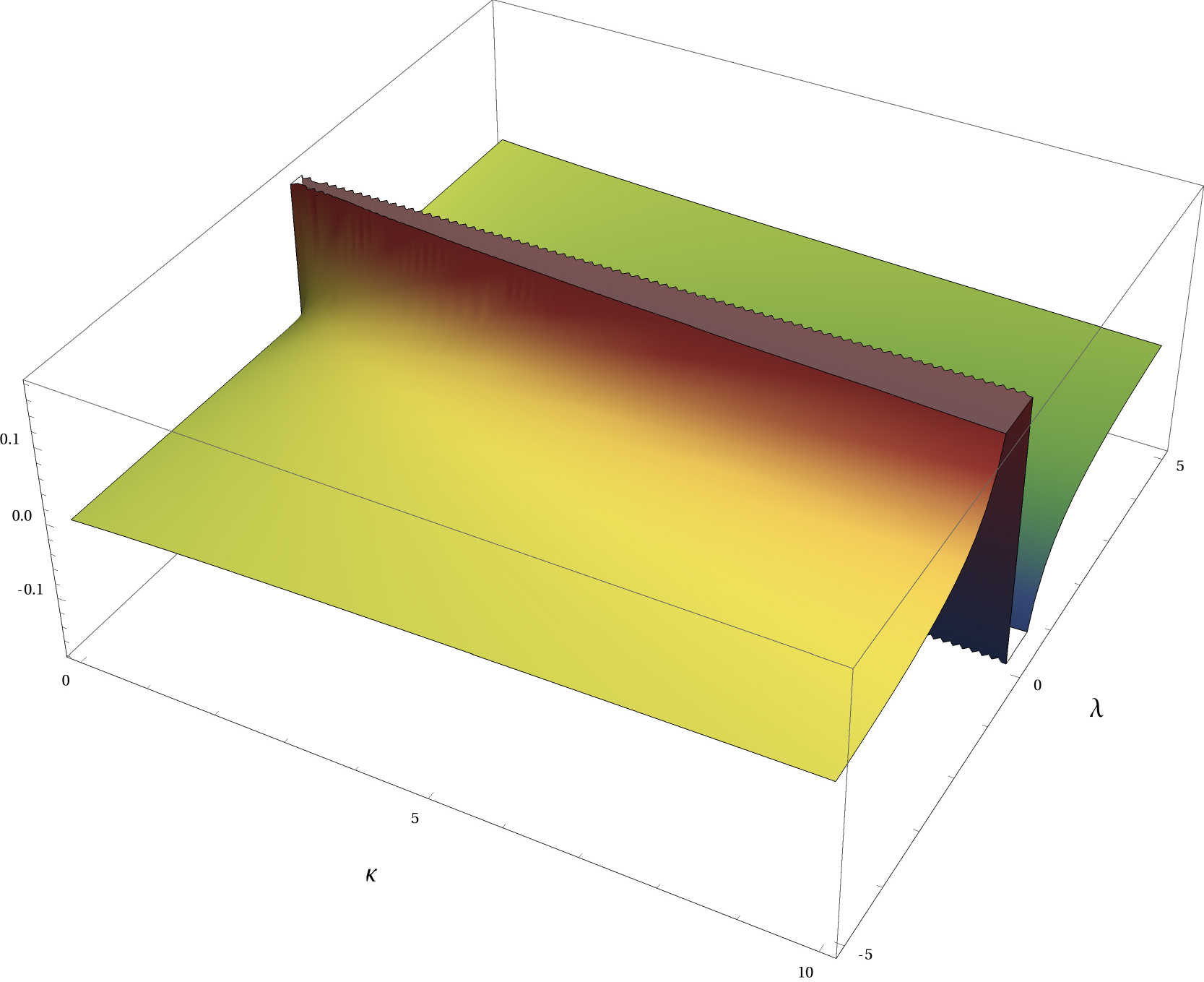}
  \label{fig:sfig12}
\end{subfigure}
\\
\begin{subfigure}{.27\textwidth}
  \centering
  \includegraphics[width=.8\linewidth]{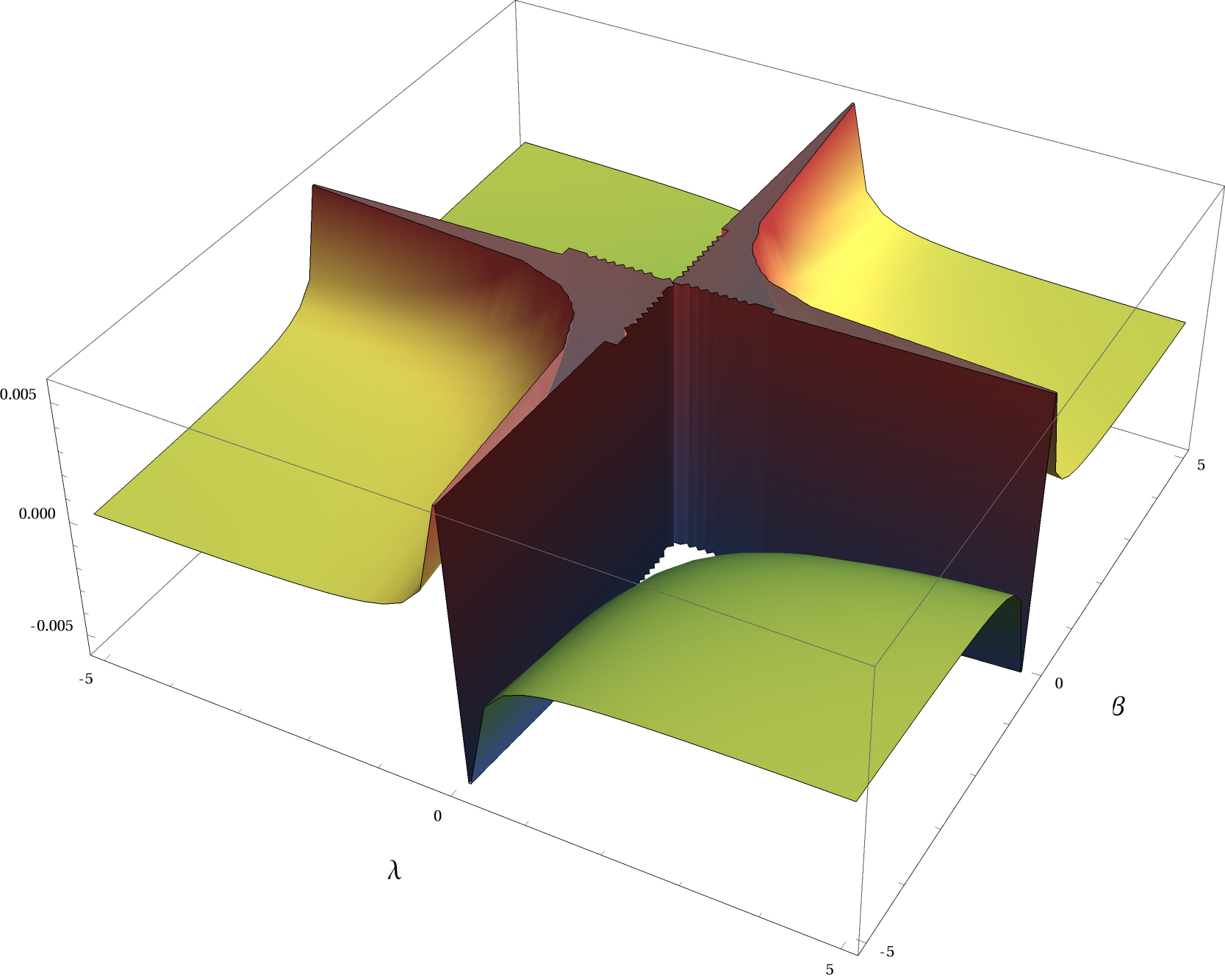}
  \label{fig:sfig13}
\end{subfigure}
\hfill
\begin{subfigure}{.27\textwidth}
  \centering
  \includegraphics[width=.8\linewidth]{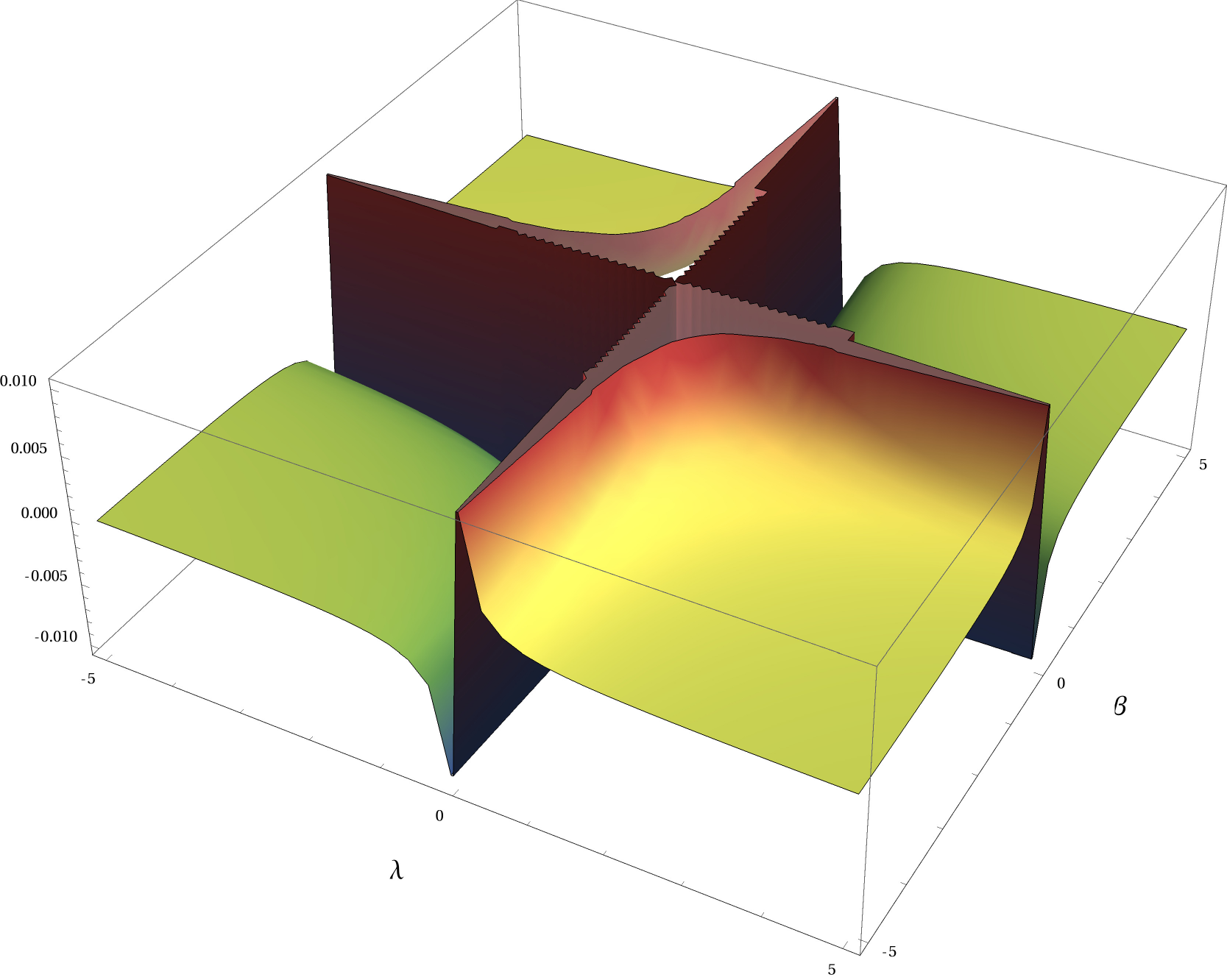}
  \caption*{$G_{\lambda\beta}(\lambda,\beta)$}
  \label{fig:sfig14}
\end{subfigure}
\hfill
\begin{subfigure}{.27\textwidth}
  \centering
  \includegraphics[width=.8\linewidth]{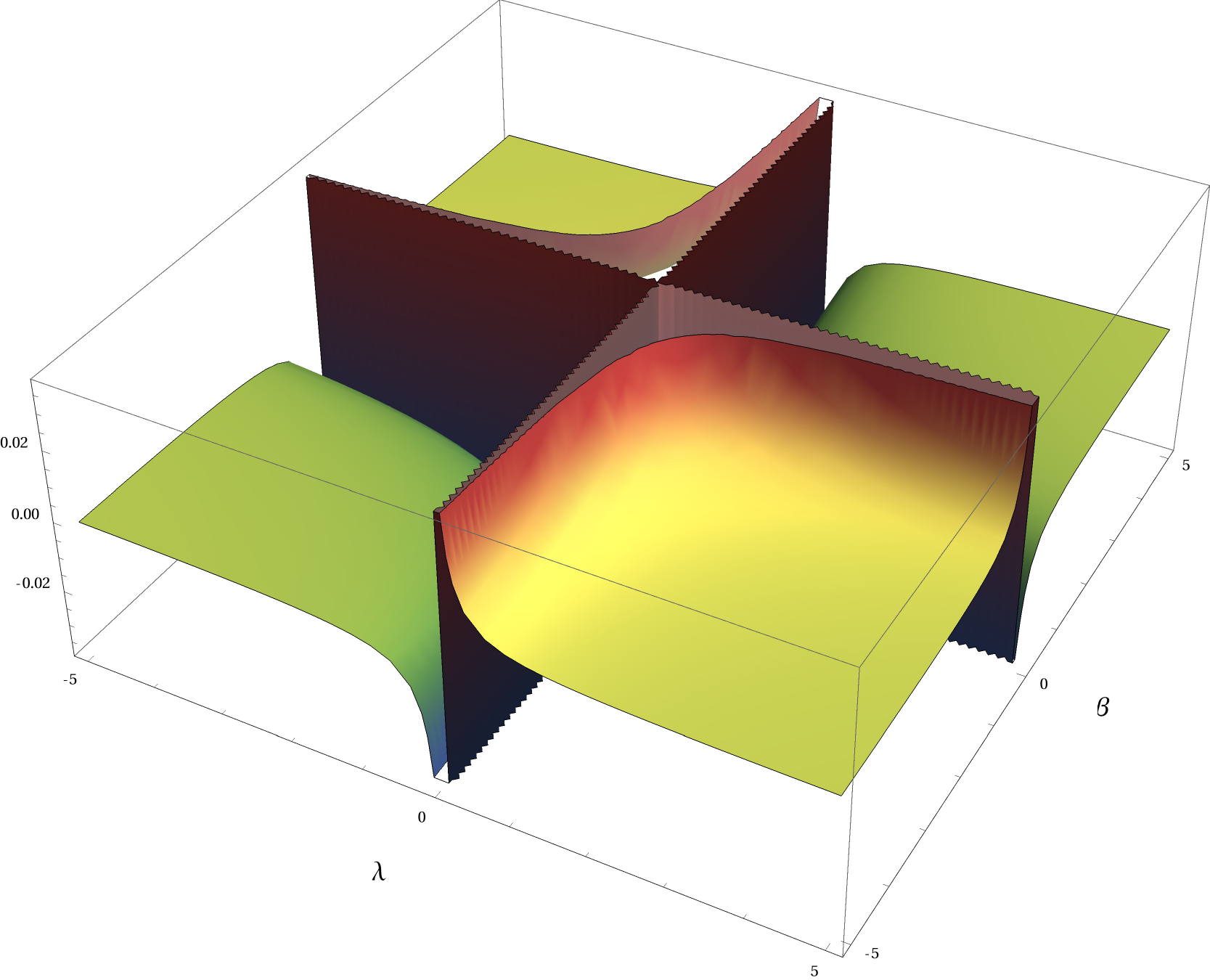}
  \label{fig:sfig15}
\end{subfigure}
\caption{Entries of the QGT of the symmetrically coupled Toda oscillators with the most predominant analytical and perturbative differences. The left column has the analytical solution, in the center the first-order approximation, and in the right column the second-order.}
\label{fig:PertCompSymToda}
\end{figure}

\begin{figure}[!h]
\begin{subfigure}{.30\textwidth}
  \centering
  \includegraphics[width=.8\linewidth]{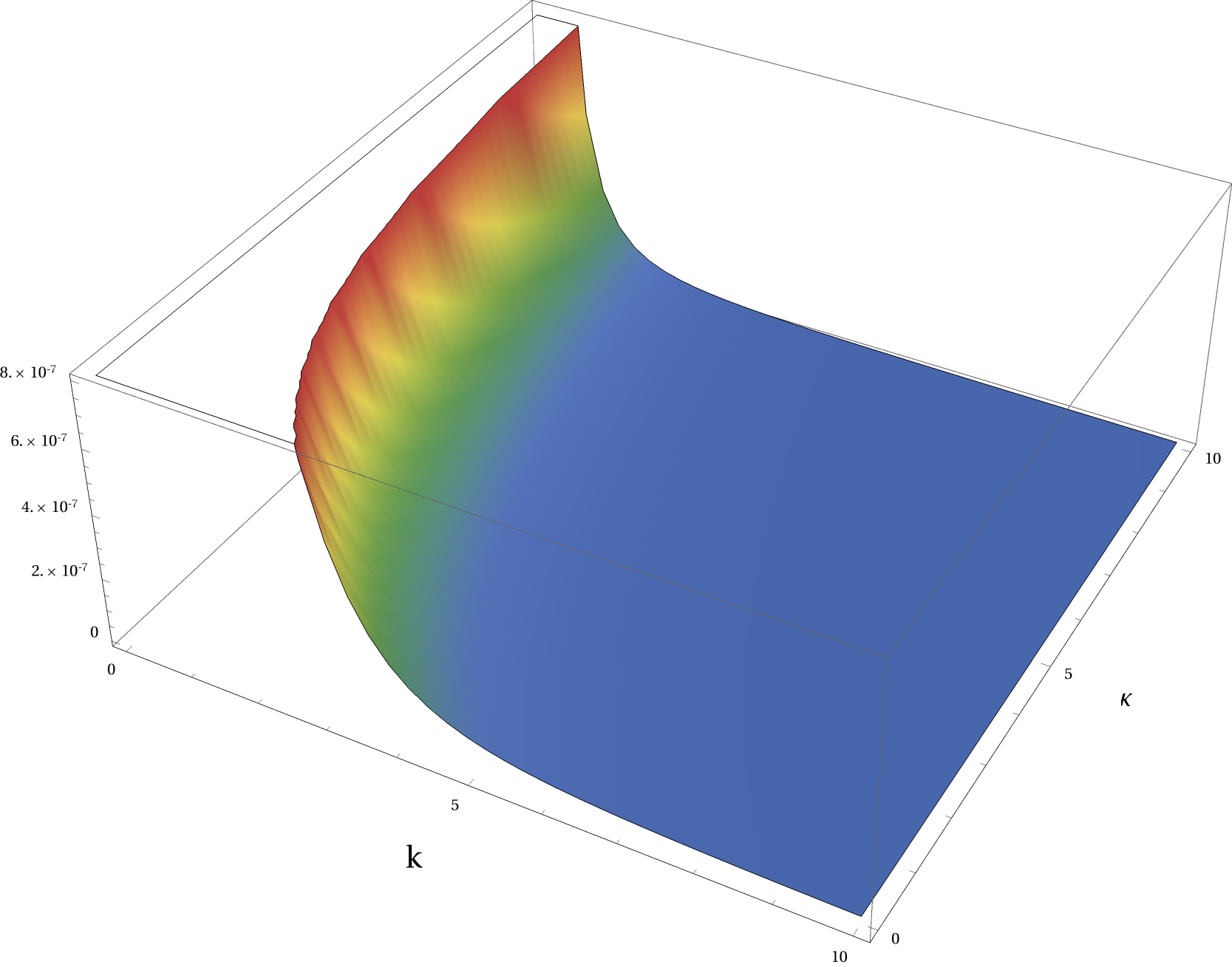}
  \label{fig:detcomp1}
\end{subfigure}
\hfill
\begin{subfigure}{.30\textwidth}
  \centering
  \includegraphics[width=.8\linewidth]{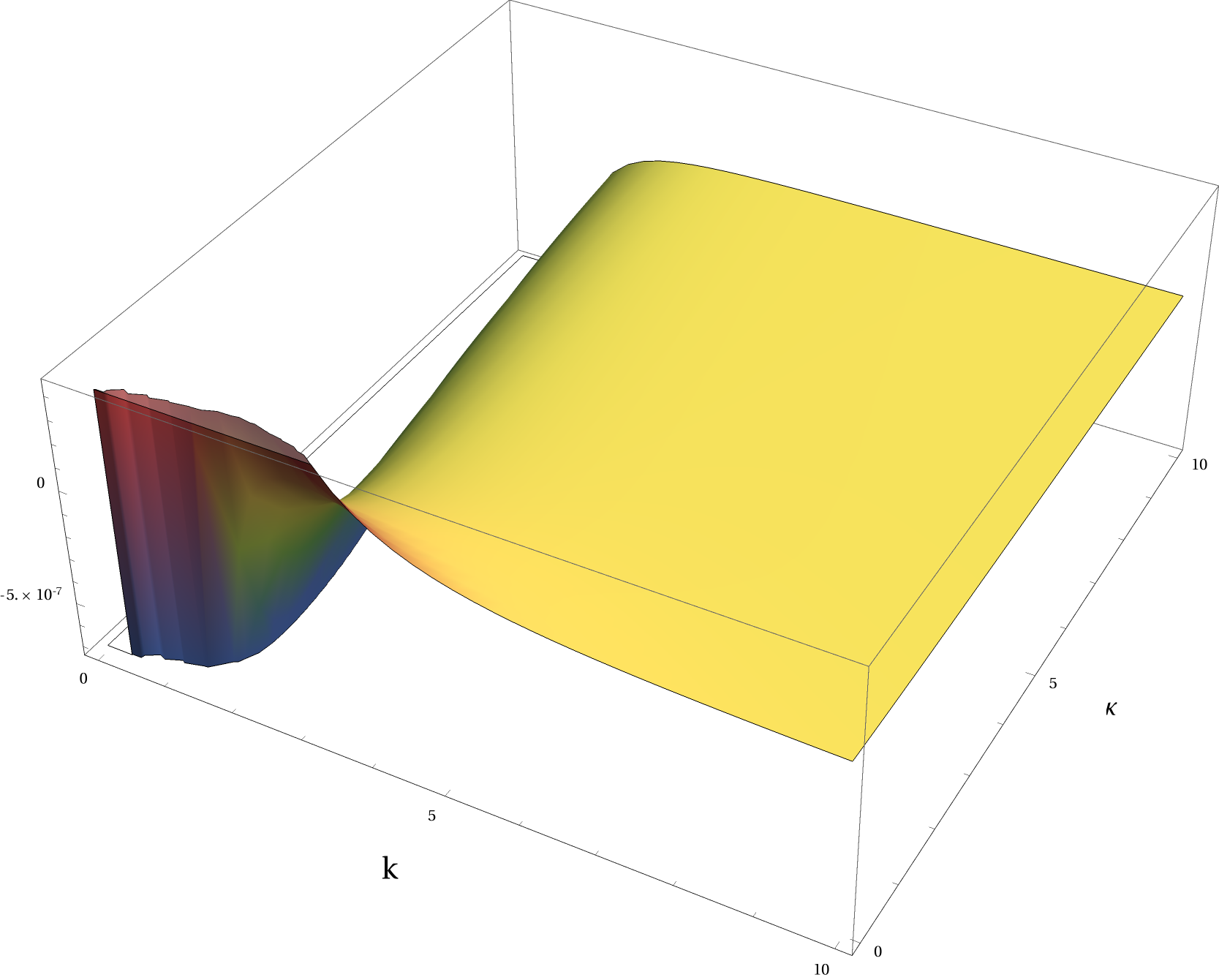}
  \caption*{$\det[G(k,\kappa)]$}
  \label{fig:detcomp2}
\end{subfigure}
\hfill
\begin{subfigure}{.30\textwidth}
  \centering
  \includegraphics[width=.8\linewidth]{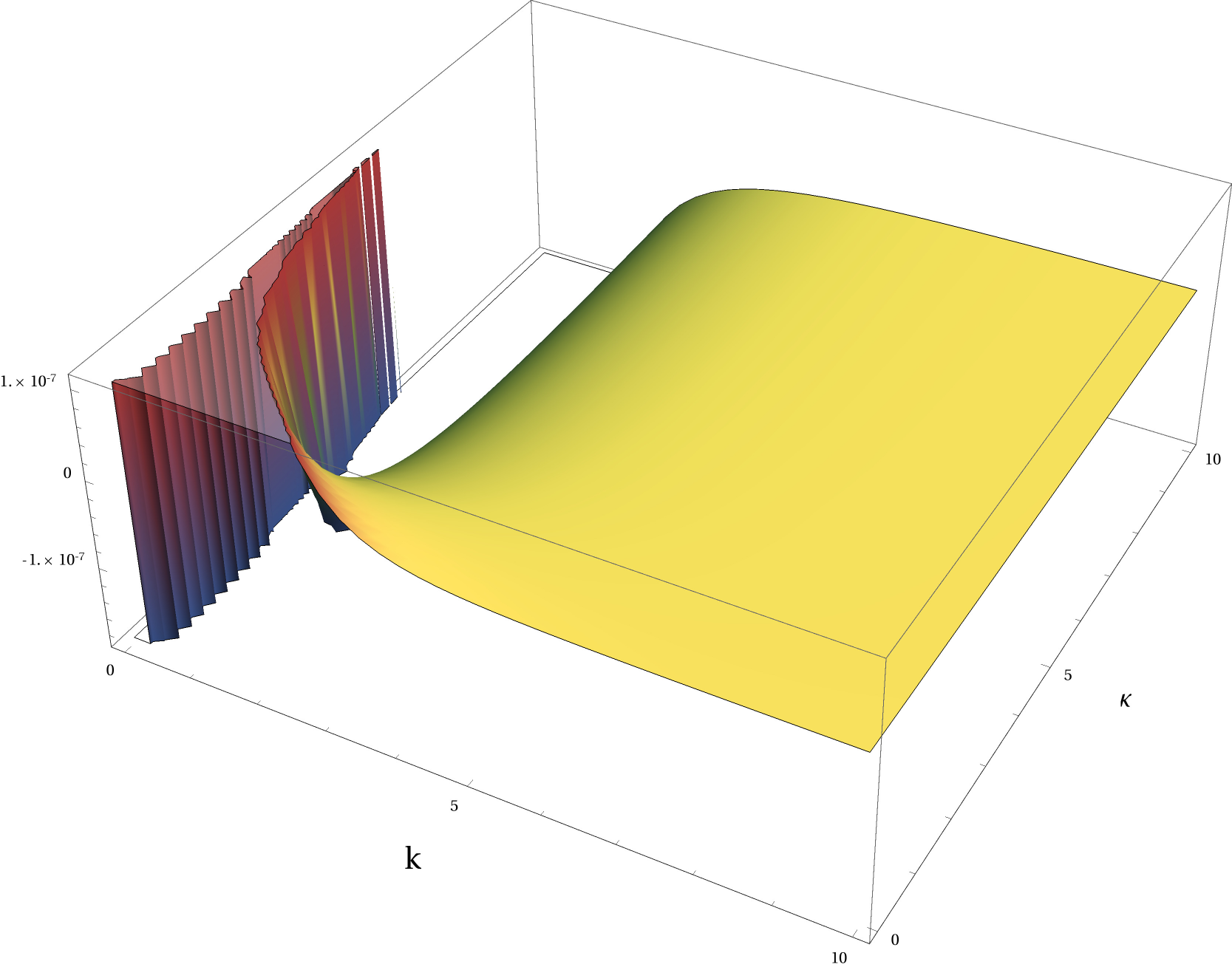}
  \label{fig:detcomp3}
\end{subfigure}
\\
\begin{subfigure}{.30\textwidth}
  \centering
  \includegraphics[width=.8\linewidth]{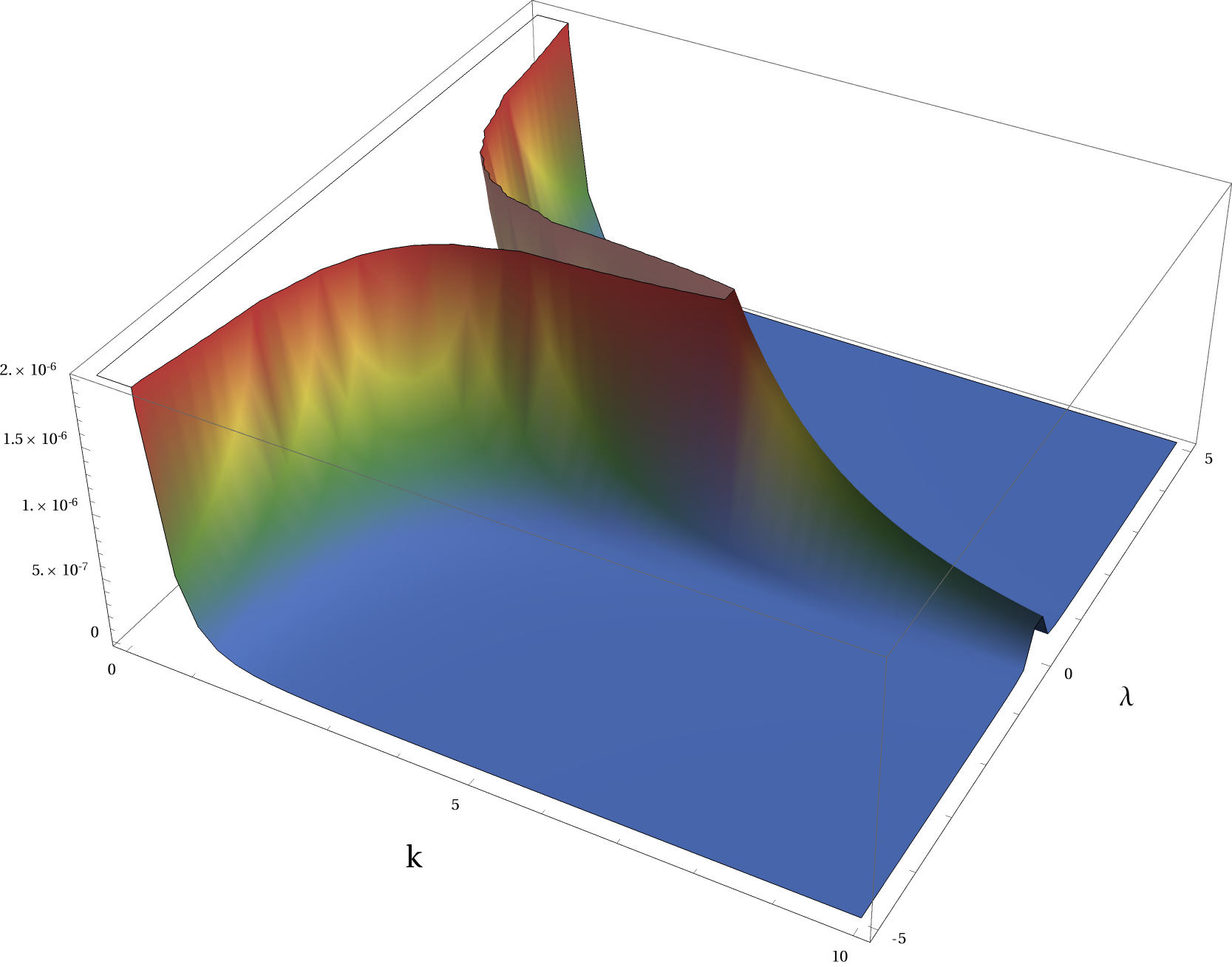}
  \label{fig:detcomp4}
\end{subfigure}
\hfill
\begin{subfigure}{.30\textwidth}
  \centering
  \includegraphics[width=.8\linewidth]{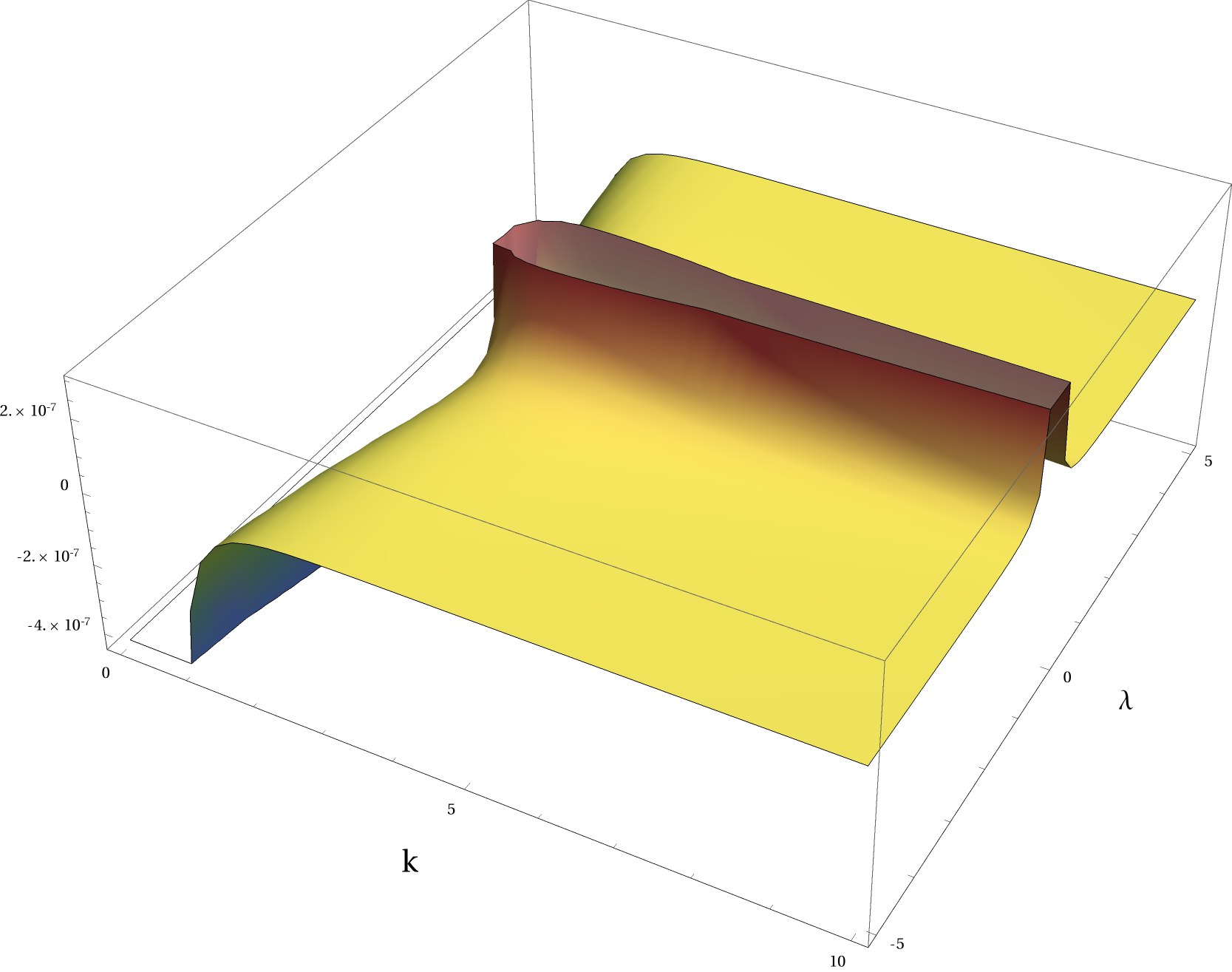}
  \caption*{$\det[G(k,\lambda)]$}
  \label{fig:detcomp5}
\end{subfigure}
\hfill
\begin{subfigure}{.30\textwidth}
  \centering
  \includegraphics[width=.8\linewidth]{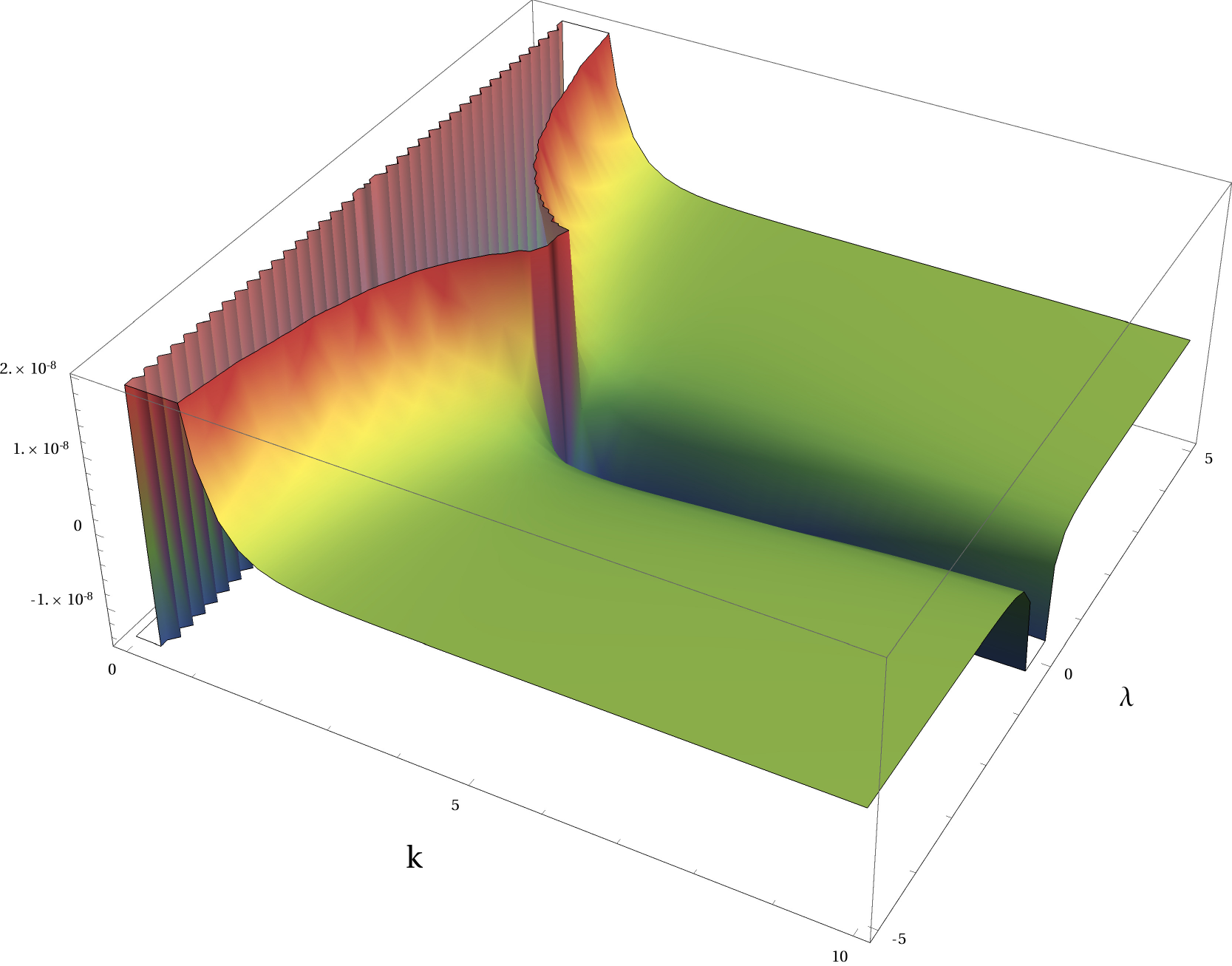}
  \label{fig:detcomp6}
\end{subfigure}
\\
\begin{subfigure}{.30\textwidth}
  \centering
  \includegraphics[width=.8\linewidth]{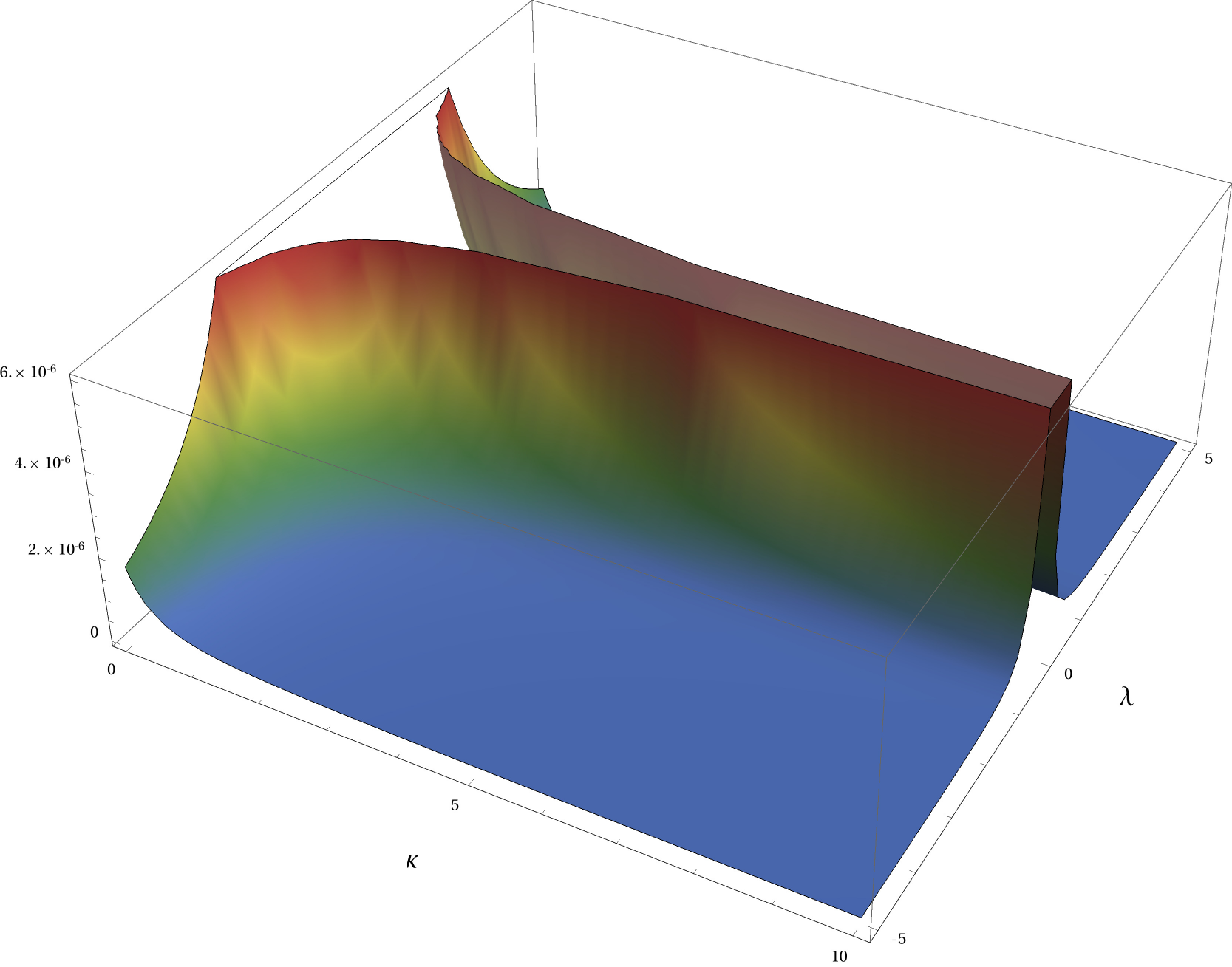}
  \label{fig:detcomp7}
\end{subfigure}
\hfill
\begin{subfigure}{.30\textwidth}
  \centering
  \includegraphics[width=.8\linewidth]{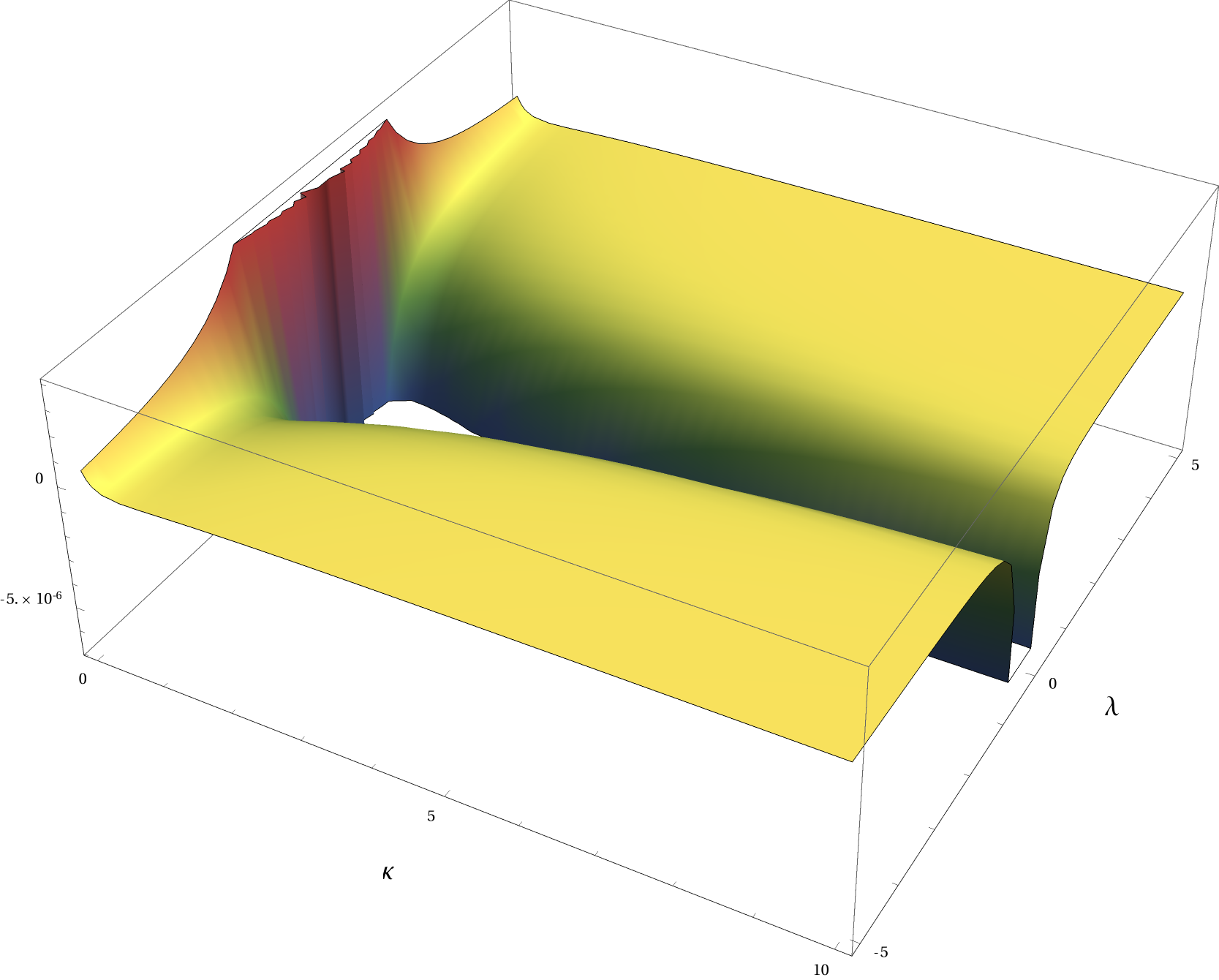}
  \caption*{$\det[G(\kappa,\lambda)]$}
  \label{fig:detcomp8}
\end{subfigure}
\hfill
\begin{subfigure}{.30\textwidth}
  \centering
  \includegraphics[width=.8\linewidth]{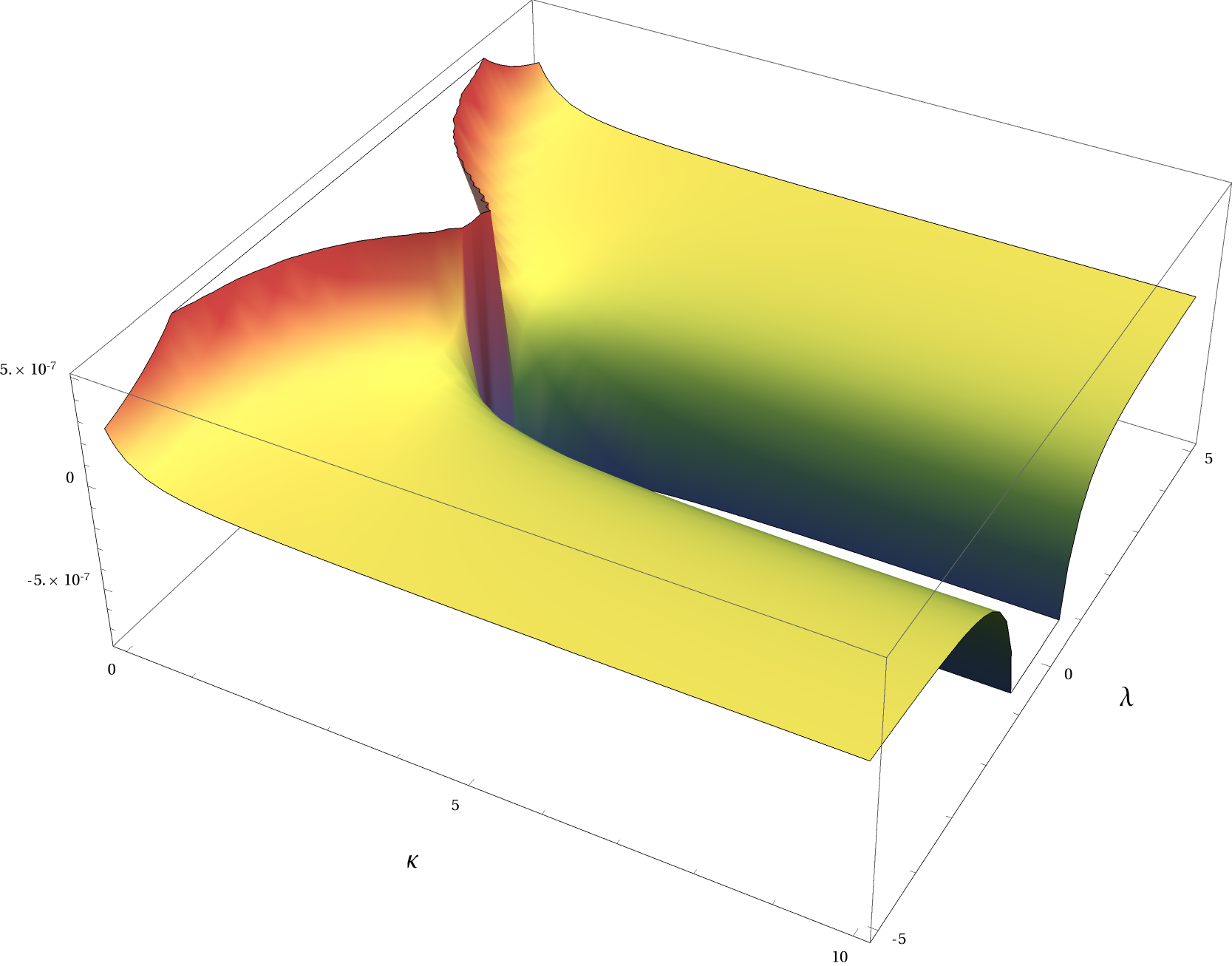}
  \label{fig:detcomp9}
\end{subfigure}
\\
\begin{subfigure}{.30\textwidth}
  \centering
  \includegraphics[width=.8\linewidth]{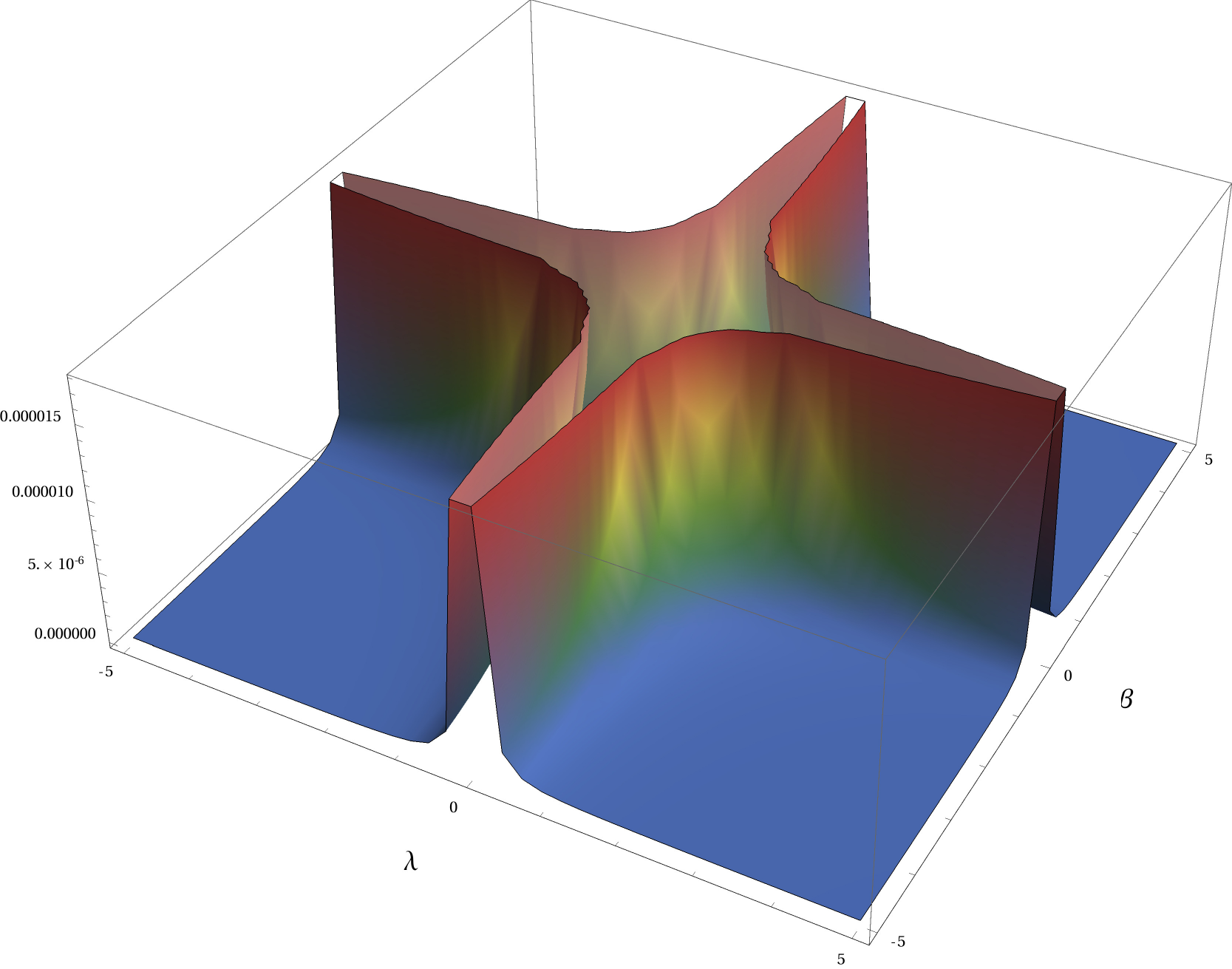}
  \label{fig:detcomp10}
\end{subfigure}
\hfill
\begin{subfigure}{.30\textwidth}
  \centering
  \includegraphics[width=.8\linewidth]{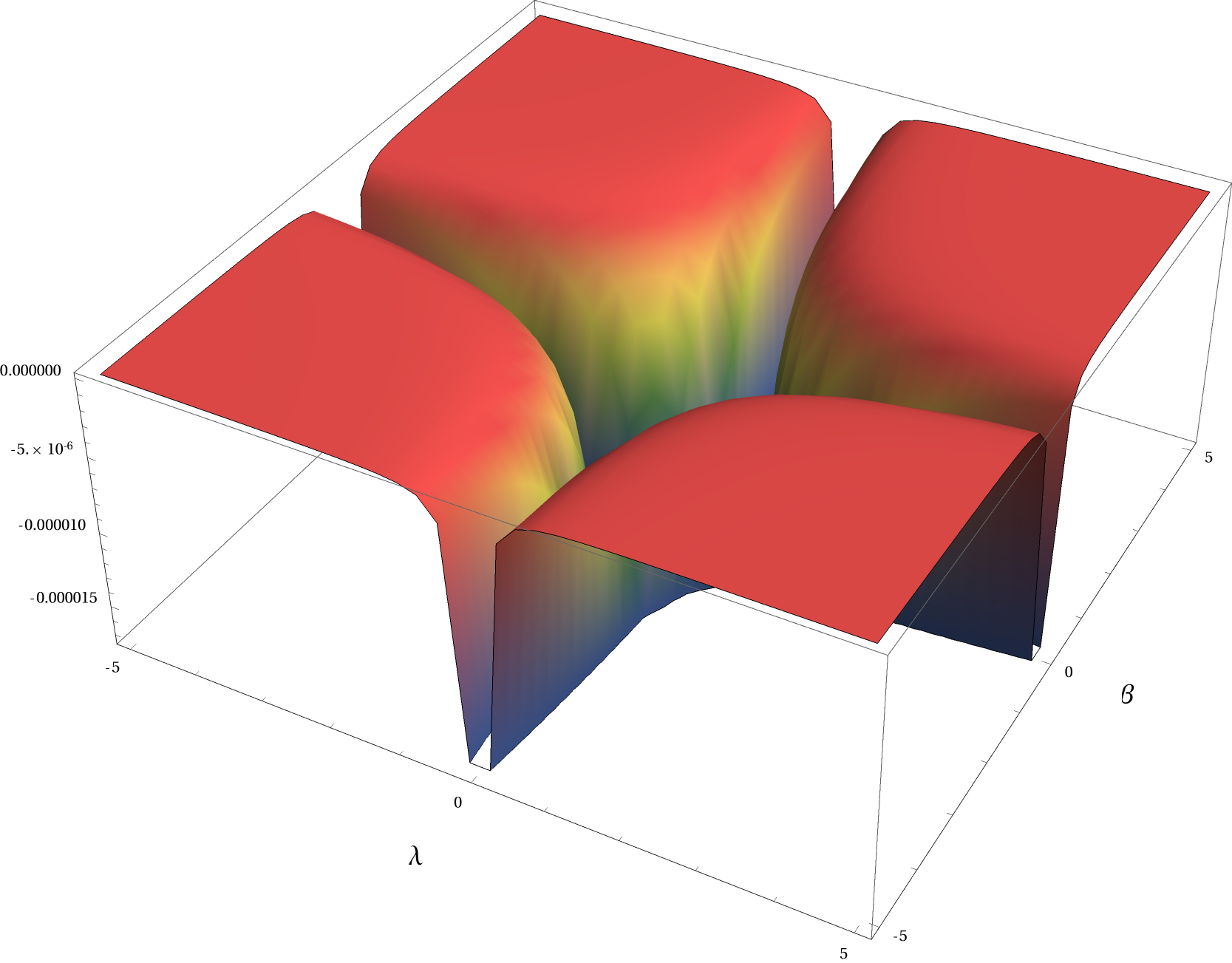}
  \caption*{$\det[G(\lambda,\beta)]$}
  \label{fig:detcomp11}
\end{subfigure}
\hfill
\begin{subfigure}{.30\textwidth}
  \centering
  \includegraphics[width=.8\linewidth]{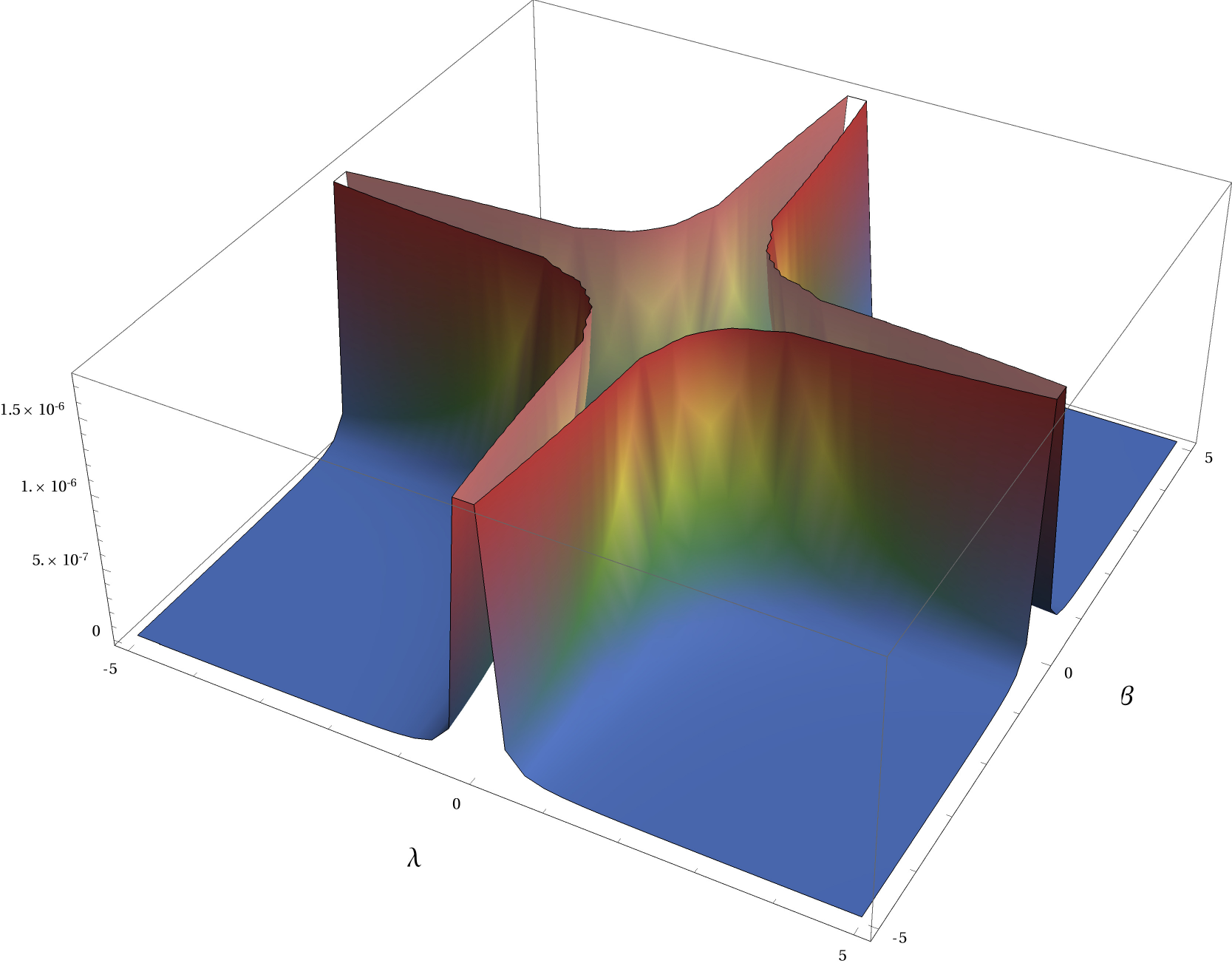}
  \label{fig:detcomp12}
\end{subfigure}
\caption{Behaviour of the determinant of the QGT of the symmetrically coupled Toda oscillators, for analytical and perturbative solutions. The left column has the analytical solution, in the center the first order approximation and the right column the second order.}
\label{fig:DetPertSymToda}
\end{figure}

In Fig. \ref{fig:PertCompSymToda}, we can see the projections of the components with the most visible difference between the analytical and the perturbative solution. Notice how several of them present an inversion in the direction of the divergence, such as the $G_{\kappa\lambda}$ component (discussed in the previous section and shown in Fig. \ref{fig:2DGkplamvarlam} and Fig. \ref{fig:2DGkplamvark}), which is particularly sensitive to the perturbative order.
\newpage
A similar analysis is made to the determinant of the QGT depicted in Fig. \ref{fig:DetPertSymToda}. In particular, the projections $\det[G(k,\lambda)]$, $\det[G(\kappa,\lambda)]$, and $\det[G(\lambda,\beta)]$ are considerably sensitive to the perturbative order. In the analytical case, every determinant diverges positively. However, in several instances, the order of the perturbation induces a negatively defined determinant, which would imply a non-positively defined distance. This is just an illusion of the perturbative analysis since it is not present at all in the exact solution.
\newpage

Due to the given symmetry between the parameters $\lambda$ and $\beta$ in \eqref{eq:Sym_Toda_L}, the quantum phase transition arising from the limit $\lambda\to 0$, as stated in \eqref{eq:lambda_0_L_SymToda} and \eqref{eq:lambda 0 H SymToda}, will similarly occur for $\beta\to 0$. In order to explore a different phase transition, it is necessary to modify the potential in the Lagrangian, as we will see in our next example.

\subsection{Anharmonic oscillator coupled with a Toda oscillator}\label{sec:Anharmonic_Toda}

As a second example, we consider an anharmonic oscillator coupled with a Toda oscillator, which again is given by a two-dimensional configuration space $\vec{x}=(x, y)$ and a four-dimensional parameter space $\vec{\lambda}=(k,\kappa,\lambda,\beta)$. So, the Lagrangian reads
\begin{equation}\label{eq:Anh_Toda_L}
\begin{aligned}
    \mathcal{L}(\vec{x},\vec{\lambda}) = \frac{1}{2}\big[\lambda^2 x^4\dot{x}^{2}& + \beta^2 e^{-2\beta y}\dot{y}^2\\ 
    &-k(\lambda^2 x^4 + e^{-2\beta y}) -\kappa \left(\lambda x^2- e^{-\beta y}\right)^2 \big]
\end{aligned}
\end{equation}
and the Hamiltonian  
\begin{equation}\label{eq:Anh_Toda_H}
H(\vec{x},\vec{\lambda})=\frac{1}{2}\left[g^{\mu\nu}p_\mu p_\nu +k\left(\lambda^2 x^4 + e^{-2\beta y}\right)+\kappa\left(\lambda x^2 - e^{-\beta y}\right)^2\right],
\end{equation}
with the diagonal metric
\begin{equation}
g_{\mu \nu}=\left(\begin{array}{cc}
4\lambda^2 x^2 & 0\\
0 & \beta^2 e^{-2 \beta y} 
\end{array}\right).
\end{equation}
and determinant $g=4\lambda^2 \beta^2 x^2 e^{-2 \beta y} $.

Then, the deformation vector \eqref{eq:sigma} is
\begin{equation}\left(\begin{array}{c}
\sigma_k \\
\sigma_\kappa \\
\sigma_\lambda\\
\sigma_\beta
\end{array}\right)=\left(\begin{array}{c}
0 \\
0 \\
-2  \lambda^{-1} \\
-2 \beta^{-1}+2y
\end{array}\right).
\end{equation}

Following the same procedure as in our previous example, we introduce the canonical transformations
\begin{equation}\label{eq:canonical_u1u2_anharmonic}
    U_1 = \lambda x^2, \qquad U_2 = e^{-\beta y}, \qquad P_1 = \frac{p_1}{2\lambda x}, \qquad P_2 = -\frac{e^{\beta y}}{\beta}p_2,
\end{equation}
which allow us to decouple the oscillators by introducing the normal coordinates \eqref{eq:can_trans_upum}. 

\begin{figure}[!h]
\begin{subfigure}{.25\textwidth}
  \centering
  \includegraphics[width=.8\linewidth]{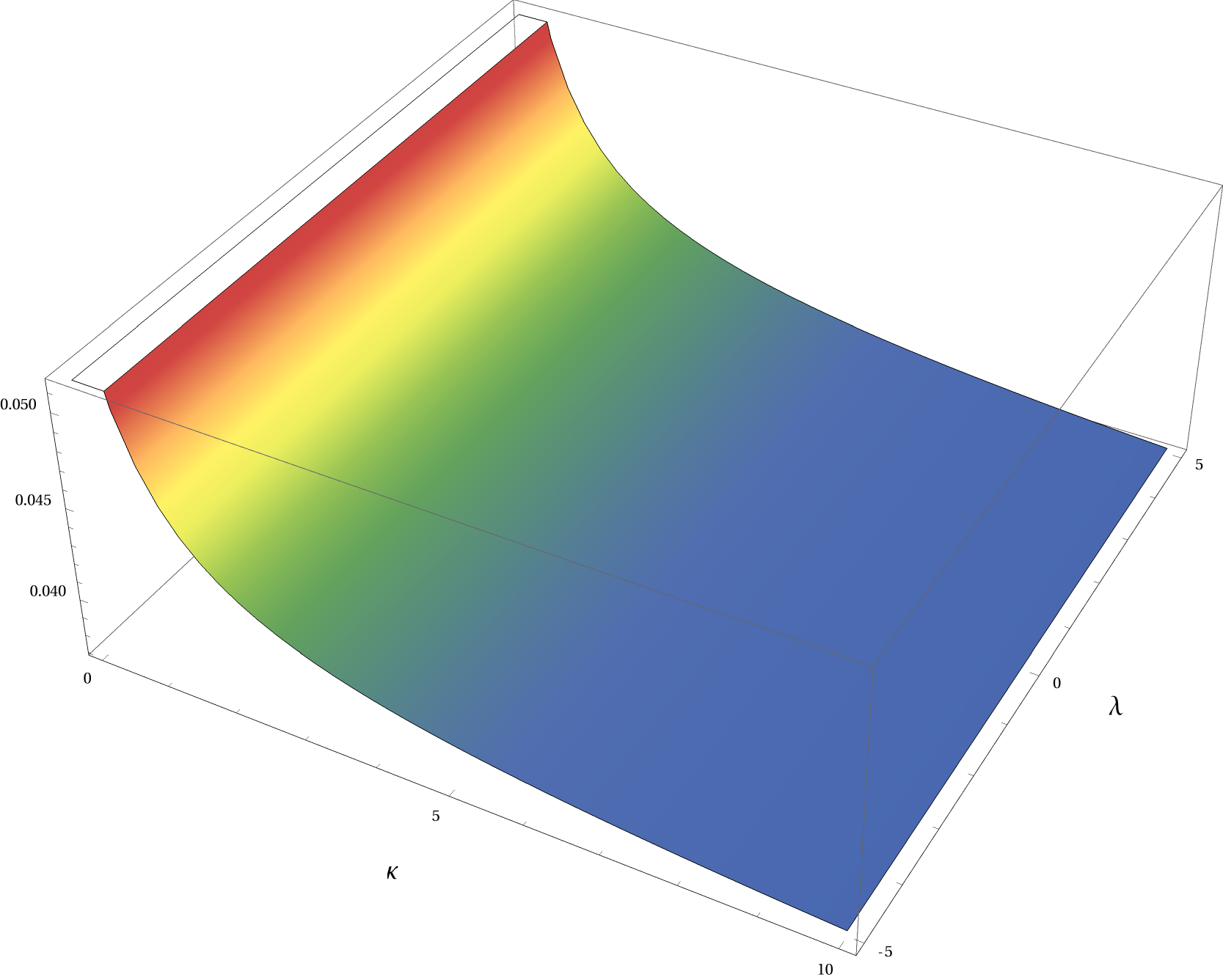}
  \caption{$G_{kk}$}
  \label{fig:combined1}
\end{subfigure}
\hfill
\begin{subfigure}{.25\textwidth}
  \centering
  \includegraphics[width=.8\linewidth]{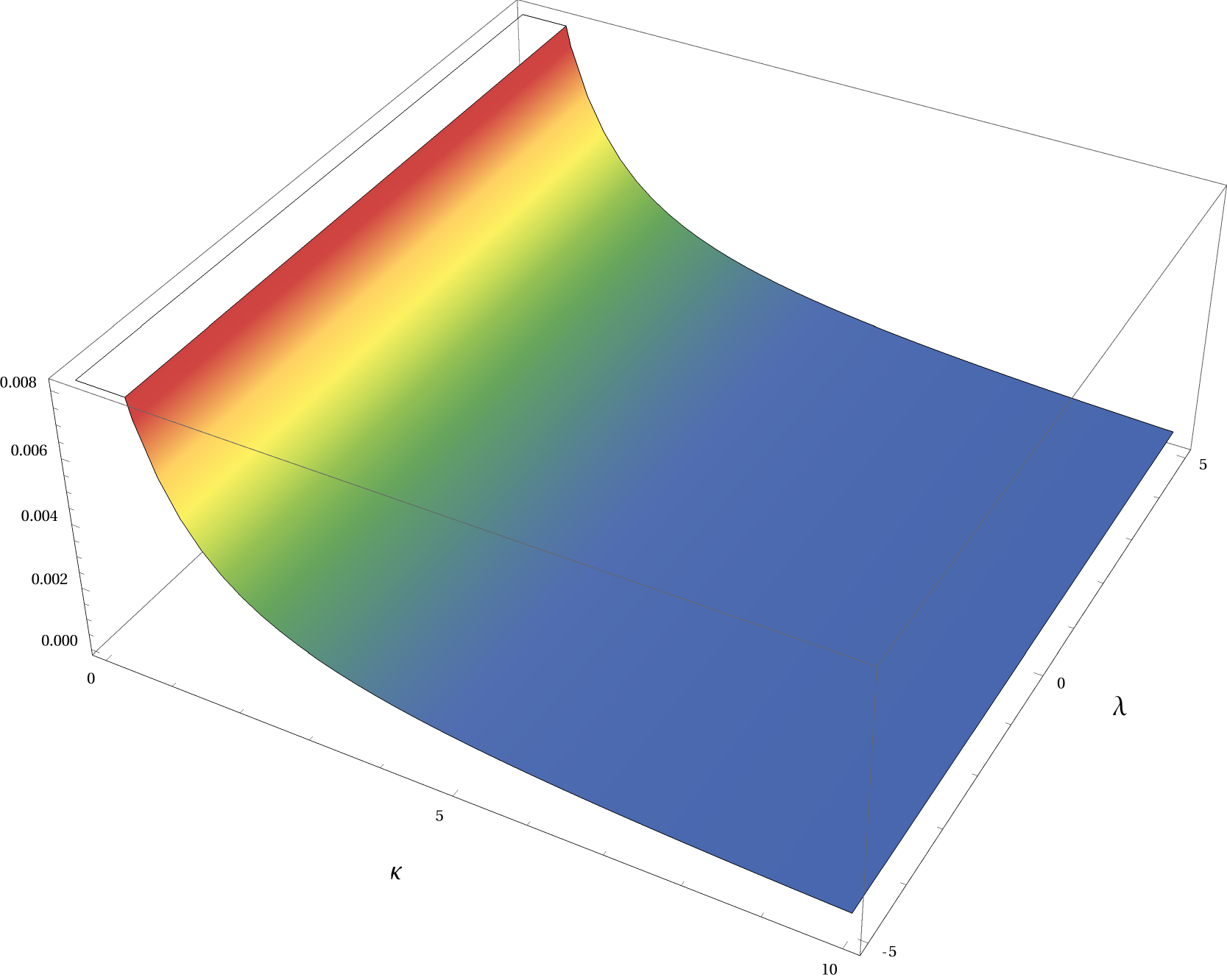}
  \caption{$G_{k\kappa}$}
  \label{fig:combined2}
\end{subfigure}
\hfill
\begin{subfigure}{.25\textwidth}
  \centering
  \includegraphics[width=.8\linewidth]{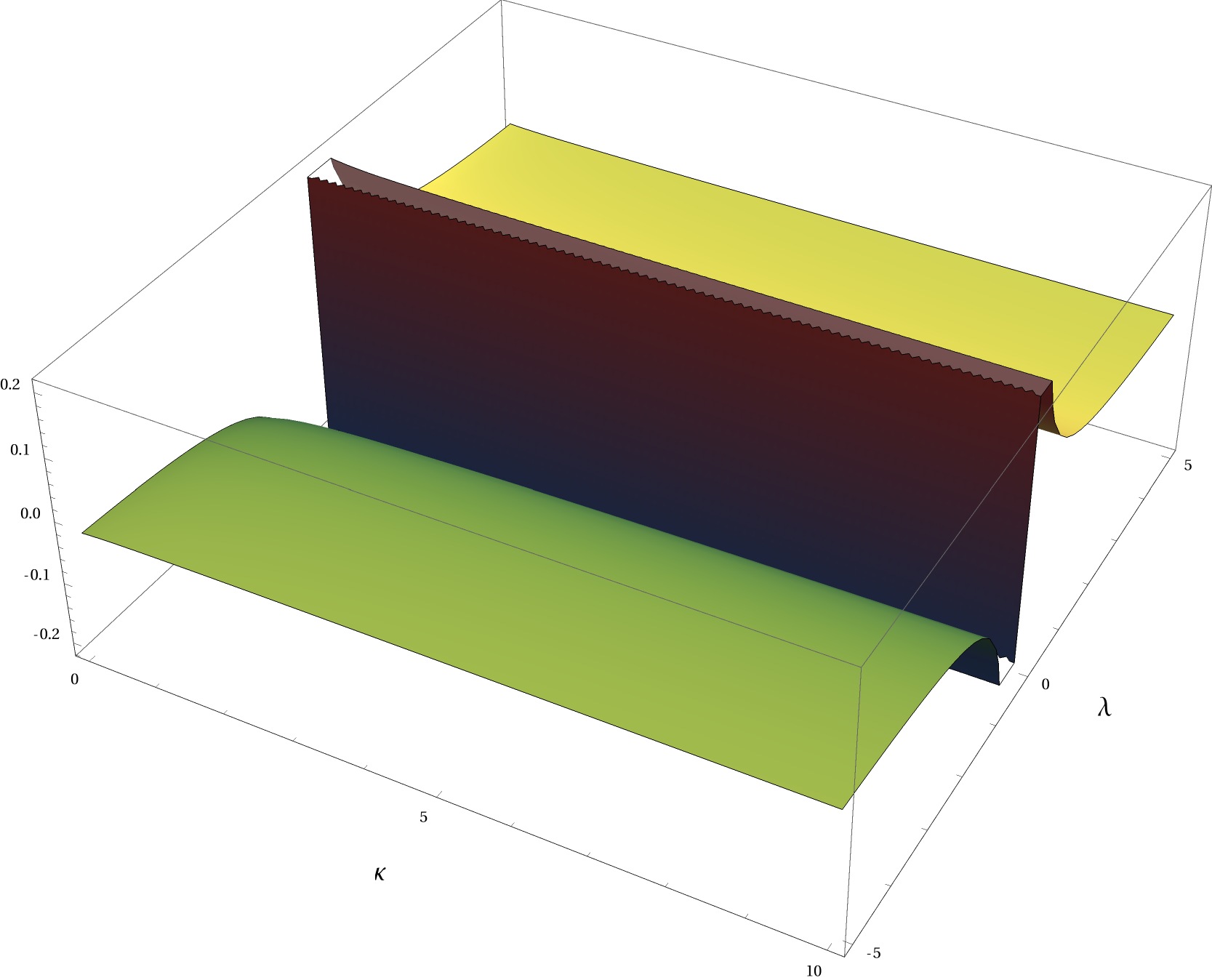}
  \caption{$G_{k\lambda}$}
  \label{fig:combined3}
\end{subfigure}
\\
\begin{subfigure}{.25\textwidth}
  \centering
  \includegraphics[width=.8\linewidth]{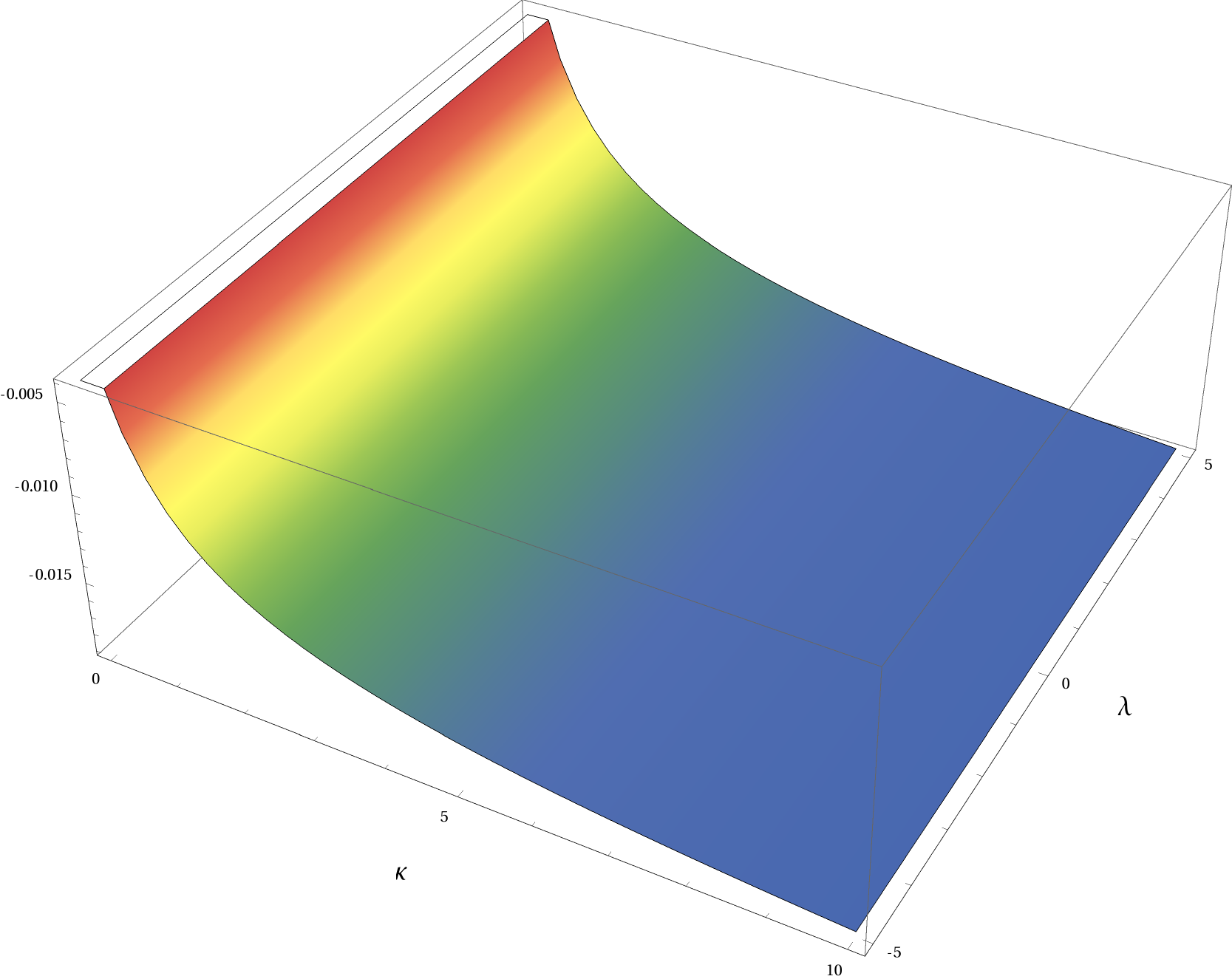}
  \caption{$G_{k\beta}$}
  \label{fig:combined4}
\end{subfigure}
\hfill
\begin{subfigure}{.25\textwidth}
  \centering
  \includegraphics[width=.8\linewidth]{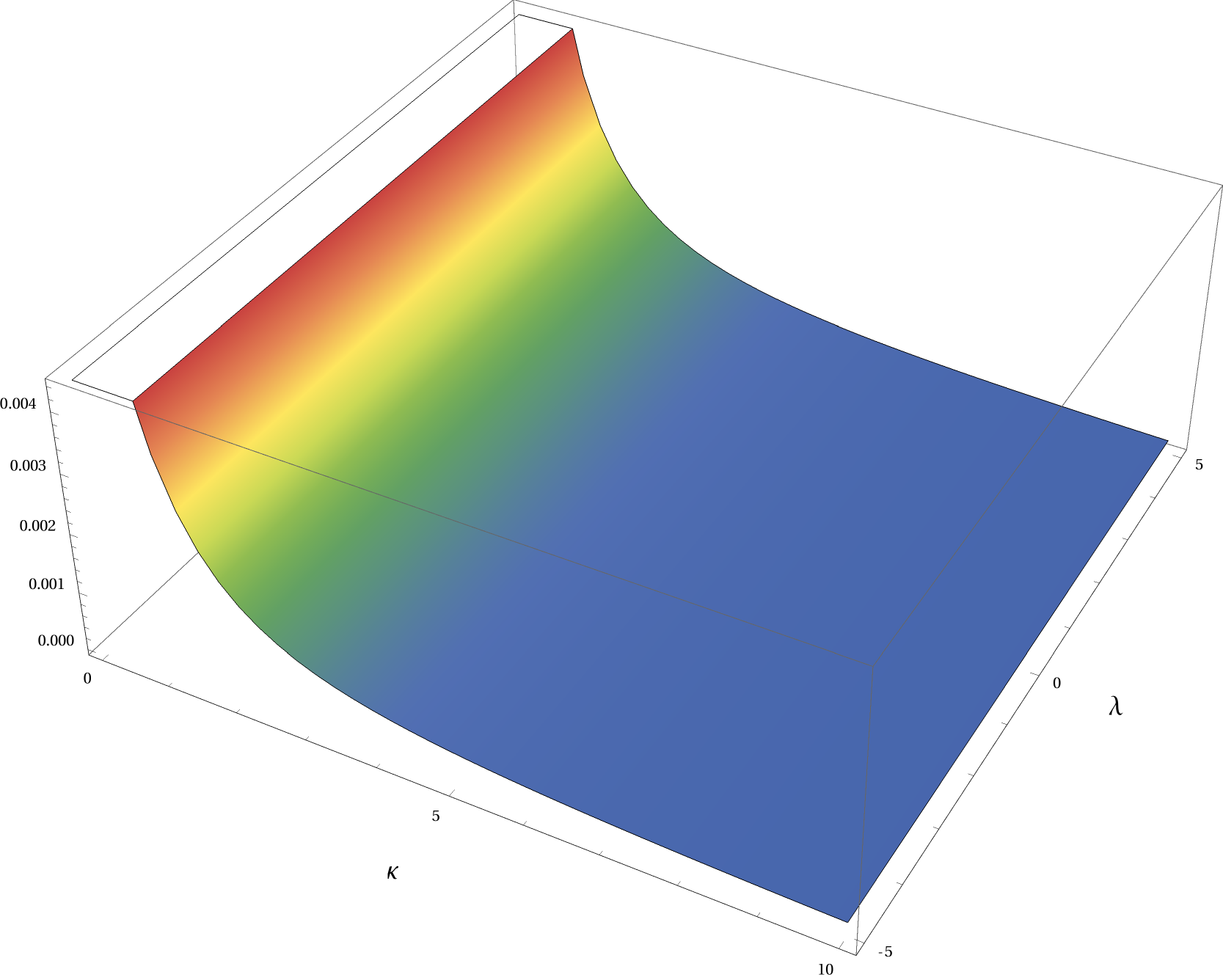}
  \caption{$G_{\kappa\kappa}$}
  \label{fig:combined5}
\end{subfigure}
\hfill
\begin{subfigure}{.25\textwidth}
  \centering
  \includegraphics[width=.8\linewidth]{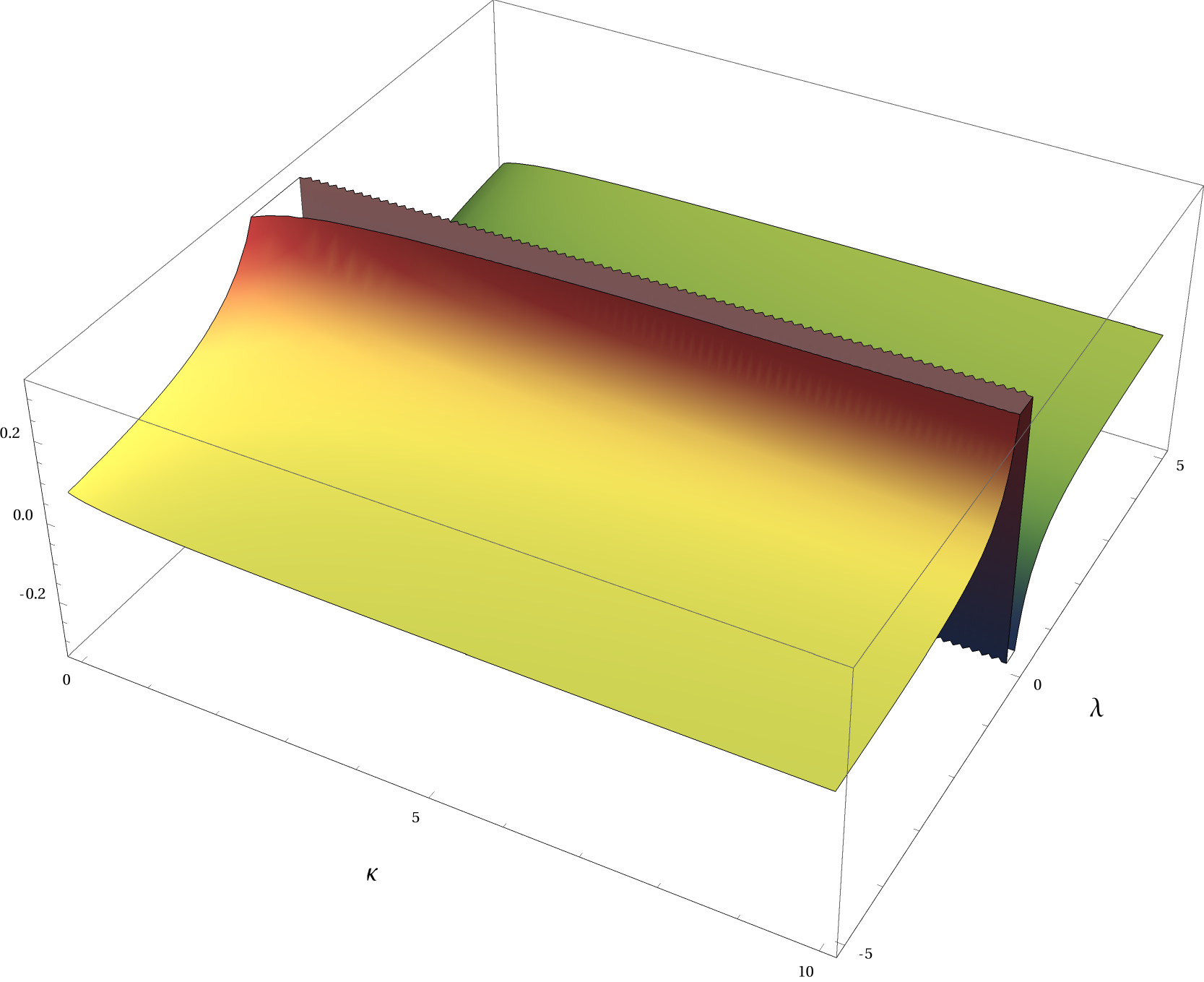}
  \caption{$G_{\kappa\lambda}$}
  \label{fig:combined6}
\end{subfigure}
\\
\begin{subfigure}{.25\textwidth}
  \centering
  \includegraphics[width=.8\linewidth]{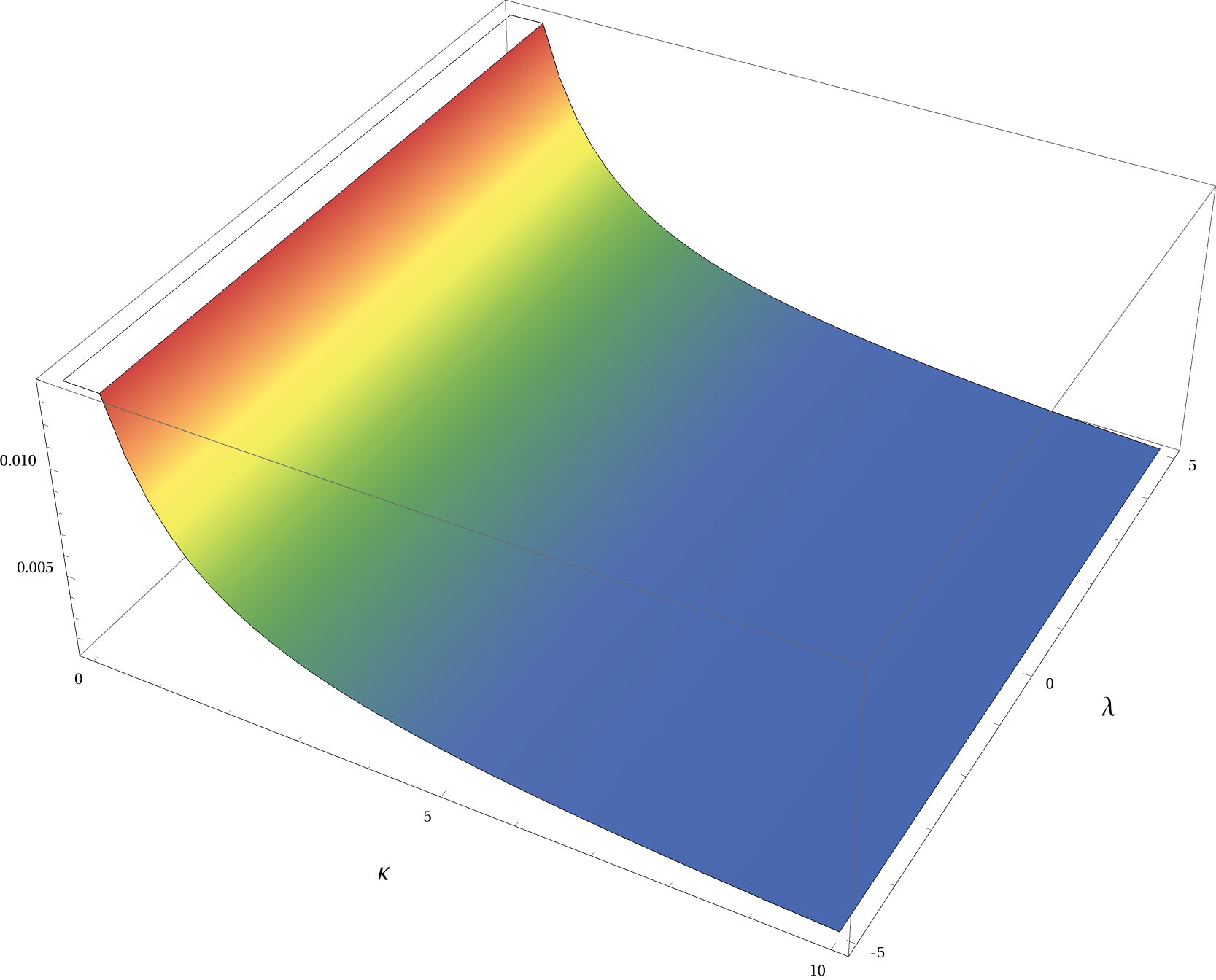}
  \caption{$G_{\kappa\beta}$}
  \label{fig:combined7}
\end{subfigure}
\hfill
\begin{subfigure}{.25\textwidth}
  \centering
  \includegraphics[width=.8\linewidth]{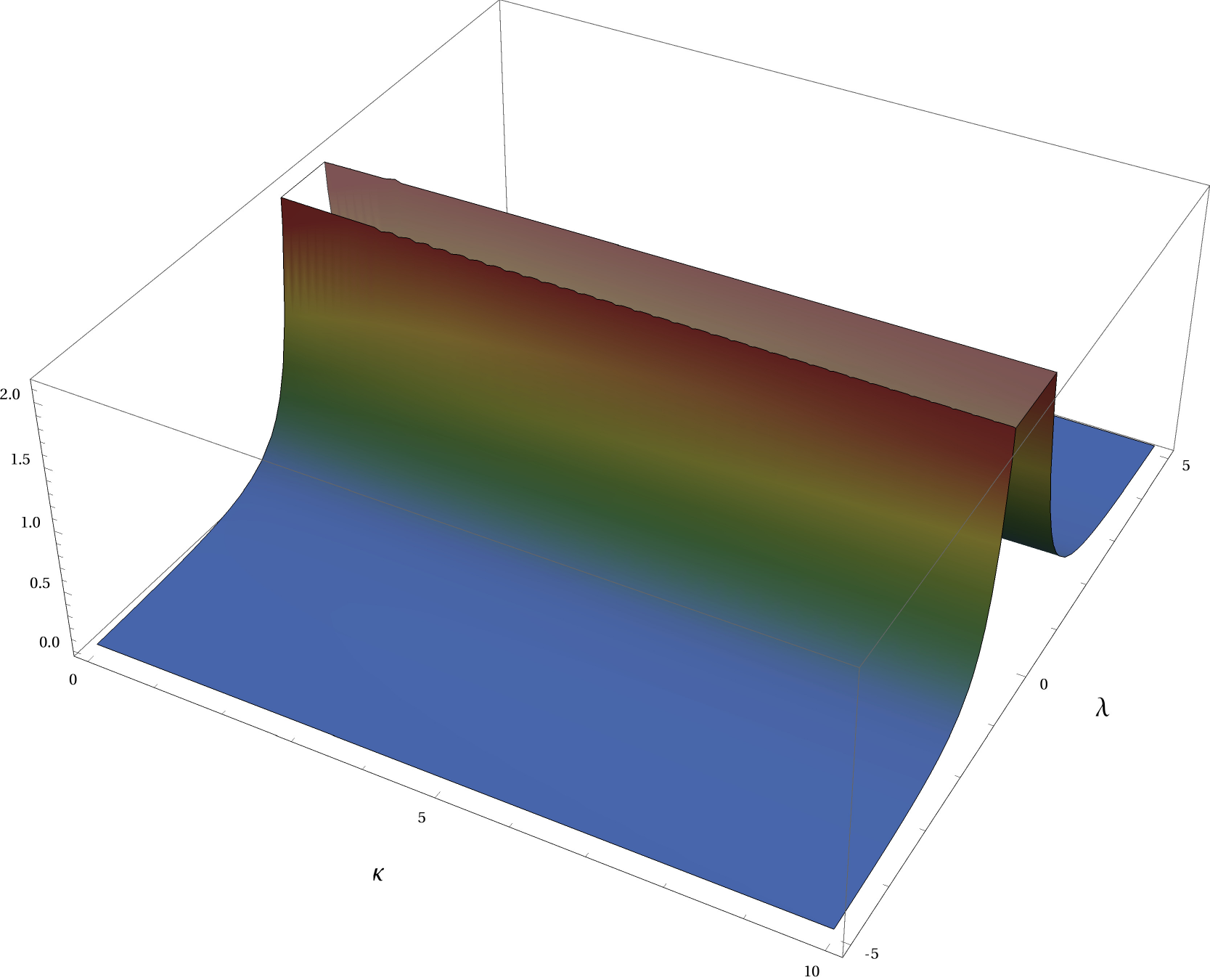}
  \caption{$G_{\lambda\lambda}$}
  \label{fig:combined8}
\end{subfigure}
\hfill
\begin{subfigure}{.25\textwidth}
  \centering
  \includegraphics[width=.8\linewidth]{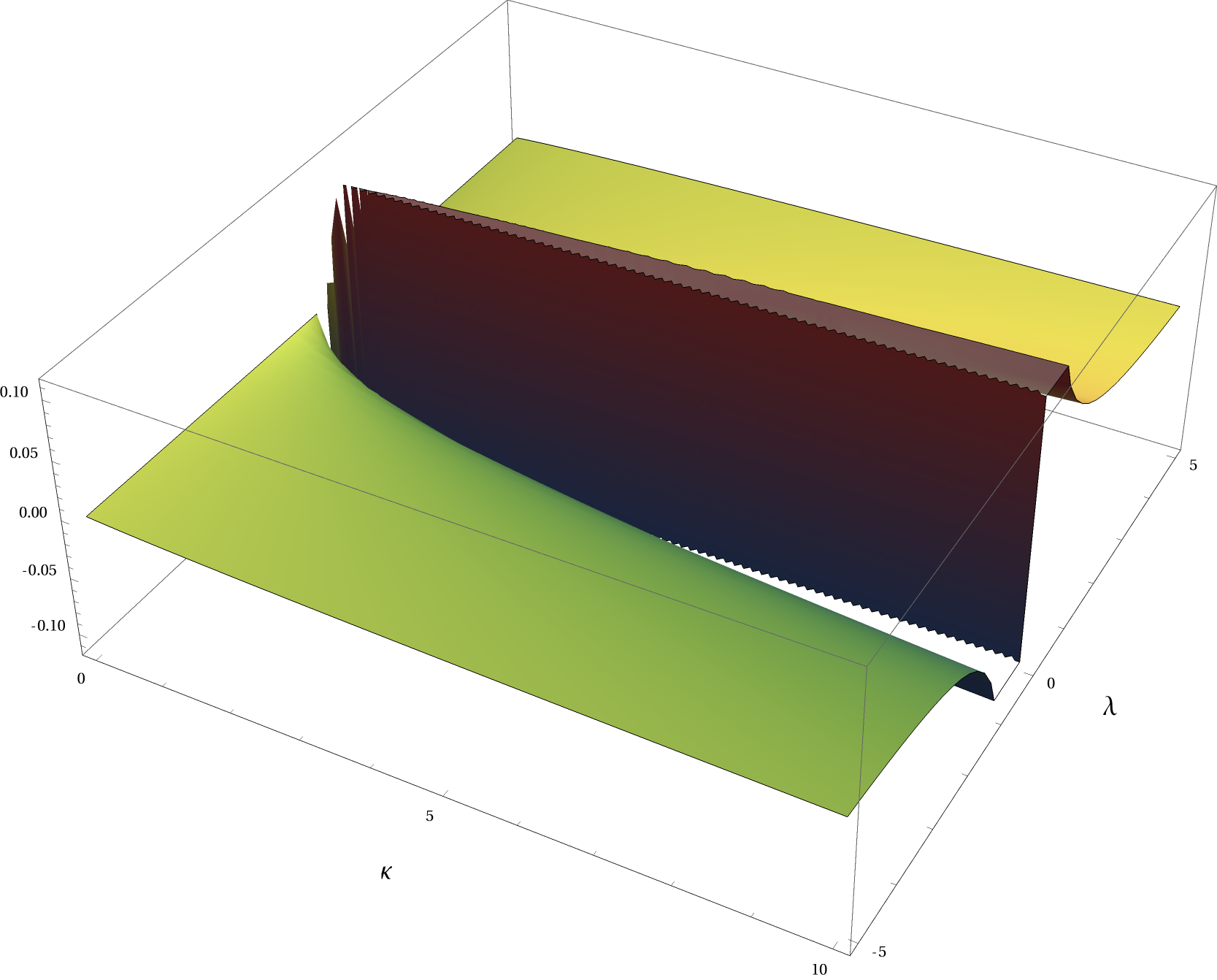}
  \caption{$G_{\lambda\beta}$}
  \label{fig:combined9}
\end{subfigure}
\\
\begin{subfigure}{.25\textwidth}
  \centering
  \includegraphics[width=.8\linewidth]{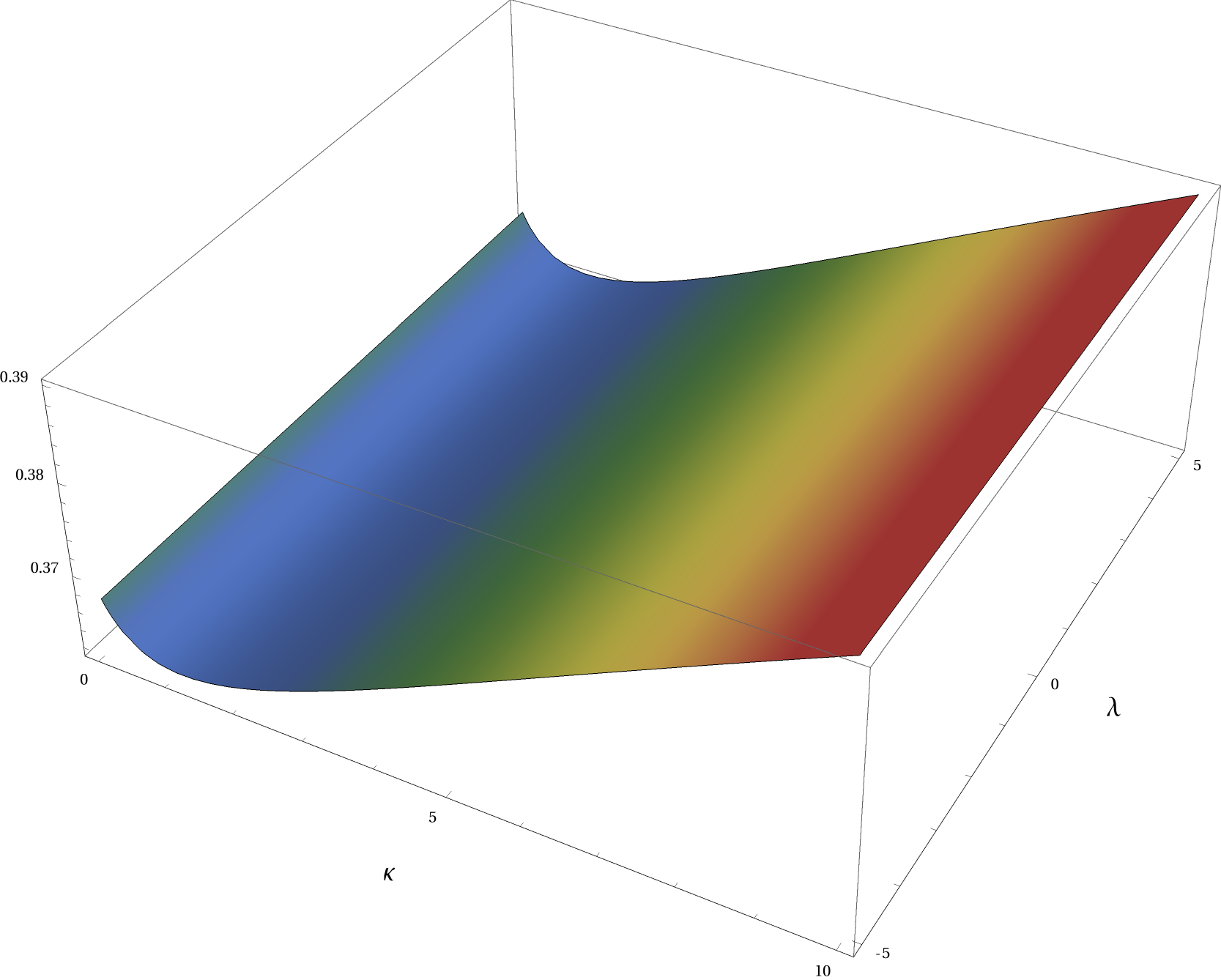}
  \caption{$G_{\beta\beta}$}
  \label{fig:combined10}
\end{subfigure}
\hfill
\begin{subfigure}{.25\textwidth}
  \centering
  \includegraphics[width=.8\linewidth]{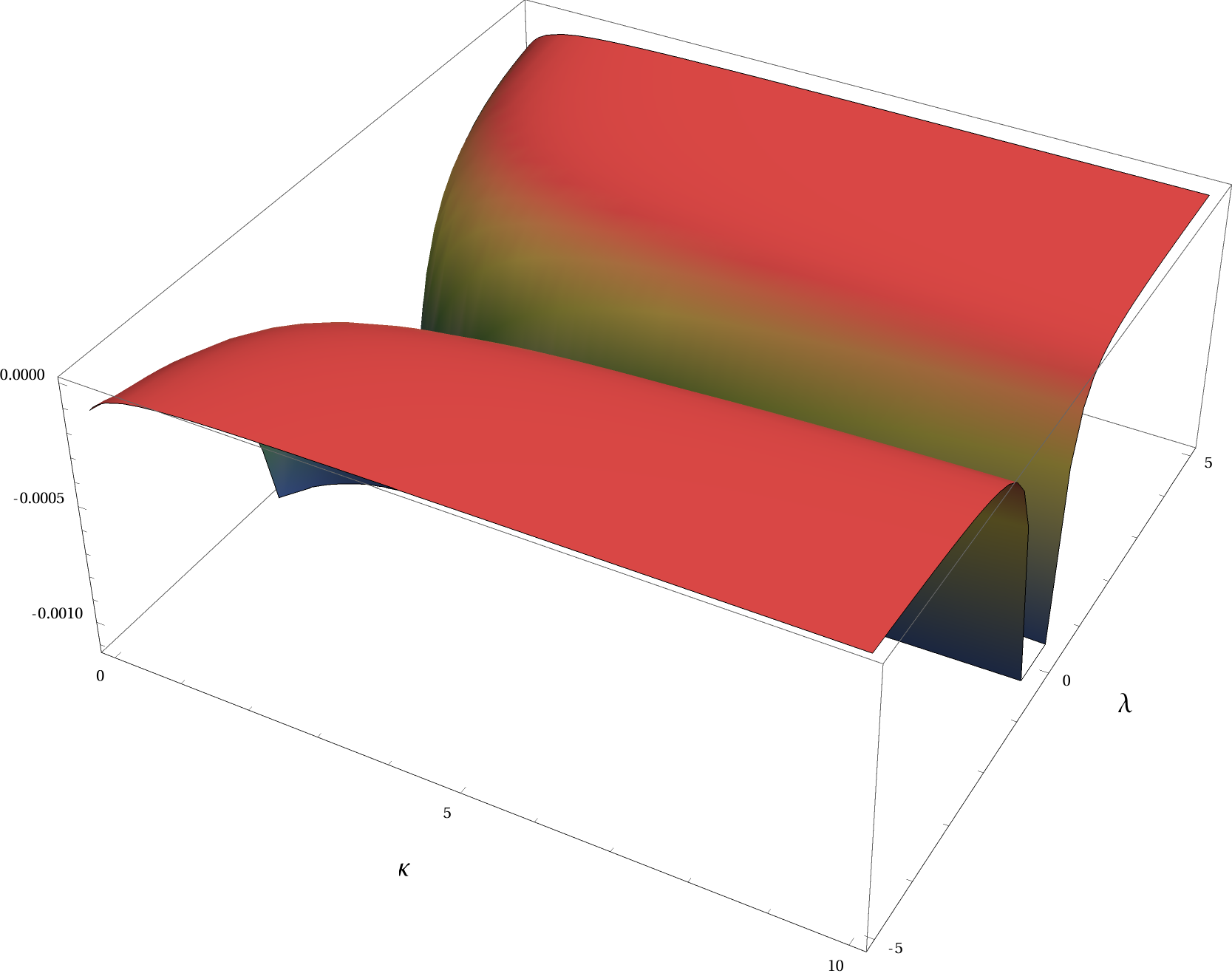}
  \caption{$\det[G]$}
  \label{fig:combined11}
\end{subfigure}
\hfill
\begin{subfigure}{.25\textwidth}
  \centering
  \includegraphics[width=.8\linewidth]{blank.pdf}
\end{subfigure}
\caption{QGT components for the coupled anharmonic and Toda oscillators in terms of $(\kappa,\,\kappa)$ with $\lambda=\beta=1$.}
\label{fig:AnhToda_QGT}
\end{figure}

\begin{figure}[!h]
\centering
  \includegraphics[width=10cm]{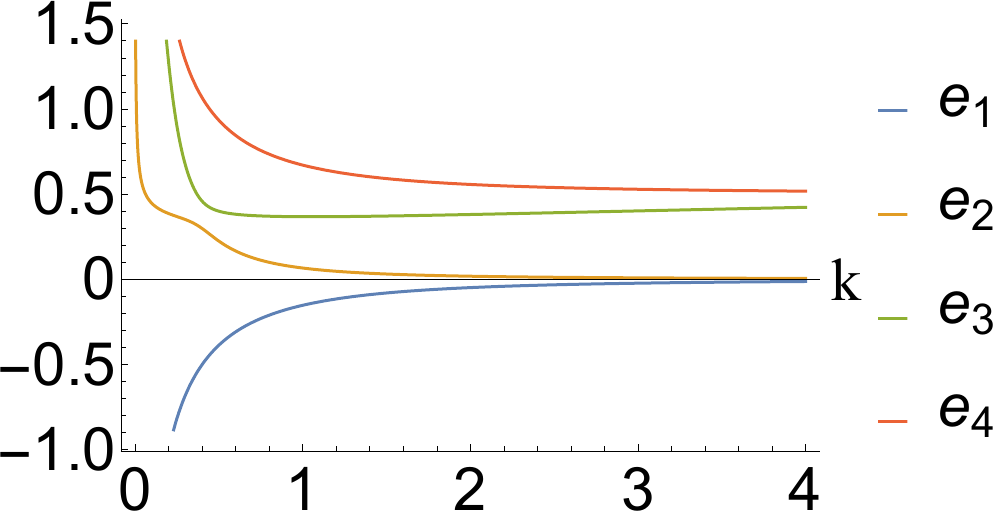}
\caption{Eigenvalues of the QGT for the anharmonic oscillator coupled to the Toda oscillator in terms of $k$ with $\kappa=0.1,\,\alpha=\beta=1$. The vertical axis is the value of the eigenvalue of the metric $G_{ij}$, and the horizontal is for $k$\ (No particular order is taken for the values of the eigenvalues). See Appendix \ref{graphicseigen} for complementary information.}
\label{fig:EigenAnhToda}
\end{figure}

Beware, as even though the transformations applied to both our previous and current examples, it might make them appear similar, but there are crucial differences in the integration range and normalization constant.
The normalization constant is 
\begin{equation}
    \mathcal{N}=\sqrt{\frac{\sqrt[4]{k (k+2 \kappa)}}{\pi -\arctan\left(\frac{\sqrt{2} \sqrt[4]{k (k+2 \kappa)}}{\sqrt{k+\kappa-\sqrt{k (k+2 \kappa)}}}\right)}}
\end{equation}
In the integration process, we use $U_1$ and $U_2$ defined in (\ref{eq:canonical_u1u2_anharmonic}). The domain for $x$ and $y$ is $\mathbb{R}$, with this $U_1$\,(an even function of $x$) is integrated through
\begin{equation}
    \int_{-\infty}^\infty\,dx\rightarrow\,2\int_0^{\infty}\,dU_1
\end{equation}
and $U_2$\,(a function of $y$) is integrated like
\begin{equation}
    \int_{-\infty}^\infty\,dy\rightarrow\,-\int_0^{\infty}\,dU_2
\end{equation}
Once again the expressions for the analytical solution of the QGT are too cumbersome to write them explicitly. In Fig. \ref{fig:AnhToda_QGT} we present the projection graphs of the entries varying $(k,\lambda)$ while fixing $(\kappa=\beta=1)$. One of the striking features of this model is the QGT determinant negativity (Fig. \ref{fig:AnhToda_QGT}k), we can argue this happened because the off-diagonal elements has bigger size than the diagonal elements ($G_{kk},\,G_{\kappa\kappa},\,G_{\lambda\lambda},\,G_{\beta\beta}>0$) and this cause the negative sign. However, when we calculate the eigenvalues of the QGT, we obtain that one of them is always negative, just like in Fig. \ref{fig:EigenAnhToda}.

Similarly to our approach in the previous example, let us study what happens to our system as $\lambda\to 0$. Then, the Lagrangian given by \eqref{eq:Anh_Toda_L} becomes
\begin{equation}\label{eq:lambda_0_L_AnhToda}
    L(y,\vec{\lambda})=\frac{1}{2}[\beta^2 e^{-2\beta y}\dot{y}^2 -(k+\kappa) e^{-2\beta y}].
\end{equation}
By introducing a canonical transformation, our system is then reduced to a standard harmonic oscillator.

On the other hand, if $\beta\to 0$, the Lagrangian is now 
\begin{equation}\label{eq:beta_0_L_AnhToda}
    \mathcal{L}(x,\vec{\lambda}) = \frac{1}{2}\big[\lambda^2 x^4\dot{x}^{2}-(k+\kappa)\lambda^2 x^4 +2 \kappa\lambda x^2  -(k+\kappa)\big].
\end{equation}
which corresponds to a double well \cite{VergaraEnPreparacion} and once the canonical transformation \eqref{eq:canonical_u1u2_anharmonic} has taken place, we return to the oscillator with the linear term, as in the previous example.

In our following example, we will construct a system to study the effects of the extra terms in the Berry curvature.
\newpage

\subsection{Exponential oscillator coupled to a generalized harmonic oscillator}\label{sec:generalizado_exponencial}
In this example, the idea is to consider a system with a non-vanishing Berry curvature. To do that, we consider a term linear in the velocity analogously to a minimal coupling between an electromagnetic field and a particle.  Following \eqref{potve}, we consider the Lagrangian

\begin{equation}
\begin{aligned}
\mathcal{L}\left(\vec{x},\dot{\vec{x}},\vec{\lambda}\right) & = \frac{1}{2}x^2\dot{x}^2 + \frac{1}{2}\left(\lambda^2e^{-2\lambda y}\right)\dot{y}^2 - Y\left[\frac{x^3 \dot{x}}{2} - \lambda e^{-2\lambda y} \dot{y}\right]\\
& \quad + \left(\frac{2Y^2-(\alpha+\alpha')}{4}\right)\left[\frac{x^4}{4}+e^{-2\lambda y}\right] - \left(\frac{\alpha-\alpha'}{2}\right)\left[\frac{x^2 e^{-\lambda y}}{2}\right]
\end{aligned}\label{eq:LagExpReg}
\end{equation}
whose corresponding Hamiltonian is
\begin{equation}
\begin{aligned}
\mathcal{H}\left(\vec{x},\vec{p},\vec{\lambda}\right) & = \frac{p_{x}^2}{2x^2} + \frac{e^{2\lambda y}}{2\lambda^2}p_{y}^2 + \frac{Y}{2}\left[ \frac{1}{x}\left(\frac{x^2}{2}p_{x} + p_{x}\frac{x^2}{2}\right) - \frac{e^{\lambda y}}{\lambda}\left(e^{-\lambda y}p_{y} + p_{y}e^{-\lambda y}\right)\right]\\
& \quad + \left(\frac{\alpha+\alpha'}{4}\right)\left[\frac{x^4}{4}+e^{-2\lambda y}\right] + \left(\frac{\alpha-\alpha'}{2}\right)\left[\frac{x^2 e^{-\lambda y}}{2}\right],
\end{aligned}\label{eq:HamExpReg}
\end{equation}
where the metric tensor is
\begin{equation}
    g_{\mu\nu}(x,y,\lambda) = \left(\begin{matrix}
        x^2 & 0\\
        0 & \lambda^2e^{-2\lambda y}
    \end{matrix} \right)\label{4.11}
\end{equation}
and the only no null contribution to the deformation vector $\sigma_i$, is
\begin{equation}
    \sigma_\lambda= 2\left(y - \frac{1}{\lambda}\right).
\end{equation}

 Then, the Schrödinger equation reads
\begin{equation}
\begin{split}
E_n\psi_n & = -\frac{1}{2}\left(\frac{1}{x}\frac{\partial}{\partial x}\right)^2\psi_n -\frac{1}{2}\left(\frac{e^{\lambda y}}{\lambda}\frac{\partial}{\partial y}\right)^2\psi_n - \frac{iY}{2}\left[x\frac{\partial \psi_n}{\partial x} - \frac{2}{\lambda}\frac{\partial \psi_n}{\partial y} + 2\psi_n\right]\\
& \quad + \left(\frac{\alpha+\alpha'}{4}\right)\left[\frac{x^4}{4} + e^{-2\lambda y}\right]\psi_n + \left(\frac{\alpha - \alpha'}{2}\right)\left[\frac{x^2e^{-\lambda y}}{2}\right]\psi_n.
\end{split}\label{4.10}
\end{equation}
Now, using the change of variables
\begin{equation}
    q_1 = \frac{x^2}{2}, \qquad q_2 = e^{-\lambda y},
\end{equation}
equation \eqref{4.10} becomes
\begin{equation}
\begin{aligned}
     E_n\psi_n & = -\frac{1}{2}\frac{\partial^2 \psi_n}{\partial q_ 1^2} -\frac{1}{2}\frac{\partial^2 \psi_n}{\partial q_2 ^2} - iY\left[q_1\frac{\partial \psi_n}{\partial q_1} + q_2\frac{\partial \psi_n}{\partial q_2} + \psi_n\right]\\
     & \quad + \left(\frac{\alpha+\alpha'}{4}\right)\left[q_1^2 + q_2^2\right]\psi_n + \left(\frac{\alpha - \alpha'}{2}\right)q_1q_2\psi_n.
\end{aligned}\label{4.20}
\end{equation}
where
\begin{equation}
\begin{aligned}
    \alpha & = k^2 + (1-2k)Y^2 + Y^4,\\
    \alpha' & = (k+2\kappa)^2 + \left(1-2(k+2\kappa)\right)Y^2 + Y^4 \quad \text{with } k>Y^2.
\end{aligned}\label{4.48}
\end{equation}
Now, defining a new set of variables as
\begin{equation}
    q_+ = \frac{q_1 + q_2}{\sqrt{2}}, \quad q_- = \frac{q_1 - q_2}{\sqrt{2}},
\end{equation}
equation \eqref{4.20} takes the form
\begin{equation}
\begin{aligned}
    E_n\psi_n   & = -\frac{1}{2}\frac{\partial^2\psi_n}{\partial q_+^2} - \frac{iY}{2}\left[2q_+\frac{\p \psi_n}{\p q_+} + \psi_n\right] + \frac{\alpha}{2}q_+^2\psi_n\\
                & \quad -\frac{1}{2}\frac{\partial^2\psi_n}{\partial q_-^2} - \frac{iY}{2}\left[2q_{-}\frac{\p \psi_n}{\p q_{-}} + \psi_n\right] + \frac{\alpha'}{2}q_-^2\psi_n,
\end{aligned}\label{4.50}
\end{equation}
Taking into account the proper region of integration its ground state solution is given by

\begin{equation}\label{eq:groundStateJoshuaProblem3}
    \begin{split}
        \psi_0\left(x,y\right)  = \mathcal{N}_0 & \exp\left[-\left( \frac{\omega_+ + \omega_-}{4}\right)\left(\frac{x^4}{4} + e^{-2\lambda y}\right) - \left(\frac{\omega_+ - \omega_-}{2}\right)\frac{x^2e^{-\lambda y}}{2}\right]  \\
            &\times\exp\left[\frac{iY}{2}\left(\frac{x^4}{4} + e^{-2\lambda y}\right)\right]
\end{split}
\end{equation}
where
\begin{equation}
    \mathcal{N}_0 = \frac{\left(\omega_+\omega_-\right)^{1/4}}{\left(\arctan\left(\sqrt{\frac{\omega_-}{\omega_+}}\right)\right)^{1/2}}, \quad \omega_+ = k-Y^2, \quad \omega_-  = k + 2\kappa - Y^2.
\end{equation}

The energy of the ground state is given by
\begin{equation}
E_0 = k+\kappa-Y^2,\label{4.E}
\end{equation}

\begin{figure}[!h]
    \centering
    \includegraphics[width=0.6\linewidth]{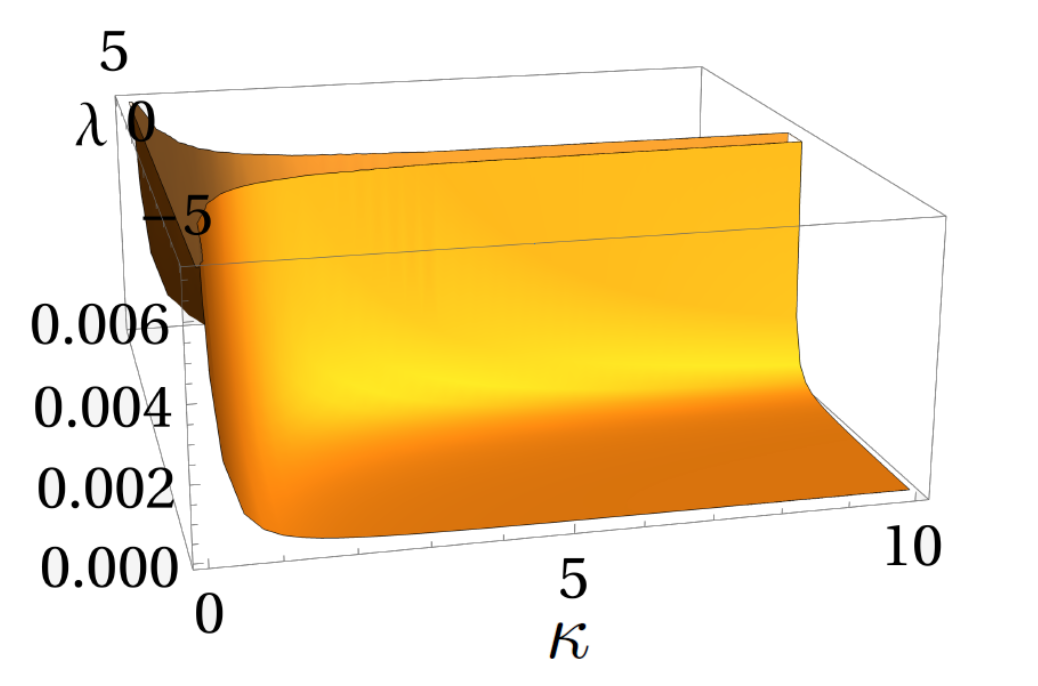}
    \caption{Projection of $\det\left[G(\kappa,\lambda)\right]$ with $Y=1$, $k = 1.5$ and $0<\kappa<10$, $-5<\lambda<5$. Singularities are observed when $\kappa \rightarrow 0$ and $\lambda\rightarrow 0$}
    \label{detMetric}
\end{figure}

\begin{figure}[h!]
\hspace{-1cm}
\begin{subfigure}{.38\textwidth}
  \centering
  \includegraphics[width=1.5\linewidth]{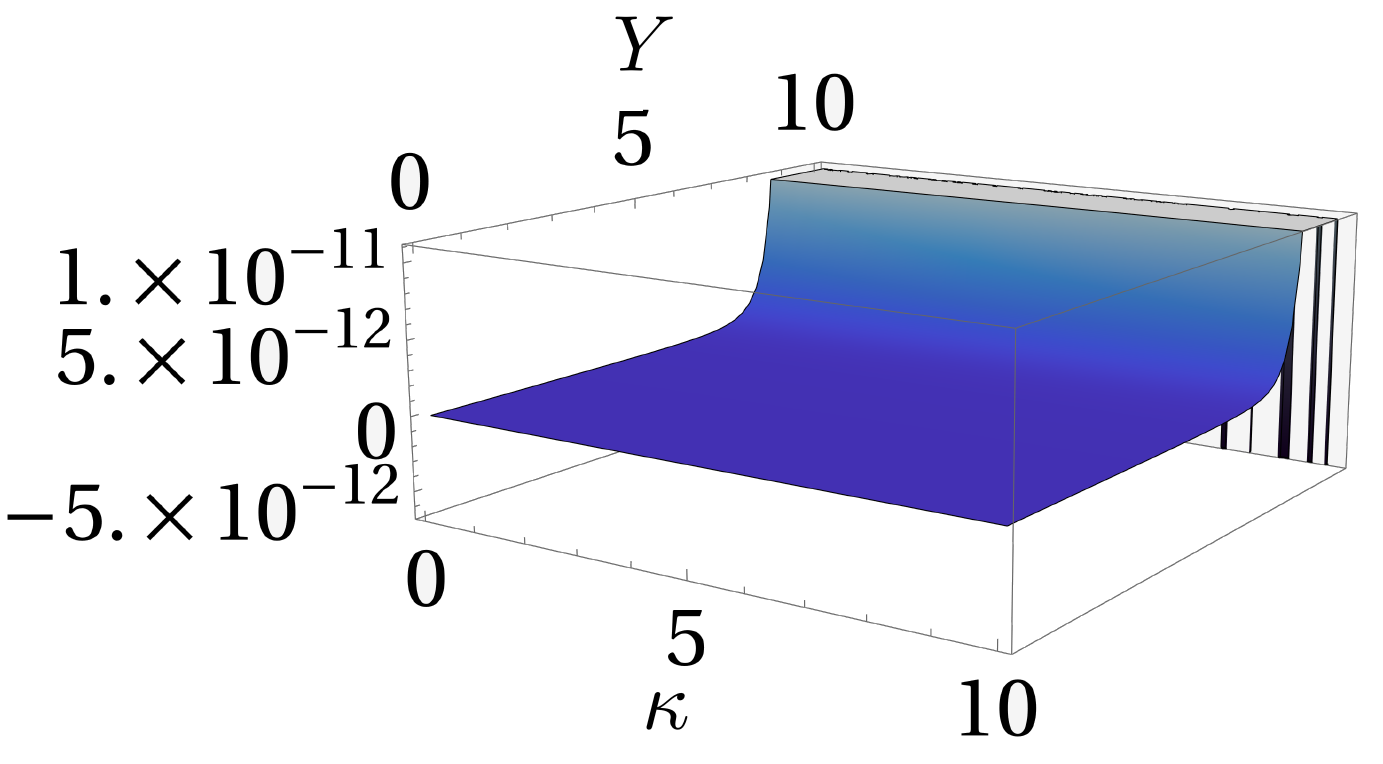}
  \caption{}
  \label{detkpY}
\end{subfigure}
\hspace{3.5cm}
\begin{subfigure}{.38\textwidth}
  \centering
  \includegraphics[width=1.4\linewidth]{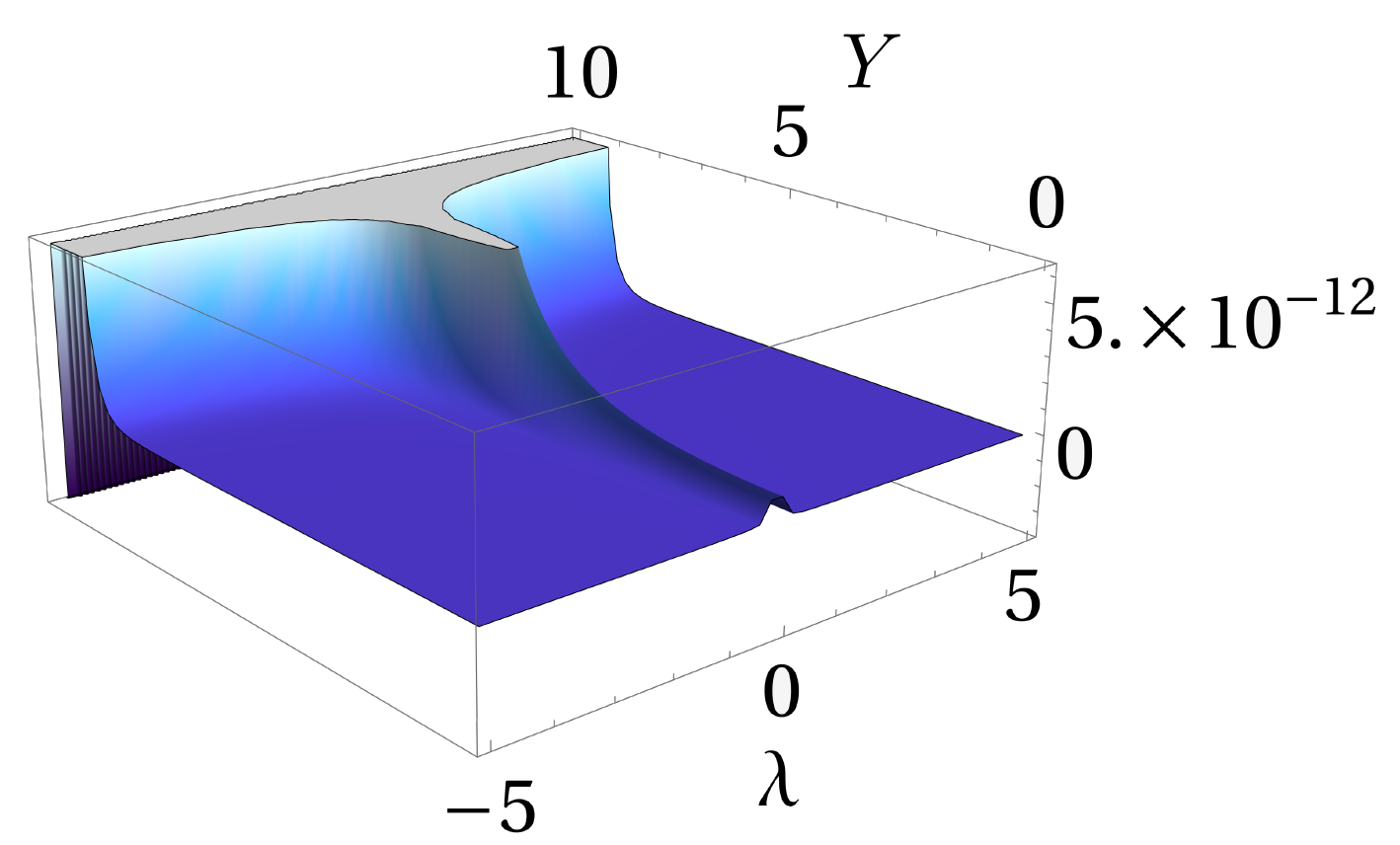}
  \caption{}
  \label{detlY}
\end{subfigure}
\par\medskip
\caption{Projection of $\det\left[G\right]$. (a) projection in the planes $k = 100$, $\lambda=1$. (b) projection in the planes $k=100$, $\kappa = 1$.}
\label{transiciónY}
\end{figure}

\begin{figure}[h!]
    \centering
    \includegraphics[width=1.1\linewidth]{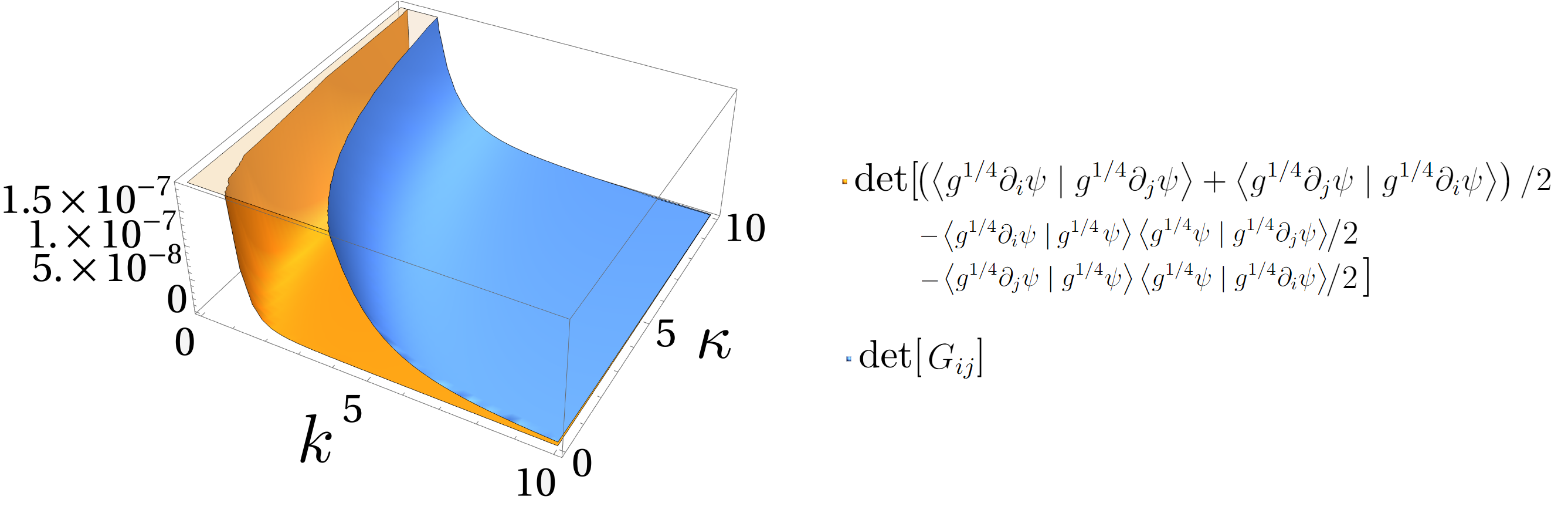}
    \caption{\small Projection of $\det\left[G\right]$ setting $Y=1$ and $\lambda = 1$. It also shows the determinant's projection without the contributions of $\sigma_i$.}
    \label{transiciónk}
\end{figure}

Quantum phase transitions for the ground state \eqref{eq:groundStateJoshuaProblem3} can be studied by analyzing the divergences of the determinant of the quantum metric tensor, as shown in Figures \ref{detMetric}, \ref{transiciónY} and \ref{transiciónk}.

First, we observe from Fig. \ref{detMetric} that $\det[G]$ tends to infinity as $\lambda$ approaches zero, which is an indication of the presence of a phase transition. When $\lambda = 0$, the ground state solution \eqref{eq:groundStateJoshuaProblem3} becomes a function only of $x$, so the system reduces to have a configuration space of just one dimension (in other words, one oscillator is turned off, but the energies \eqref{4.E} are not affected in this regime because they are independent of $\lambda$. This means, that the system retains a trace of the no longer moving oscillator.
\\
\\
On the other hand, another phase transition seems to sprout when $\kappa = 0$. At this value, $q_1$ and $q_2$ are no longer coupled in the solution \eqref{eq:groundStateJoshuaProblem3} and the oscillation frequencies become equal; that is, the system is now described by two independent oscillators in terms of the coordinates $q_1$ and $q_2$, each one with the same angular frequency. Furthermore, the energies do change, taking the value $E = k-Y^2 = \omega_+ = \omega_-$.
\newpage
The graphs in Fig. \ref{transiciónY} show the presence of a singularity when $Y^2\rightarrow k$. It follows from equation \eqref{4.48} that in this limit $\alpha\rightarrow k$, which tells us that the system splits into a harmonic oscillator and a free particle. Then, the normalization constant $\mathcal{N} = 0$, due to the fact that we require a normalizable system and the free particle is not. Thus, the system presents a phase transition to this new system.

\hfill\break
\normalsize
From the Fig. \ref{transiciónk}, it can be asserted that:
\begin{itemize}
    \item[1.] The difference between the case with $\sigma_i$ contributions and without them is remarkable and, therefore, it should not be omitted. 
    \item[2.]There is a phase transition when $k\rightarrow 0$. In this case, Schrödinger equation for $q_+$ becomes the equation of a free particle (remember that $k>Y^2$ so, if $k$ approaches zero, so do $Y$). However, $q_-$ keeps as a generalized oscillator with $E = \kappa$ and $\omega_- = 2\kappa$.
\end{itemize}
\newpage
Another feature of this system is the existence of a non-vanishing Berry curvature:
\begin{figure}[!ht]
\begin{subfigure}{.28\textwidth}
  \centering
  \includegraphics[width=\linewidth]{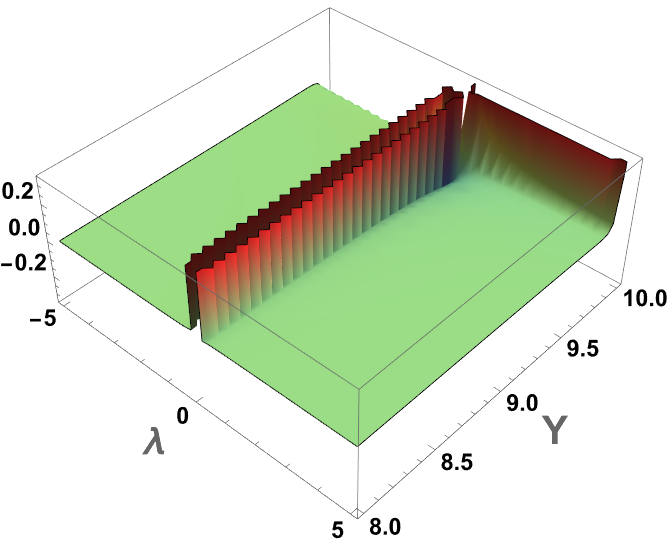}
  \end{subfigure}
\hfill
\begin{subfigure}{.28\textwidth}
  \centering
  \includegraphics[width=\linewidth]{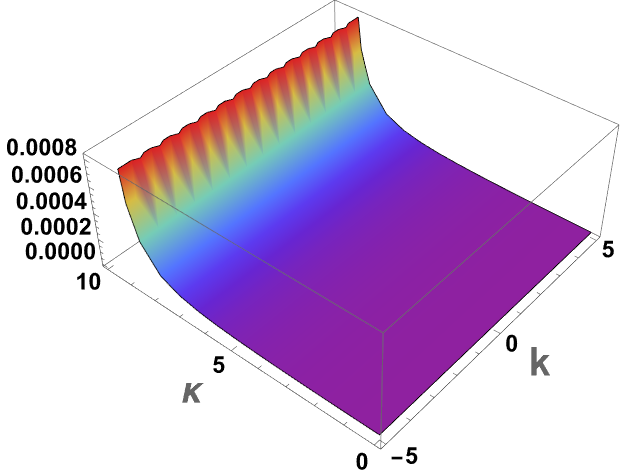}
  \end{subfigure}
\hfill
\begin{subfigure}{.28\textwidth}
  \centering
  \includegraphics[width=\linewidth]{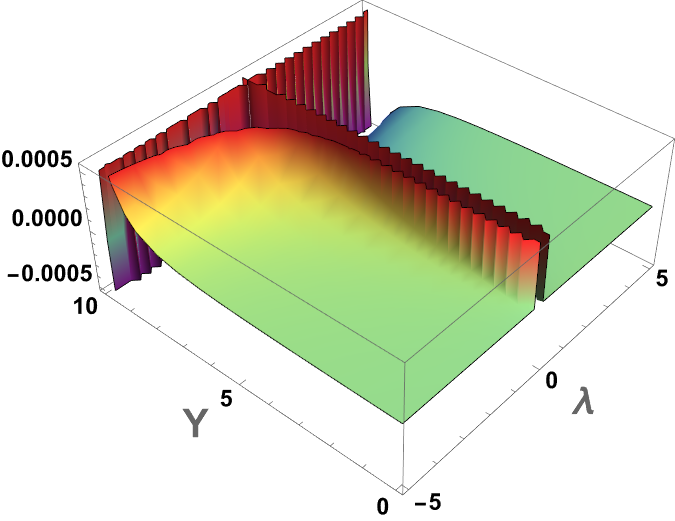}
  \end{subfigure}
\caption{Projections of: (a) $\mathcal{F}_{k\lambda}$, (b) $\mathcal{F}_{kY}$ and (c) $\mathcal{F}_{\kappa\lambda}$, with $k=100, \kappa = 1$.}
\label{fig:detOscExpAcop}
\end{figure}
\\
Actually, is easy to direct our attention to the Berry connection in the usual case and in this modification in order to compare the two Berry curvatures.

Because $\langle \sigma_\lambda \rangle\neq 0$, the modified Berry connection $\beta_\lambda$ exhibits a clear difference from its usual definition. Below are displayed graphics with the comparison of the modified Berry connection's module $|\beta_\lambda|$ with $|-i\left\langle g^{1/4}\psi|g^{1/4}\partial_\lambda\psi\right\rangle|$. In every plot, we see that
\[
|\beta_\lambda| \leq \left|-i\langle g^{1/4}\psi|g^{1/4}\partial_\lambda\psi \rangle\right|.
\]
\begin{figure}[!h]
\begin{subfigure}{.3\textwidth}
  \centering
  \includegraphics[width=1.2\linewidth]{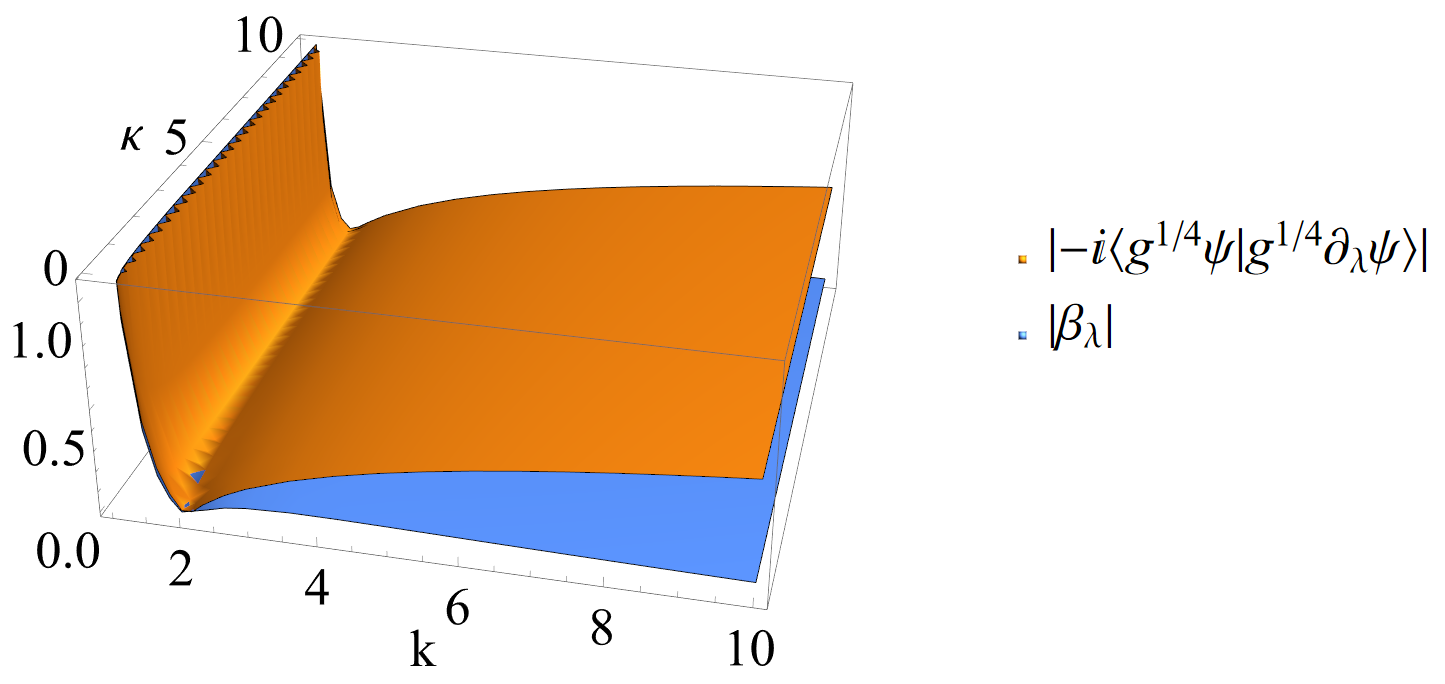}
  \caption{}
  \label{berryl_a}
\end{subfigure}
\hspace{4cm}
\begin{subfigure}{.3\textwidth}
  \centering
  \includegraphics[width=1.2\linewidth]{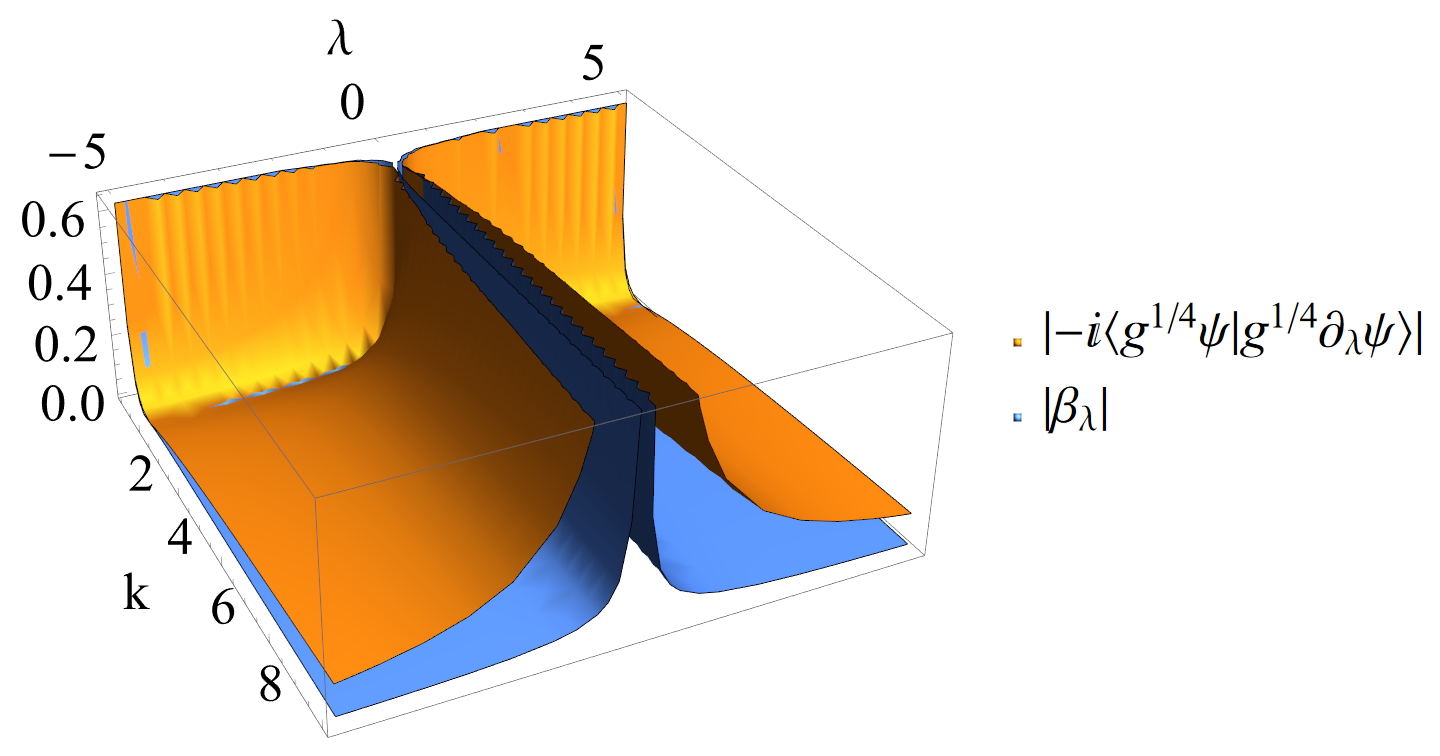}
  \caption{}
  \label{berryl_b}
\end{subfigure}
\par\medskip
\begin{subfigure}{.3\textwidth}
  \centering
  \includegraphics[width=1.2\linewidth]{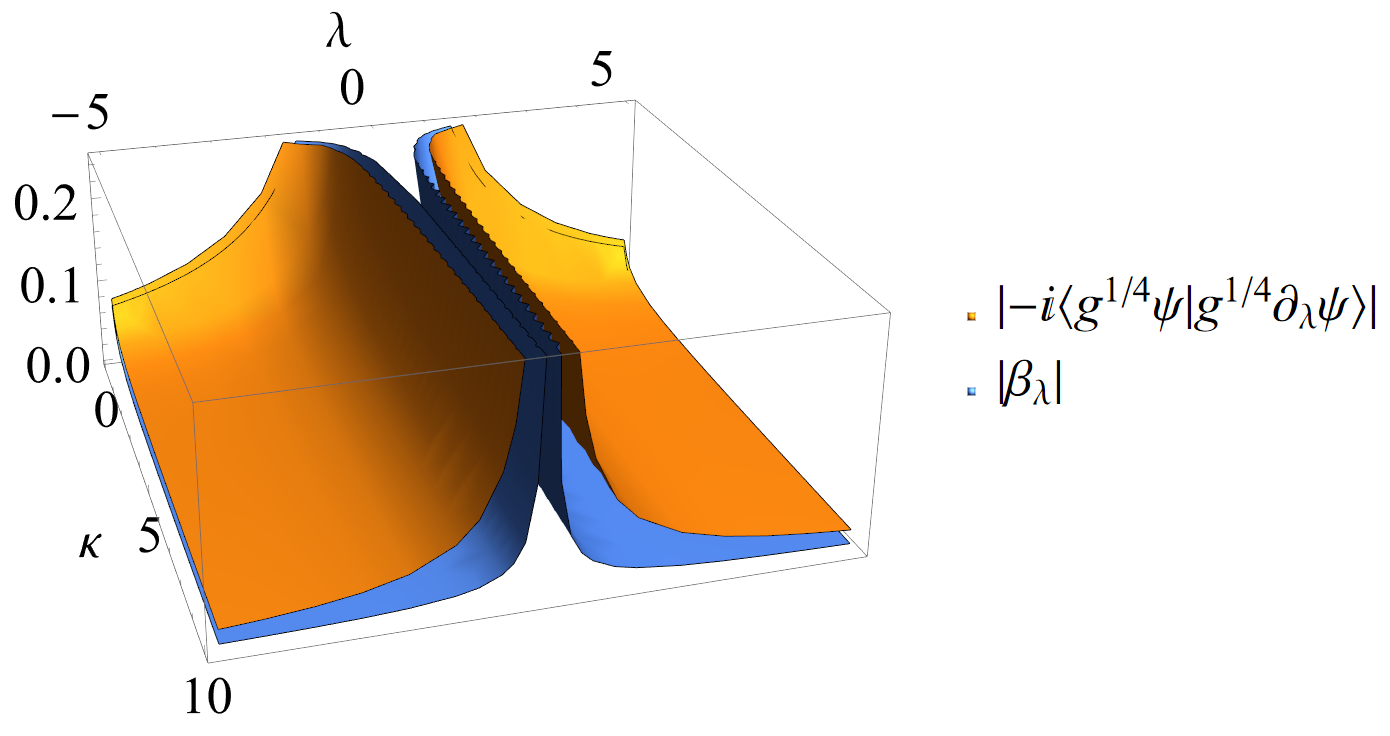}
  \caption{}
  \label{berryl_c}
\end{subfigure}
\hspace{4cm}
\begin{subfigure}{.3\textwidth}
  \centering
  \includegraphics[width=1.2\linewidth]{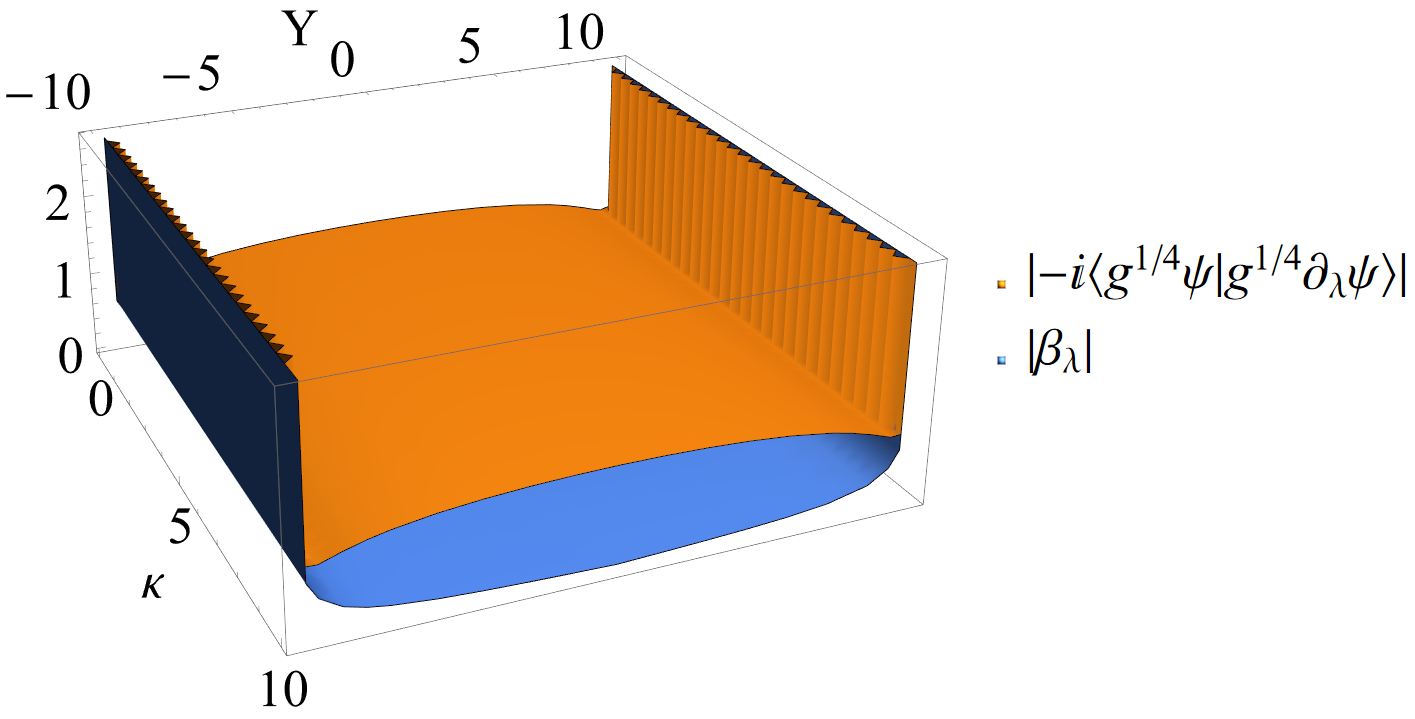}
  \caption{}
  \label{berryl_d}
\end{subfigure}
\par\medskip
\begin{subfigure}{.3\textwidth}
  \centering
  \includegraphics[width=1.2\linewidth]{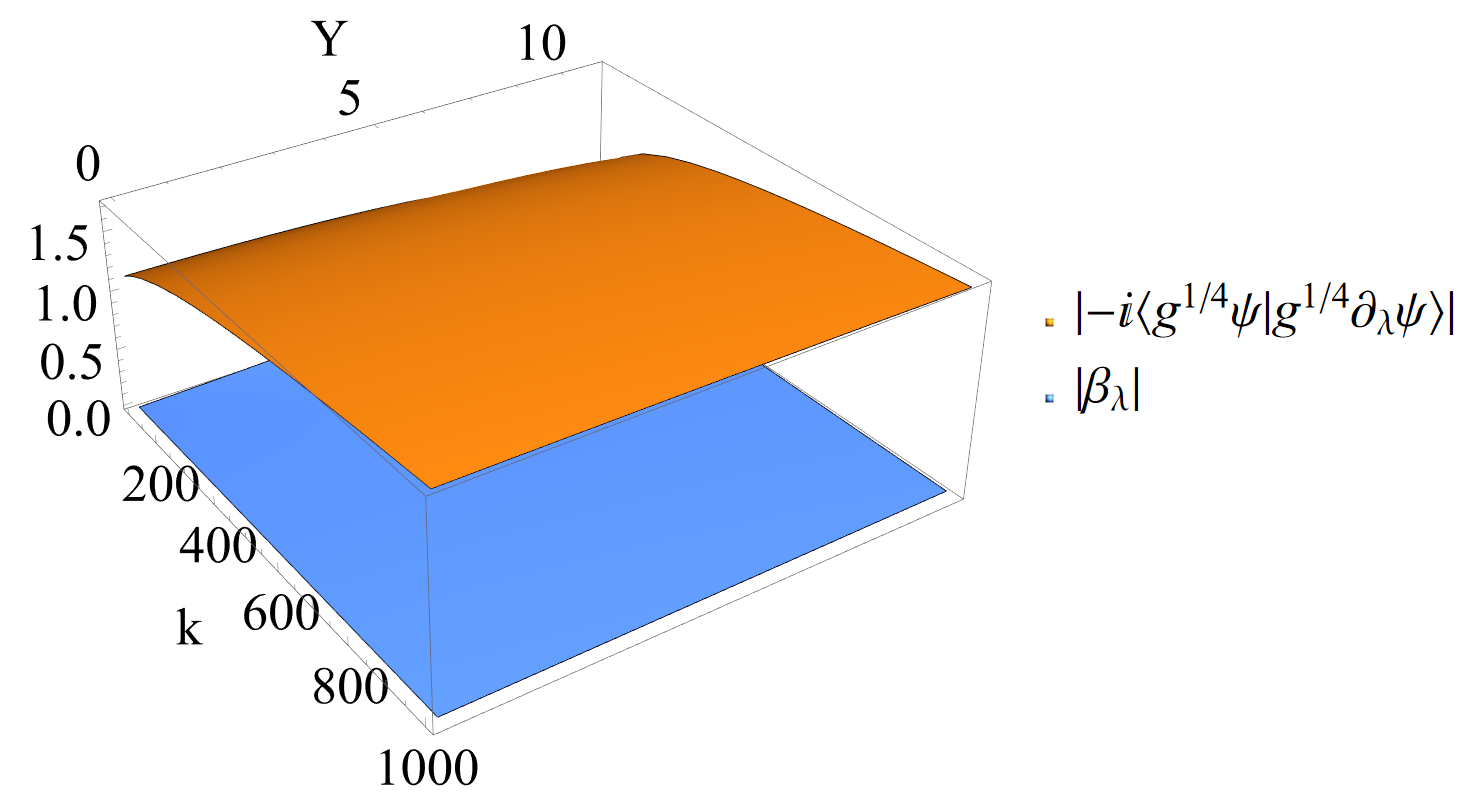}
  \caption{}
  \label{berryl_e}
\end{subfigure}
\hspace{4cm}
\begin{subfigure}{.3\textwidth}
  \centering
  \includegraphics[width=1.2\linewidth]{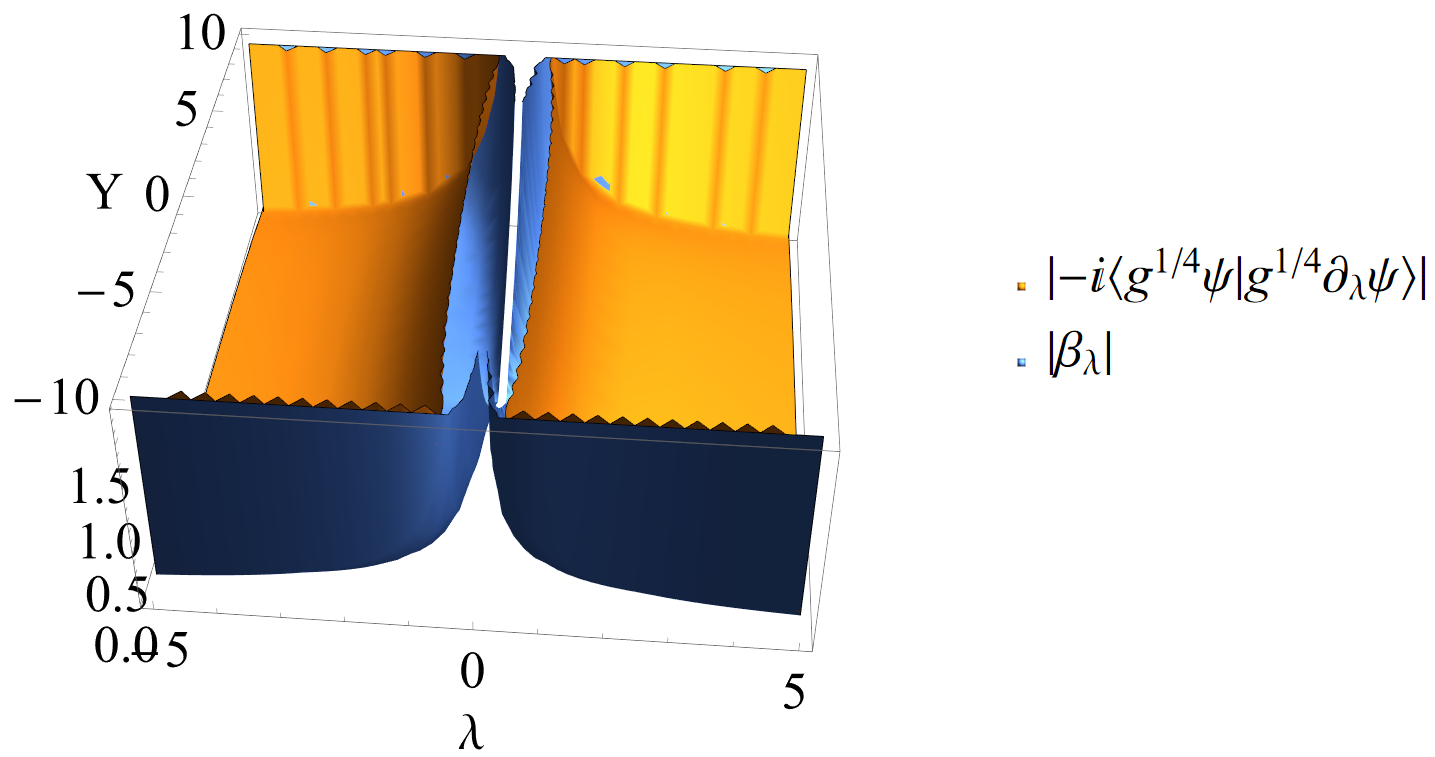}
  \caption{}
  \label{berryl_f}
\end{subfigure}
\caption{\footnotesize Comparison of $|\beta_\lambda|$ and $|-i\left\langle g^{1/4}\psi|g^{1/4}\partial_\lambda\psi\right\rangle|$. In (a), $Y=\lambda = 1$, in (b) $Y= \kappa = 1$, in (c) $\kappa = \lambda = 1$, in (d) $Y = 1, k = 1.5$, in (e) $\lambda = 1, k = 100$, and in (f) $k=100, \kappa = 1$.}
\label{berryl}
\end{figure}
\newpage

Furthermore, we can observe that the modified Berry connection \eqref{eq:modBerryconn} is effectively real, whereas the usual Berry connection $\langle \psi| \partial_\lambda \psi$ is not. This shows that our corrections introduced in the formalism are crucial to consider the curvature of the physical space.

\section{Conclusions}
In this paper, several goals were achieved. First, a perturbative extension of the quantum geometric tensor to the case of a curved space with a parameter-dependent metric was proposed. This new approach involves considering a modification of the completeness condition, and new terms appear in the new formula, which are related to the commutation of the Hamiltonian with the metric and the deformation vector. It is expected that this formula can be very useful from a numerical point of view since, given a known basis, it is possible to express all the new matrix elements and, from there, evaluate the corresponding expression. Subsequently, several examples were analyzed, in which several types of couplings were taken into account showing several phase transitions. One of the main characteristics of the additional terms is that phase transitions are observed more quickly for a given selection of parameters. Additionally, in the example of Section 4.3, it is shown that the analogous Berry connection has an imaginary part if the corresponding corrections are not considered, which clearly shows the usefulness of our formalism. In order to evaluate the quantum geometric tensor given the considered wave functions, it was necessary to implement a technique to analytically evaluate the integrals of the type \eqref{eq:integralTodaModel}. We compared the results with a perturbative method and showed that there was non-perturbative information that was taken into account when performing the analytical calculation. This work settles the path for this tensor to be used in systems such as bilayered graphene.

An aspect worth highlighting is that the determinant of the complete QGT is greater than the determinant of the term analogous to the QGT given originally by Provost and Vallee (both in Toda system, see Fig. \ref{fig:detTodaSym} and in example of Section \ref{sec:generalizado_exponencial}. However, in Section \ref{sec:generalizado_exponencial} the modulus of the whole Berry connection of one of the parameters is less than the term analogous to the Berry connection of Provost and Vallee. 

As future work, one could consider analyzing the case of curved materials as in \cite{Siu2018}, using the perturbative approach of Section \ref{pera}. In addition to cases such as the relativistic harmonic oscillator with curvature as in \cite{Gazeau} or in spaces such as the sphere $S^2$ or the hyperbolic plane $H^2$ as in \cite{Ranada} in order to discover new physical features of these systems that appear as a result of the movement of the parameters. Alternatively, consider systems like \cite{Reuter}, where the statistical information obtained by the quantum geometric tensor can give more information about critical points in systems of interest.


\subsection*{\textbf{Apendix A: Tips for the symmetrically coupled Toda harmonic oscillators}}\label{tipsntricks}

Although the quantum geometric tensor for the examples in sections \ref{sec:osc_acop_expo} and \ref{sec:generalizado_exponencial} can be analytically obtained, this proved to be a computational excruciating and difficult task. Since the exact expressions are far too big to write explicitly in this appendix, we give more explicit instructions on how to calculate them.

 To get the explicit ground state wave function of this system, it is simpler to use the $U_1 = \exp[-\lambda x]$ and $U_2 = \exp[-\beta y]$ variables since the jacobian of this transformation is 
\begin{equation}
    J = \det{\frac{\partial(x,y)}{\partial(U_1,U_2)}} = \frac{1}{\lambda \beta} e^{(\lambda x + \beta y)} \tag{A.1}
\end{equation}

which cancels out the factor $\sqrt{g}$ of the inner product, i.e.
\begin{equation}
    \int_{-\infty}^\infty     \int_{-\infty}^\infty \sqrt{g} dx dy  =     \int_{0}^\infty     \int_{0}^\infty  dU_1 dU_2. \tag{A.2}
\end{equation}
As usual, beware that this coordinate transformation changes the limits of integration.

Now to get the normalization constant $\mathcal{N}_0$ in (\ref{eq:OAESA_psi0}) we integrate
\begin{equation}
    \mathcal{N}_0^2 \int_{0}^\infty     \int_{0}^\infty   dU_1 dU_2 \left[\exp\left(-\frac{\omega_1 U_1^2 + \omega_2 U_2^2}{2} - \gamma U_1 U_2\right)\right]^2=1, \tag{A.3}
\end{equation}
or equivalently
\begin{equation}
    \mathcal{N}_0^2 \int_{0}^\infty dU_+\left[    \int_{-U_+}^{U_+}  dU_- \left[\exp\left(-\frac{\omega_+ U_+^2 + \omega_- U_-^2}{2}\right)\right]^2\right]  =1, \tag{A.4}
\end{equation}
yielding
\begin{equation}
    \mathcal{N}_0 =
    \left[\frac{4 \left(k (k+2 \kappa)\right)^{1/4}}{2 \operatorname{arccot}\left(\frac{\sqrt{2} \left(k (k+2 \kappa)\right)^{1/4}\sqrt{\sqrt{k (k+2 \kappa)}+k+\kappa}}{\kappa}\right)+\pi }\right]^{1/2} =\left[\frac{[k(k+2\kappa)]^{1/4}}{\arctan\left([\frac{2\kappa}{k} + 1]^{1/4}   \right)}  \right]^{1/2} \tag{A.5}
\end{equation}
and thus
\begin{equation}
    \psi_0 (U_1,U_2) = \left[\frac{[k(k+2\kappa)]^{1/4}}{\arctan\left([\frac{2\kappa}{k} + 1]^{1/4}   \right)}  \right]^{1/2}\exp\left(-\frac{\omega_1 U_1^2 + \omega_2 U_2^2}{2} - \gamma U_1 U_2\right), \tag{A.6}
\end{equation}
\begin{equation}
    \psi_0 (U_+,U_-) = \left[\frac{[k(k+2\kappa)]^{1/4}}{\arctan\left([\frac{2\kappa}{k} + 1]^{1/4}   \right)}  \right]^{1/2}\exp\left(-\frac{\omega_+ U_+^2 + \omega_- U_-^2}{2}\right). \tag{A.7}
\end{equation}

\subsection*{\textbf{Apendix B: Projections of the QGT components for the symmetrically coupled Toda oscillators}}\label{moregraphics}
\begin{figure}[H]
\begin{subfigure}{.25\textwidth}
  \centering
  \includegraphics[width=.8\linewidth]{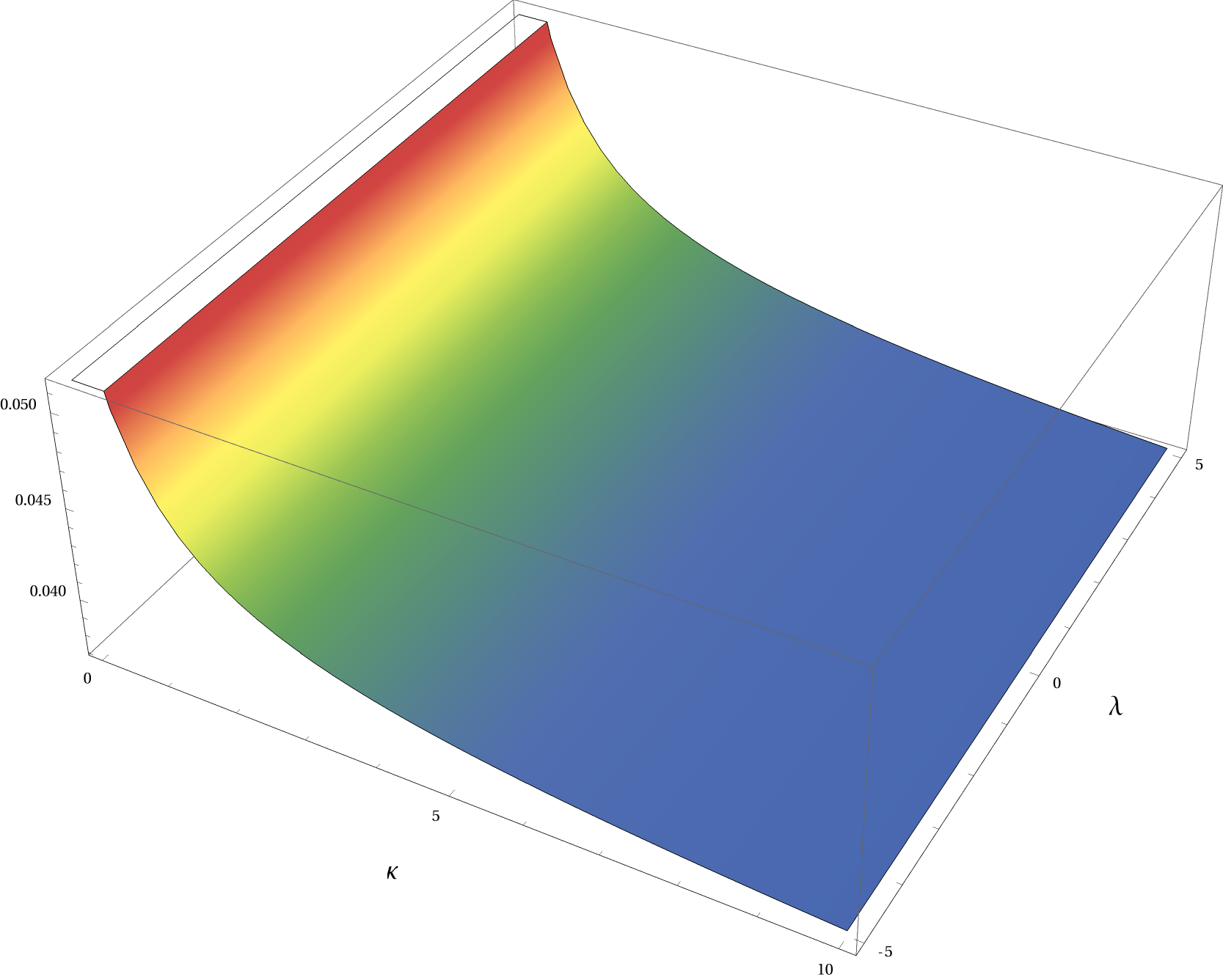}
  \caption{$G_{kk}$}
  \label{fig:coupledkpa1}
\end{subfigure}
\hfill
\begin{subfigure}{.25\textwidth}
  \centering
  \includegraphics[width=.8\linewidth]{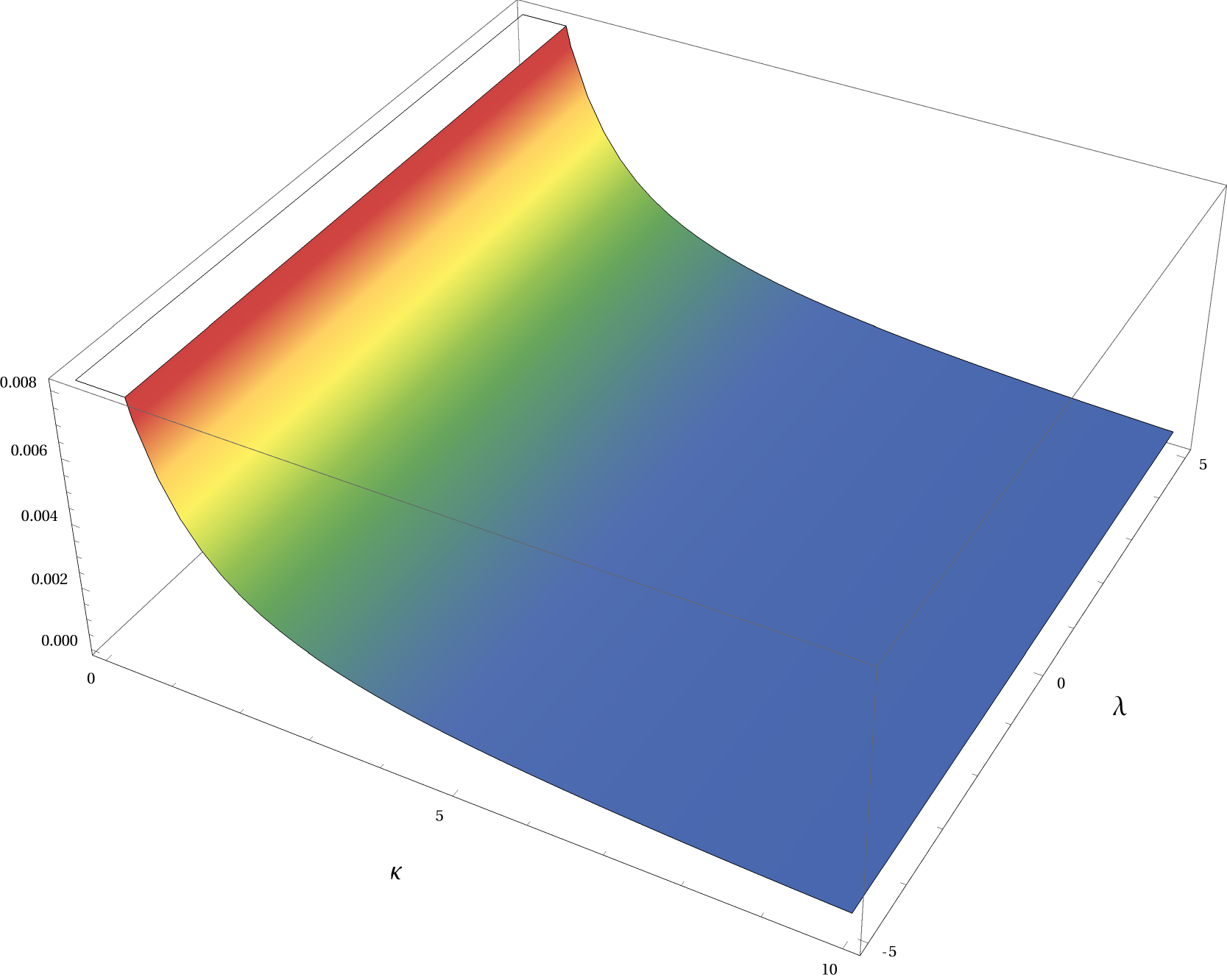}
  \caption{$G_{k\kappa}$}
  \label{fig:coupledkpa2}
\end{subfigure}
\hfill
\begin{subfigure}{.25\textwidth}
  \centering
  \includegraphics[width=.8\linewidth]{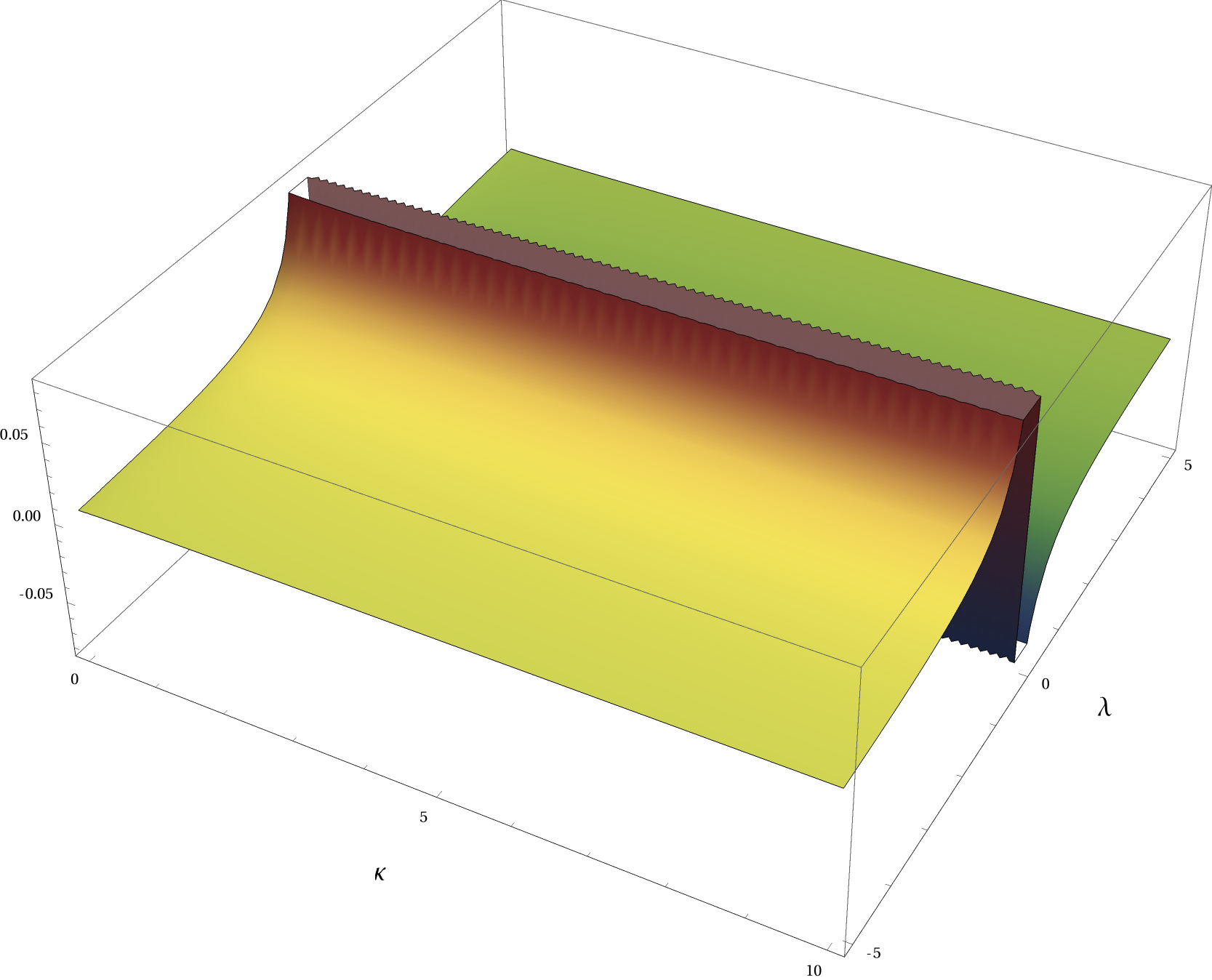}
  \caption{$G_{k\lambda}$}
  \label{fig:coupledkpa3}
\end{subfigure}
\\
\begin{subfigure}{.25\textwidth}
  \centering
  \includegraphics[width=.8\linewidth]{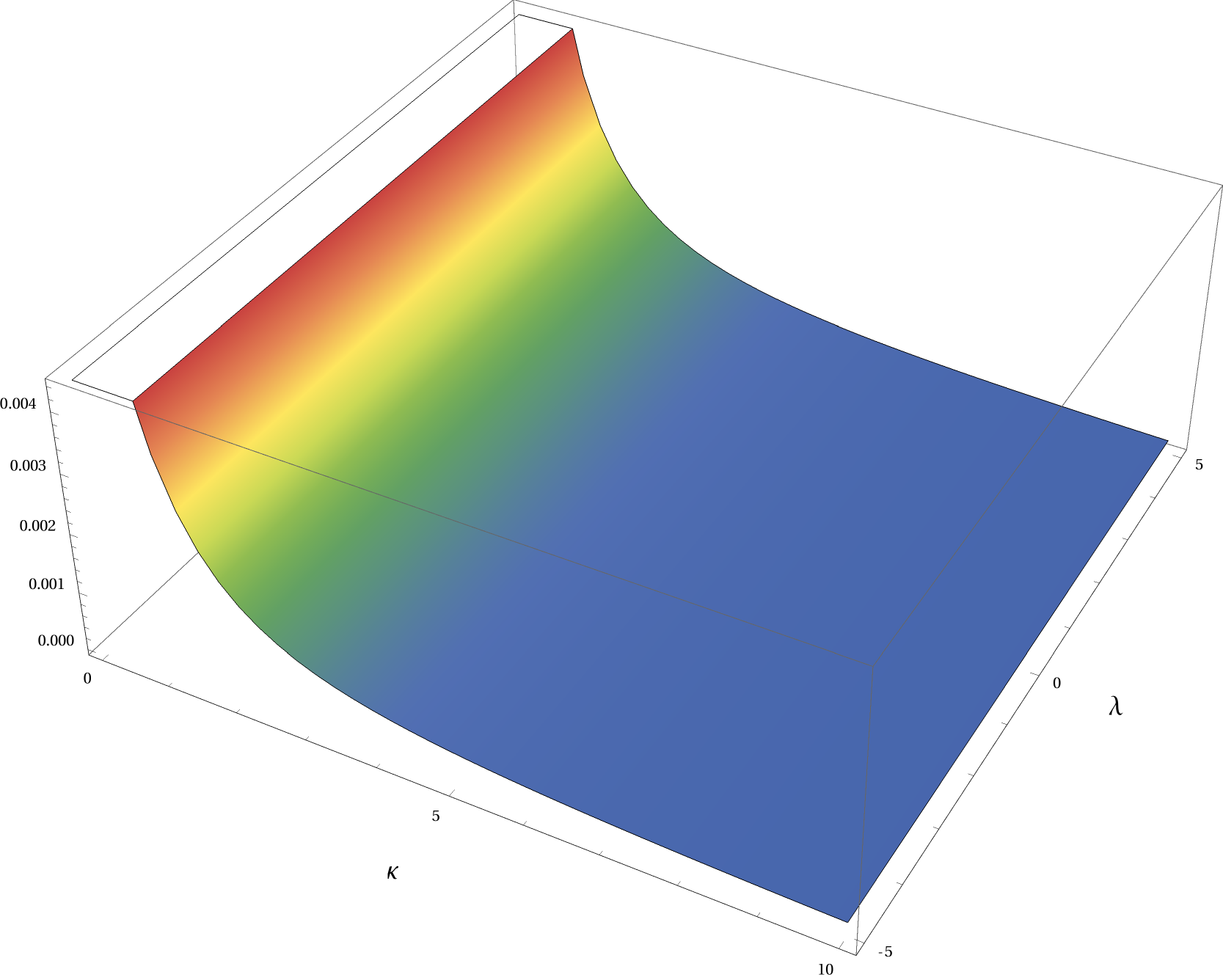}
  \caption{$G_{\kappa\kappa}$}
  \label{fig:coupledkpa4}
\end{subfigure}
\hfill
\begin{subfigure}{.25\textwidth}
  \centering
  \includegraphics[width=.8\linewidth]{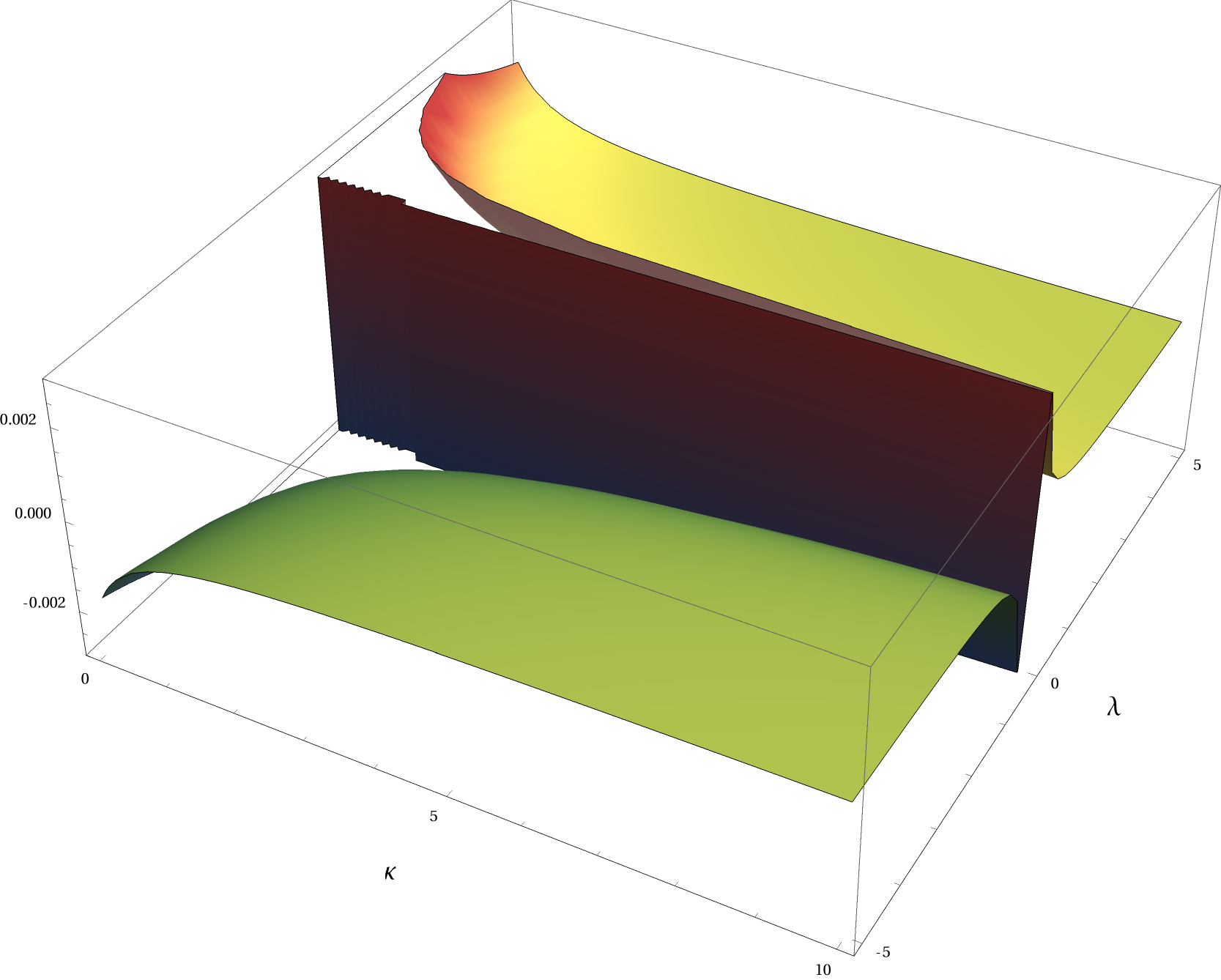}
  \caption{$G_{\kappa\lambda}$}
  \label{fig:coupledkpa5}
\end{subfigure}
\hfill
\begin{subfigure}{.25\textwidth}
  \centering
  \includegraphics[width=.8\linewidth]{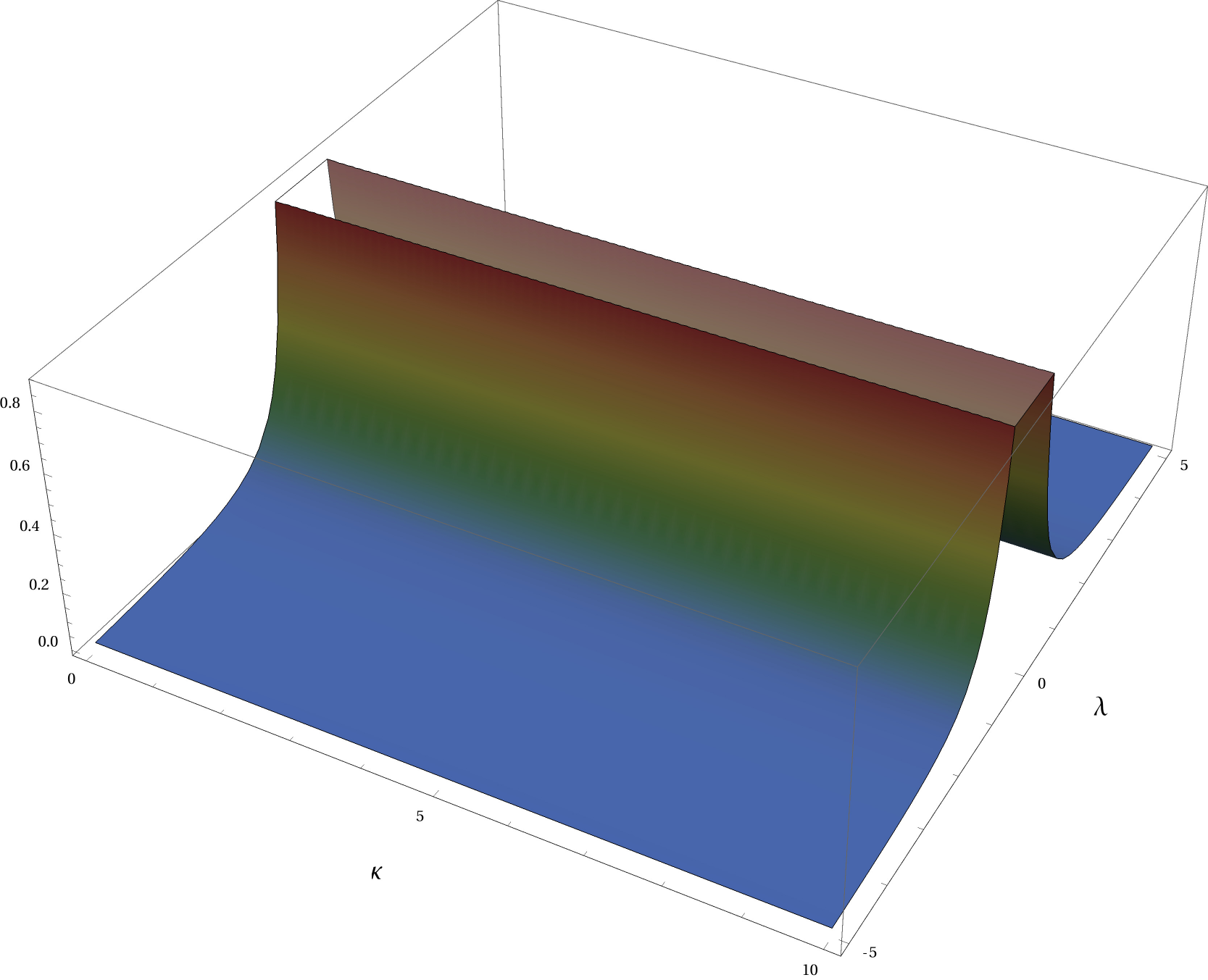}
  \caption{$G_{\lambda\lambda}$}
  \label{fig:coupledkpa6}
\end{subfigure}
\\
\begin{subfigure}{.25\textwidth}
  \centering
  \includegraphics[width=.8\linewidth]{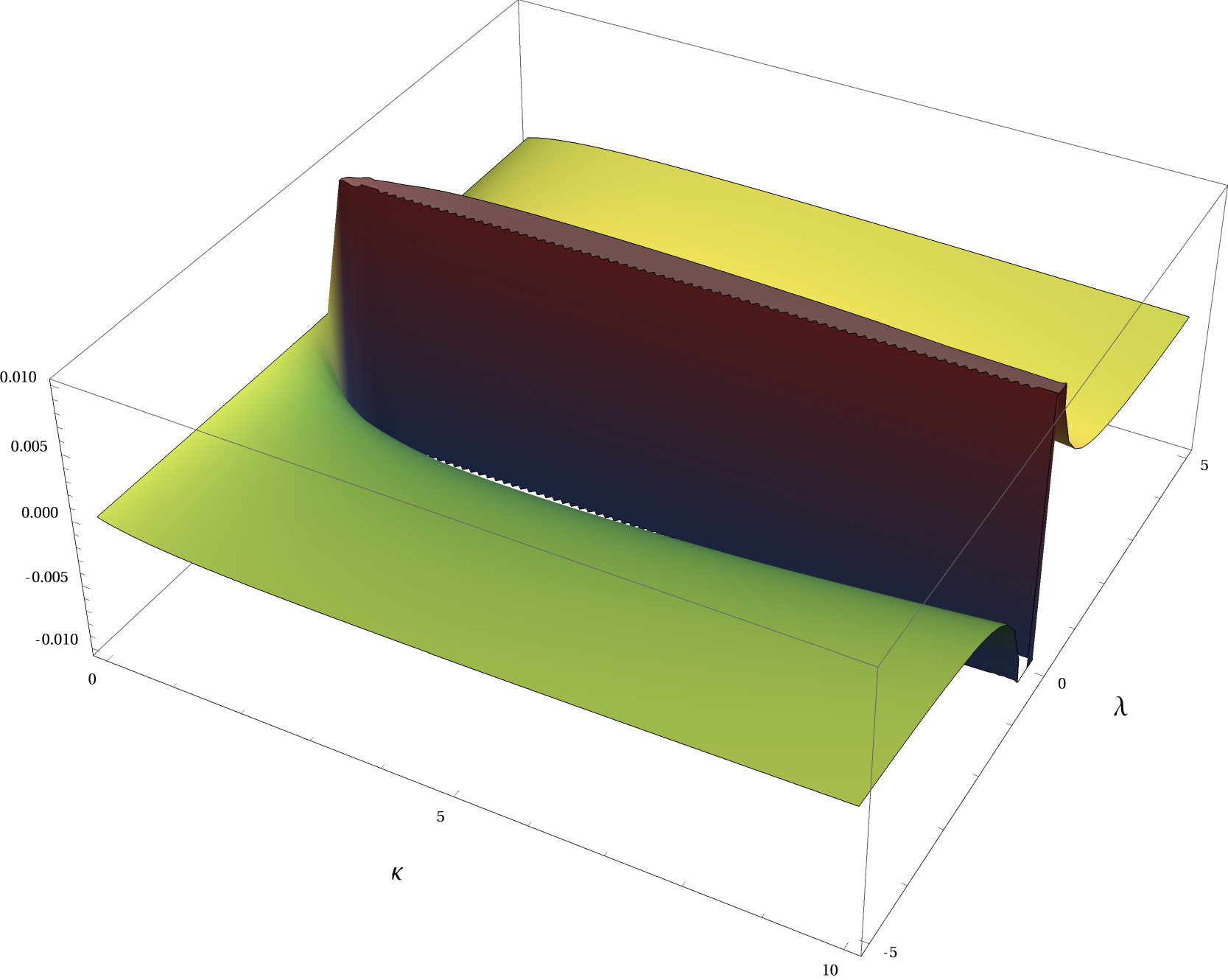}
  \caption{$G_{\lambda\beta}$}
  \label{fig:coupledkpa7}
\end{subfigure}
\hfill
\begin{subfigure}{.25\textwidth}
  \centering
  \includegraphics[width=.8\linewidth]{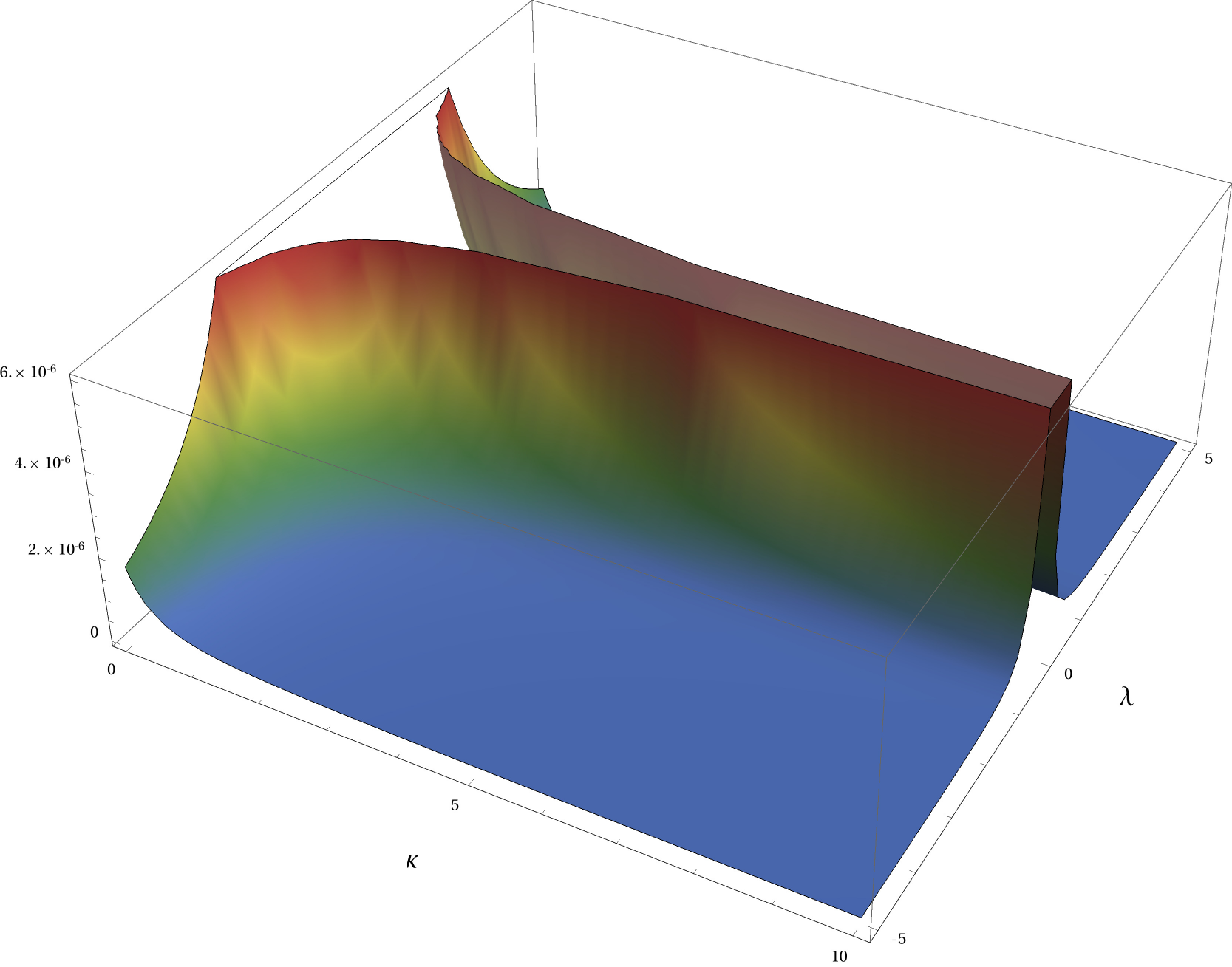}
  \caption{$\det G$}
  \label{fig:coupledkpa8}
\end{subfigure}
\hfill
\begin{subfigure}{.25\textwidth}
  \centering
  \includegraphics[width=.8\linewidth]{blank.pdf}
\end{subfigure}
\caption{QGT components in terms of $(\kappa,\,\lambda)$ with $k=\beta=1$.}
\label{fig:coupledkpa}
\end{figure}
\begin{figure}[H]
\begin{subfigure}{.25\textwidth}
  \centering
  \includegraphics[width=.8\linewidth]{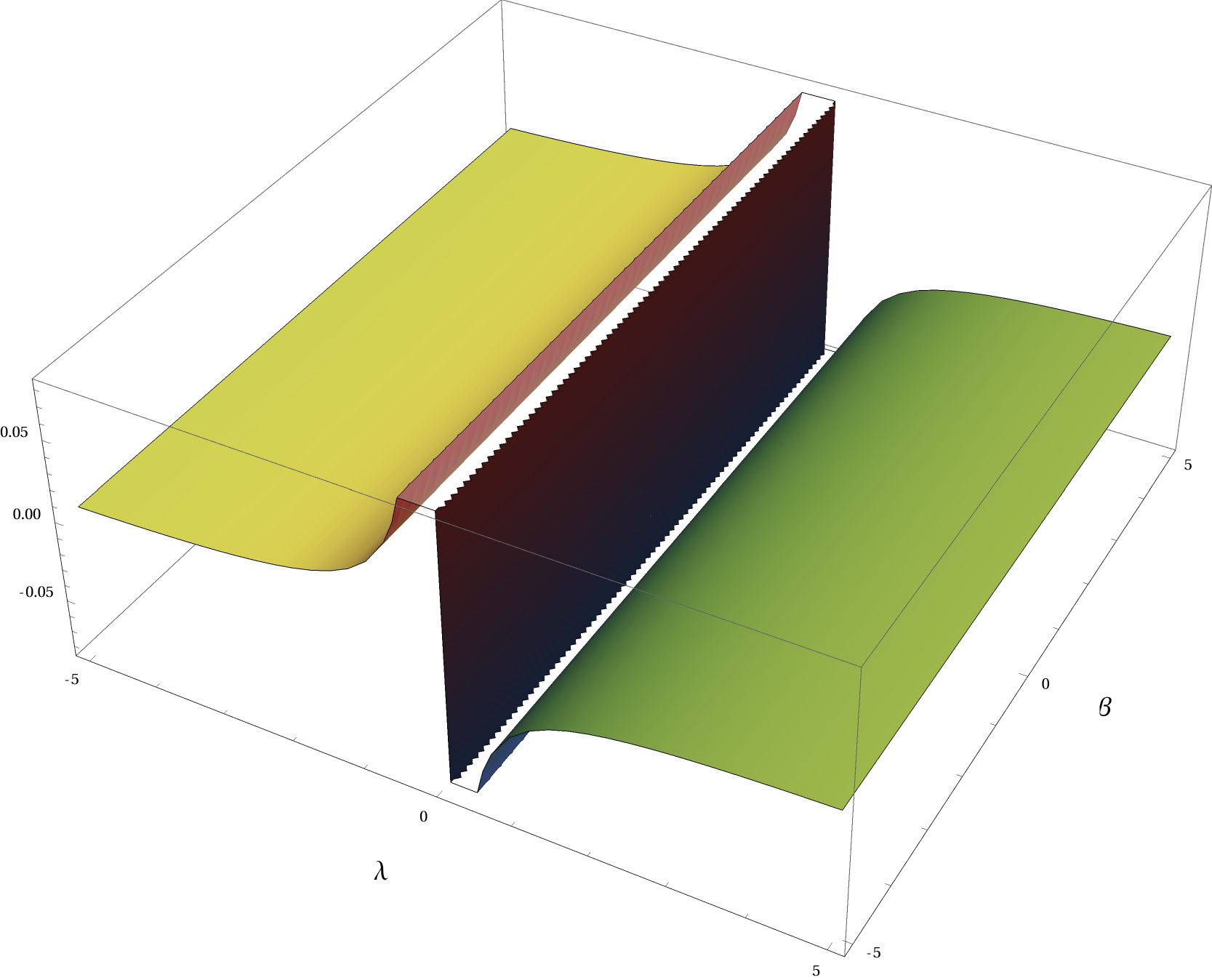}
  \caption{$G_{k\lambda}$}
  \label{fig:coupledaa3}
\end{subfigure}
\hfill
\begin{subfigure}{.25\textwidth}
  \centering
  \includegraphics[width=.8\linewidth]{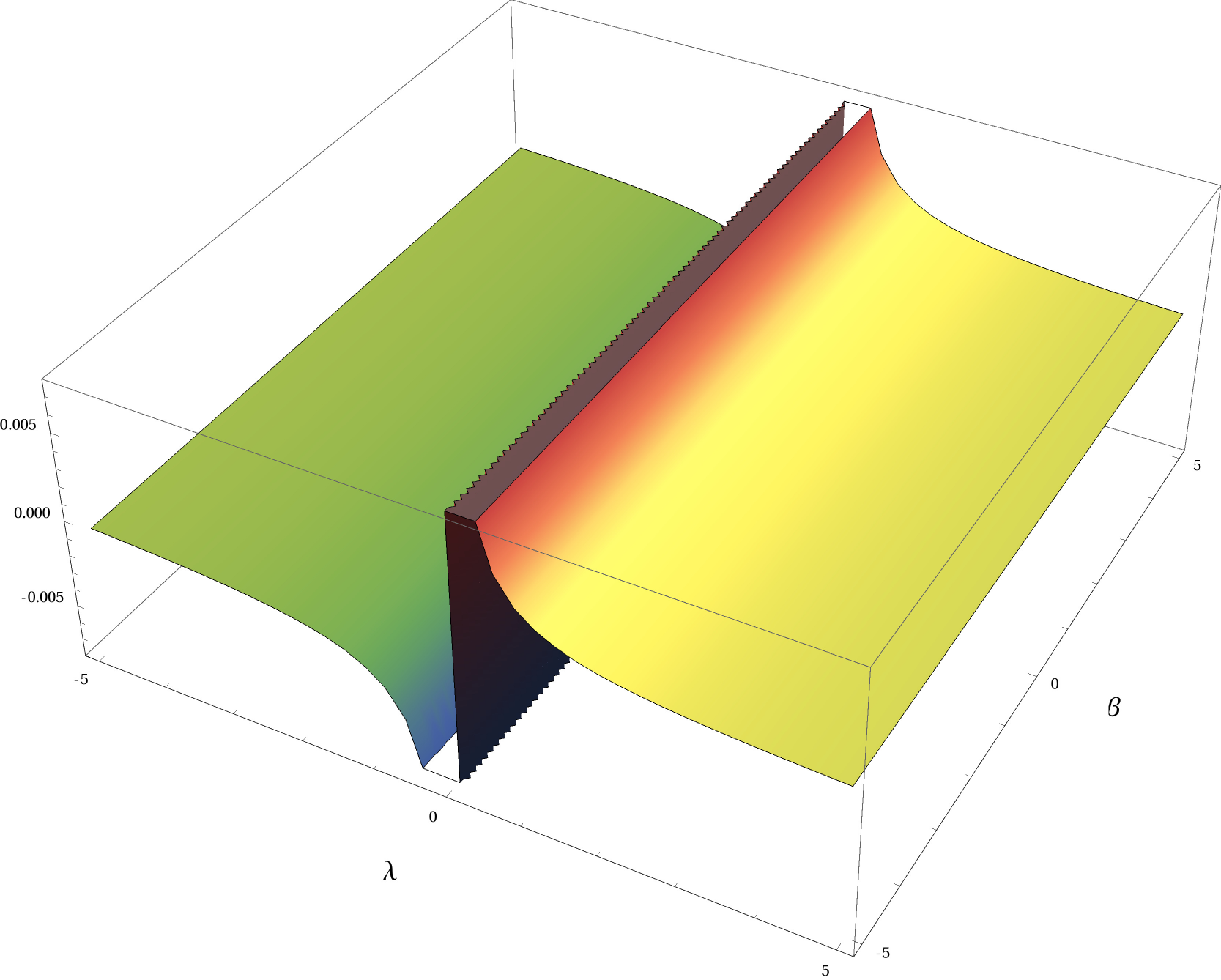}
  \caption{$G_{\kappa\lambda}$}
  \label{fig:coupledaa5}
\end{subfigure}
\hfill
\begin{subfigure}{.25\textwidth}
  \centering
  \includegraphics[width=.8\linewidth]{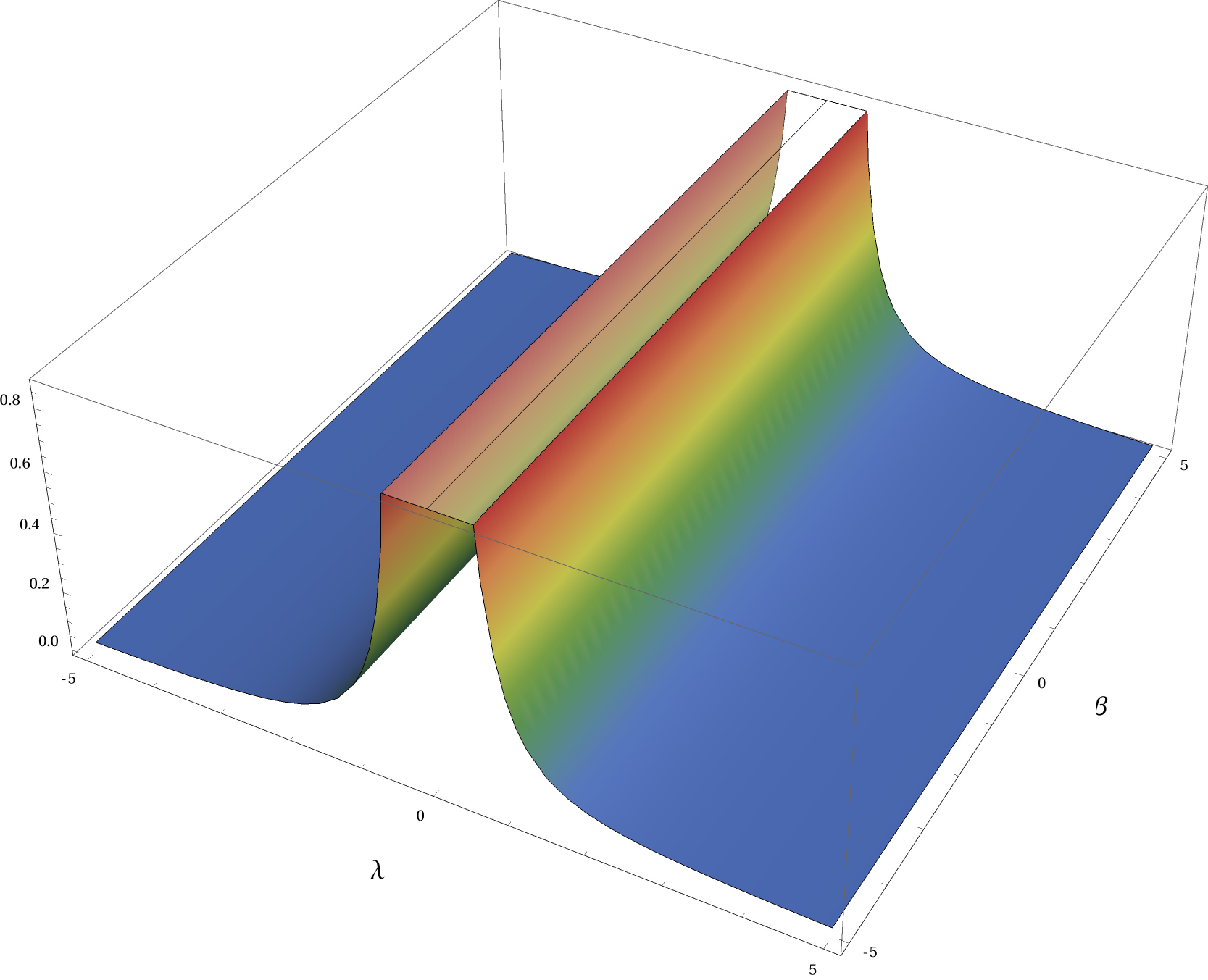}
  \caption{$G_{\lambda\lambda}$}
  \label{fig:coupledaa6}
\end{subfigure}
\\
\begin{subfigure}{.25\textwidth}
  \centering
  \includegraphics[width=.8\linewidth]{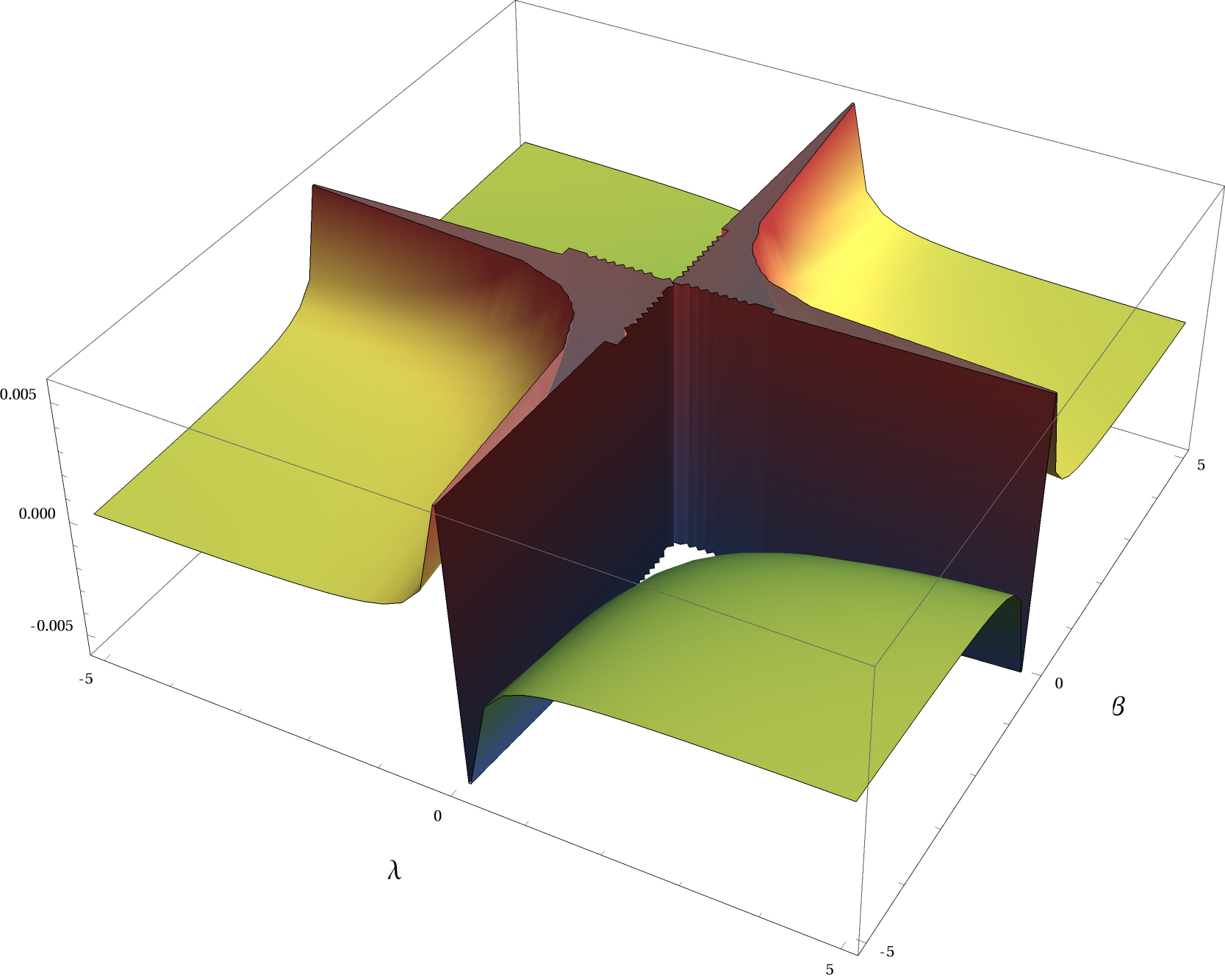}
  \caption{$G_{\lambda\beta}$}
  \label{fig:coupledaa7}
\end{subfigure}
\hfill
\begin{subfigure}{.25\textwidth}
  \centering
  \includegraphics[width=.8\linewidth]{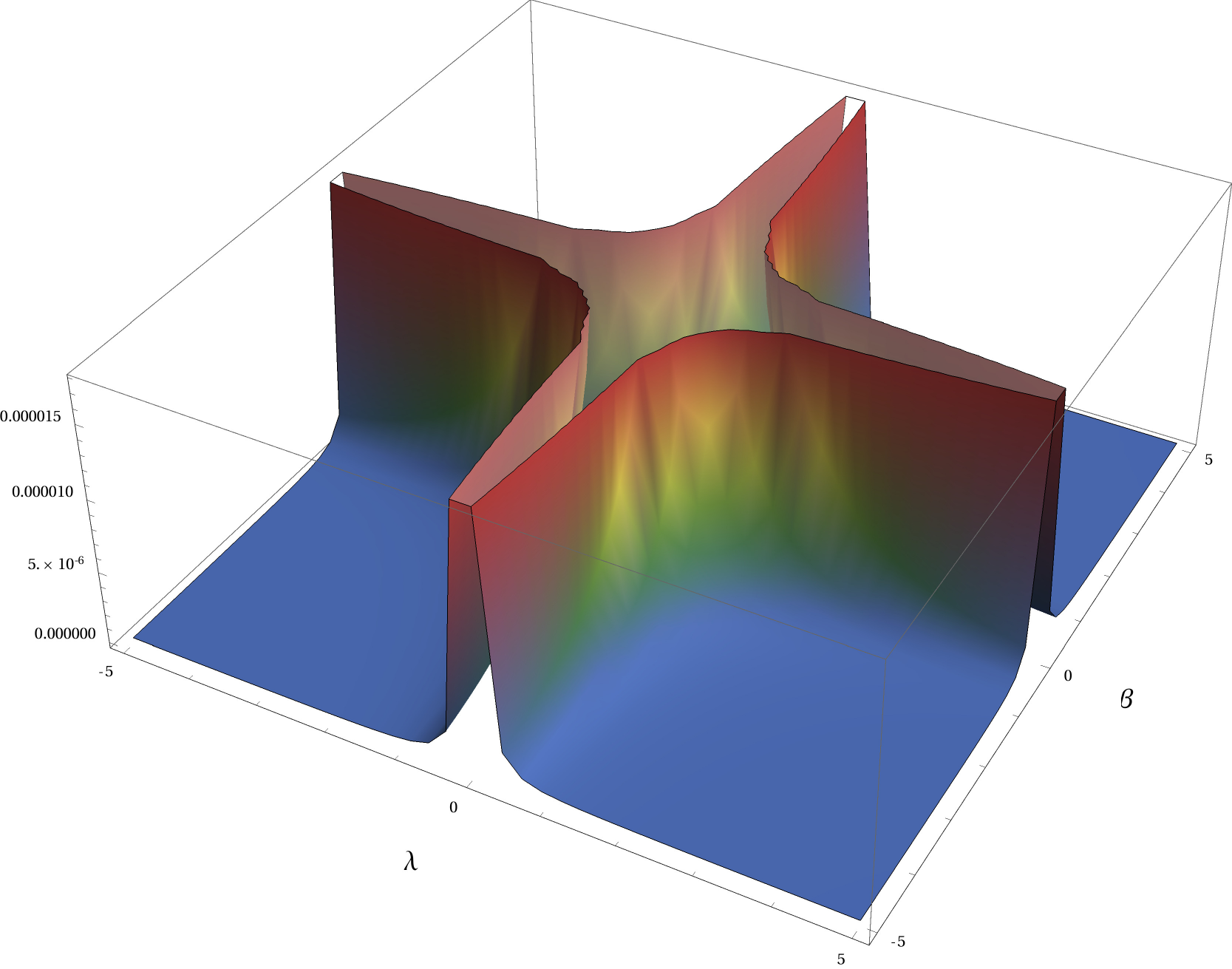}
  \caption{$\det[G]$}
  \label{fig:coupledaa8}
\end{subfigure}
\hfill
\begin{subfigure}{.25\textwidth}
  \centering
  \includegraphics[width=.8\linewidth]{blank.pdf}
\end{subfigure}
\caption{QGT components in terms of $(\lambda,\,\beta)$ with $k=\kappa=1$. The components $G_{kk},\,G_{k\kappa}$ and $G_{\kappa\kappa}$  are zero.}
\label{fig:coupledaa}
\end{figure}

\section*{\textbf{Eigenvalues of the QGT for the anharmonic oscillator coupled with a Toda oscillator}}\label{graphicseigen}
\begin{figure}[H]
\begin{subfigure}{.48\textwidth}
  \centering
  \includegraphics[width=.9\linewidth]{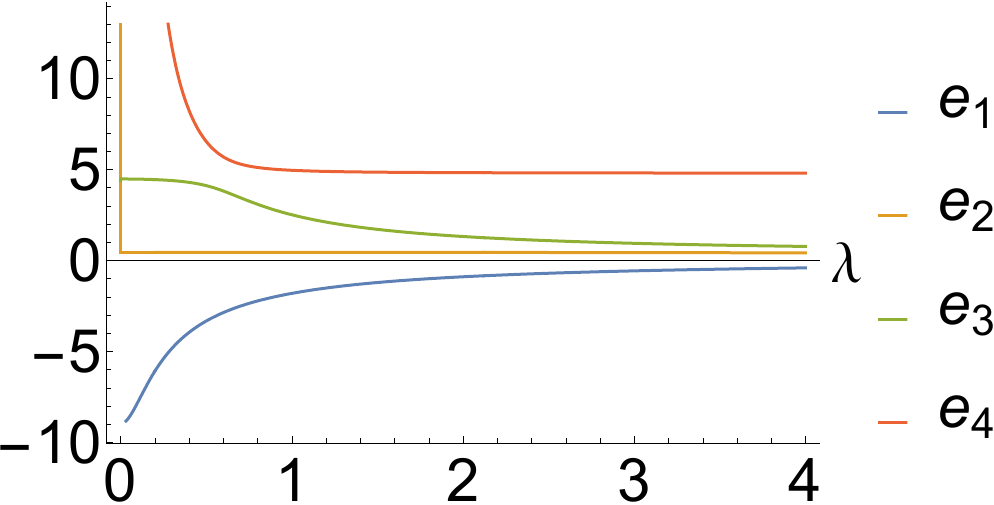}
  \caption{}
  \label{fig:eigen_a}
\end{subfigure}
\hfill
\begin{subfigure}{.48\textwidth}
  \centering
  \includegraphics[width=.9\linewidth]{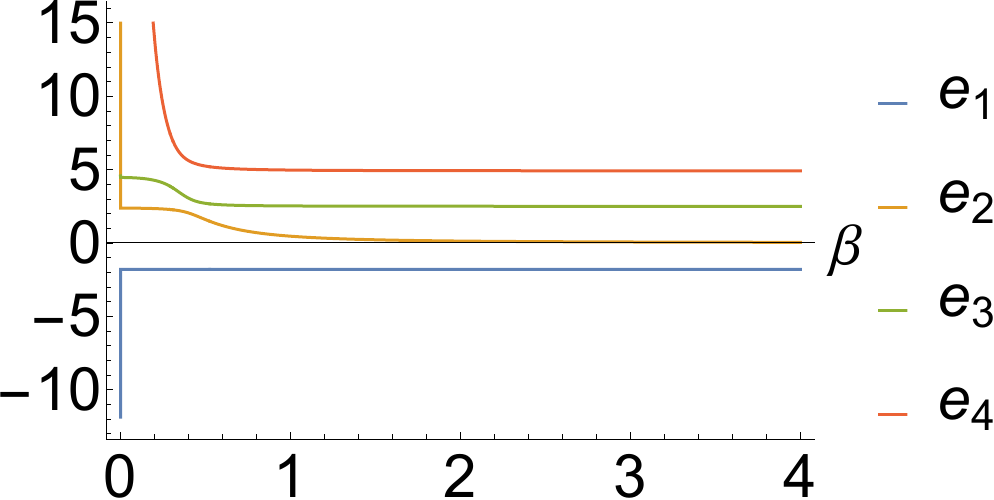}
  \caption{}
  \label{fig:eigen_b}
\end{subfigure}
\\
\begin{subfigure}{.48\textwidth}
  \centering
  \includegraphics[width=.9\linewidth]{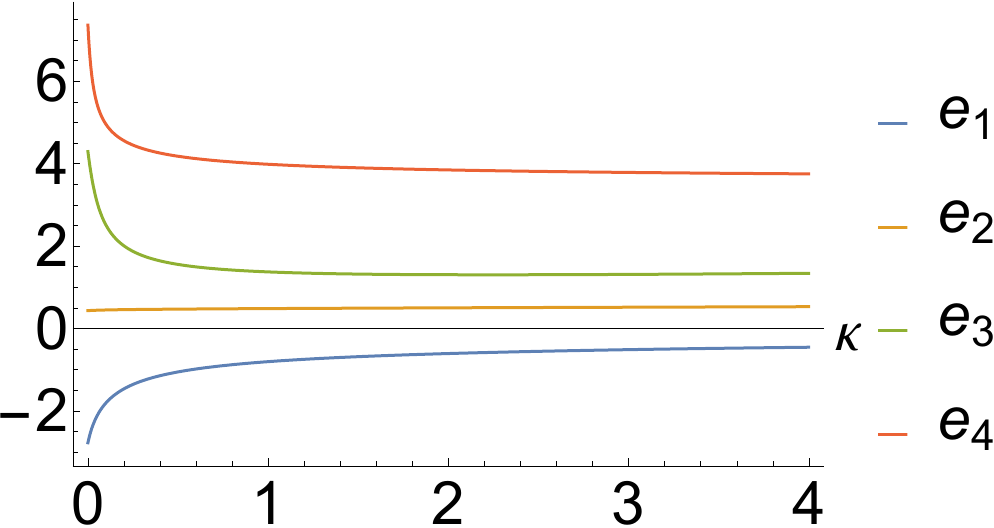}
  \caption{}
  \label{fig:eigen_kp}
\end{subfigure}
\begin{subfigure}{.48\textwidth}
  \centering
  \includegraphics[width=.9\linewidth]{blank.pdf}
\end{subfigure}
\caption{Behaviour of the QGT eigenvalues. For $(a)$ we use $\lambda$ like variable and $k=0.1,\,\kappa=0.1,\,\beta=1$, for $(b)$ we use $\beta$ like variable and $k=0.1,\,\kappa=0.1,\,\lambda=1$ and for $(c)$ we use $\kappa$ like variable and $k=0.1,\,\lambda=1,\,\beta=1$.}
\end{figure}
\section*{Acknowledgments}

This work was partially supported by DGAPA-PAPIIT Grant No. IN105422.

\bibliographystyle{ws-ijqi}
\bibliography{References.bib}
\end{document}